\newcommand{\bom}{\boldmath}
\newcommand{\be}{\begin{equation}}
\newcommand{\ee}{\end{equation}}
\newcommand{\ber}{\begin{eqnarray}}
\newcommand{\eer}{\end{eqnarray}}
\def\fm3{fm$^{-3}$}

\def\ni{\noindent}

\documentstyle[12pt,aps,epsf]{revtex}

\begin{document}
\draft
\preprint{ SUNY-NTG-98-35 }
\title{Effects of Strong and Electromagnetic Correlations on \\
Neutrino Interactions in Dense Matter}
\author {Sanjay Reddy$^{1,2}$, Madappa Prakash $^1$, James M. Lattimer $^1$,
and Jose A. Pons$^3$}
\address{$^1$Department of Physics \& Astronomy, SUNY at Stony Brook,
Stony Brook, New York 11794-3800 \\
$^2$ Institute For Nuclear Theory, University of Washington, Seattle,
WA 98195 \\
$^3$ Department d'Astronomia, Universitat de Val\`encia,
E-46100 Burjassot, Val\`encia, Spain \\}
\date{\today}
\maketitle
\begin{abstract}
An extensive study of the effects of correlations on both charged and
neutral current weak interaction rates in dense matter is performed.
Both strong and electromagnetic correlations are considered.  The
propagation of particle-hole interactions in the medium plays an
important role in determining the neutrino mean free paths.  The
effects due to Pauli-Blocking and density, spin, and isospin
correlations in the medium significantly reduce the neutrino cross
sections.  Due to the lack of experimental information at high
density, these correlations are necessarily model dependent. For
example, spin correlations in nonrelativistic models are found to lead
to larger suppressions of neutrino cross sections compared to those of
relativistic models.  This is due to the tendency of the
nonrelativistic models to develop spin instabilities.  Notwithstanding
the above caveats, and the differences between nonrelativistic and
relativistic approaches such as the spin- and isospin-dependent
interactions and the nucleon effective masses, suppressions of order
2--3, relative to the case in which correlations are ignored, are
obtained.  Neutrino interactions in dense matter are especially
important for supernova and early neutron star evolution calculations.
The effects of correlations for protoneutron star evolution are
calculated.  Large effects on the internal thermodynamic properties of
protoneutron stars, such as the temperature, are found.  These
translate into significant early enhancements in the emitted neutrino
energies and fluxes, especially after a few seconds.  At late times,
beyond about 10 seconds, the emitted neutrino fluxes decrease more
rapidly compared to simulations without the effects of correlations,
due to the more rapid onset of neutrino transparency in the
protoneutron star.\\

\ni PACS numbers(s): 13.15.+g, 26.60.+c, 97.60.Jd
\end{abstract}
\newpage
\def\gsim{\lower 2pt \hbox{$\, \buildrel {\scriptstyle >}\over
{\scriptstyle \sim}\,$}}
\def\lsim{\lower 2pt \hbox{$\, \buildrel {\scriptstyle <}\over
{\scriptstyle \sim}\,$}}

\newcommand\simgreater{\buildrel > \over \sim}
\newcommand\simless{\buildrel < \over \sim}

\newpage
\section{INTRODUCTION}

The calculation of neutrino interactions in hot, dense matter is
highly relevant to the study of supernovae and protoneutron stars.
Neutrinos are thought to be intimately involved in the supernova
explosion mechanism.  In addition, the timescales over which
protoneutron stars deleptonize and cool, which are crucial in
predicting the neutrino light curve from a supernova, are determined
by neutrino opacities.  In an earlier paper \cite{RPL}, we developed a
formalism to determine both neutral and charged current opacities for
interacting matter, accounting for in-medium mass and energy shifts
given by its underlying equation of state (EOS).  The effects of
degeneracy and relativity were incorporated, and nuclear interactions
modelled by both nonrelativistic potential and relativistic
field-theoretical approaches were studied.  In this paper, we 
extend this treatment to include important sources of suppression and
enhancement due to in-medium correlations.  In particular, we 
include effects due to both strong and electromagnetic
correlations in the system. While neutrinos can couple to
many-particle states and induce multi-pair excitations, both of which
could modiy the response function developed in Ref. \cite{RPL},
single-pair excitations dominate over multi-pair excitations for the
kinematics of interest in neutrino scattering and absorption. Thus, we
focus on the effects of correlations on the single pair excitation
spectrum.

Pioneering work in calculating neutrino mean free path in uniform
nuclear matter was performed by Sawyer \cite{S1} who showed that the
effects due to strong interactions are important and that the relation
between the EOS and long wavelength excitations of the system may be
exploited to compute the weak interaction rates consistent with the
underlying dense matter model. Iwamoto and Pethick \cite{IP} computed
the neutrino mean free paths in pure neutron matter for the densities
of relevance for neutron stars by calculating the dynamical response
within the framework of Fermi-liquid theory
\cite{PB,L1,N,BP}. Horowitz and Wehrberger \cite{HW} studied the
influence of correlations on the neutrino scattering in dense matter
in a relativistic field-theoretical model.  More recently, Fabbri and
Matera \cite{FM} have calculated neutrino scattering in asymmetric
nuclear matter including the effects of the exchange interaction.

These earlier works on neutrino interactions in dense matter have
shaped our approach.  First accounts of our results may be found in
Refs. \cite{RPPL,PR}.  We urge the reader to consult the simultaneous
and independent work of Burrows and Sawyer in Ref. \cite{BS}, with
which our work has some overlap.  However, important differences, both
in the microphysical inputs to the opacity calculations and in the
underlying EOS models, distinguish our approach and results from
theirs.  We will present a unified approach to calculate both the
neutral current scattering and charged current absorption reactions
for the composition and temperatures of relevance to the PNS
evolution.  A systematic analysis of the various nuclear physics
inputs such as the particle-hole (abbreviated p-h) interactions in
density, spin and isospin channels is undertaken keeping in view the
known nuclear ground state and excited state properties. We will
isolate the important sources of enhancement and/or suppression due to
in-medium correlations and highlight the important role of spin and
isospin dependent p-h interactions.  The need for further efforts to
pin down the poorly known spin-dependent interactions is clearly
brought out by our investigations.

The neutrino cross sections and the EOS are intimately related.  This
relationship is most transparent in the long-wavelength or static limit, in
which the response of a system to a weak external probe is completely
determined by the ground state thermodynamics (EOS). Thus, in this limit,
neutrino opacities which are consistent with a given  EOS \cite{S1} can be
calculated. However, when the energy and momentum transferred by the neutrinos
are large, full consistency is more difficult to achieve.  Particle-hole and
particle-particle  interactions \cite{RPPL,PR,BS}, multi-pair excitations
\cite{HR}, and the renormalization of the axial charge \cite{WR,BRHO} are
examples of effects that may influence both the EOS and the opacity.  In this
work, we investigate the influence of particle-hole excitations utilizing the
Random Phase Approximation (RPA), in which ring diagrams are summed to all
orders.  This may be viewed as the first and minimal step towards achieving
self-consistency between the response of matter to neutrino-induced
perturbations and the underlying EOS, similar to what has long been established
in the study of collective excitations in nuclei (multipole giant resonances)
through electromagnetic probes \cite{BT}.

Throughout this paper, we employ both nonrelativistic potential and
relativistic field-theoretical models for the EOS and the evaluation of the
neutrino-matter cross sections.  This is done in order to evaluate effects that
are generic as well as effects that are due to the precise description of the
nucleon-nucleon interactions.  In Sec. II, we study the response of pure
neutron matter and calculate the neutrino cross sections  to gauge the effects
due to RPA correlations.  In Sec. III, we calculate the influence of many-body
correlations on the scattering and absorption reactions in both nuclear matter
and multi-component  stellar matter (charge neutral, $\beta$-equilibrated
matter).  In Sec. IV, we compute protoneutron star evolutions in order to
evaluate the effects of  correlations.  Our conclusions and outlook are
contained in Sec. V. Explicit formulae for the various p-h interactions and
polarization functions needed to calculate the neutrino cross sections are
collected in Appendices A and B, respectively.

\section{NEUTRAL CURRENT NEUTRINO CROSS SECTIONS IN NEUTRON MATTER}
We begin with pure neutron matter
to illustrate the various strong interaction
correlation effects that influence the neutrino mean free paths in
dense matter.
\subsection{Nonrelativistic Models}
In the nonrelativistic limit for the neutrons, the cross section per
unit volume
or the inverse mean free path
for the neutral current reaction $\nu + n \rightarrow \nu + n$ is
given by \cite{IP}
\begin{eqnarray}
\frac{1}{V}\frac{d^3\sigma(E_1)}{d\Omega^2~dq_0} &=& \frac{G_F^2}{8\pi^3}
\quad E_3^2\quad [1-f_{\nu}(E_3)] \nonumber \\
&\times&\left[c_V^2(1+\cos{\theta})S_{00}(q_0,q) +
c_A^2(3-\cos{\theta})S_{10}(q_0,q)\right] \,,
\label{nrdsig}
\end{eqnarray}
where $c_V=0.5$ and $c_A=0.615$ are the neutral current vector and
axial vector couplings of the neutron and $G^2_F=6.89\times
10^{-10}~{\rm MeV}^{-5}{\rm m}^{-1}$, $E_1$ and $E_3$ are the incoming
and outgoing neutrino energies, $q_0=E_1-E_3$ and
$\vec{q}=\vec{k_1}-\vec{k_3}$ are the energy and momentum transfers in
the scattering reaction.  The neutron matter response, which depends
on the kinematical variables $q_0$ and $q=|\vec{q}|$, is characterized
by the function $S(q_0,q)$ and is called the dynamic form
factor. Since neutrinos couple to the vector and axial vector currents
of the neutron, the response is given in terms of the density-density
correlation function $S_{00}(q_0,q)$ and the spin-density correlation
function $S_{10}(q_0,q)$.  The subscripts appearing in the response
function $S_{\sigma \tau}$ represent the total spin and isospin
transferred along the p-h channel and are labelled as follows:
$\tau=0$ and $\sigma=0$ for the neutral current density response, $\tau=0$
and $\sigma=1$ for the neutral current spin-density response, $\tau=1$ and
$\sigma=0$ for the charged current density response, and $\tau=1$ and
$\sigma=1$ for the charged current spin-density response. In the absence
of strong interaction correlations, $S_{00}=S_{10}=S^0(q_0,q)$. The
function $S^0(q_0,q)$ is directly related to the lowest order
polarization function $\Pi^0$ (also known as the Lindhard function \cite{Lind})
through the principle of detailed balance, and accounts for
correlations arising due to Pauli blocking \cite{FW,DS}. Explicitly,
\begin{eqnarray}
S^0(q_0,q)&=&\frac{{\rm Im}~\Pi^o(q_0,q)}{1-\exp{(-q_0/T)}} \nonumber \\
\Pi^0(q_0,q)&=&\frac{2}{(2\pi)^3}\int d^3p \,
\left[\frac{f(E_p)[1-f(E_{|\vec{p}+\vec{q}|})]}{q_0+E_p-E_{|\vec{p}+\vec{q}|}
+i\epsilon} -\frac{[1-f(E_p)]f(E_{|\vec{p}+\vec{q}|})}{q_0+E_p-E_{|\vec{p}
+\vec{q}|}-i\epsilon}\right] \nonumber \,,\\
\end{eqnarray}
where $f(E)=(1+\exp{(E-\mu_n)/T})^{-1}$ is the Fermi-Dirac
distribution function, $\mu_n$ is the neutron chemical potential, and
$E_p=p^2/2M$ is the noninteracting nonrelativistic dispersion
relation for the neutrons. The imaginary part of the polarization
function may be explicitly evaluated when $q < 2k_F$ and $q_0 < q$,
and is given by \cite{RPL,KG}
\begin{eqnarray}
{\rm Im}~\Pi^0(q_0,q) &=&
\frac{M^2T}{2\pi q} \quad \left[
 \frac{q_0}{T} +
 \ln\left(\frac{1+\exp[(e_--\mu_2)/T]}{1+\exp[(e_-+q_0-\mu_4)/T]}
\right) \right]\,,
\nonumber \\
e_- &=& \frac{p_-^2}{2M}
= \frac 14 \frac {(q_0-q^2/2M)^2}{q^2/2M}\,.
\label{impi}
\end{eqnarray}
The total cross section per unit volume or the inverse
collision mean free path of neutrinos due to scattering is
obtained by integrating over the $q_0-q$ space.

The simplest modification to $S(q_0,q)$ due to strong interactions arises due
to the medium modification of the single particle dispersion relation
\cite{RPL}.  In the
mean field approximation (in which particles are assumed to move independently
of each other in a common potential), the single particle spectrum may be cast
into the form
\begin{eqnarray}
E(p)={p^2}/{2M^*}+U \,,
\label{nrspec}
\end{eqnarray}
in schematic models which are designed to reproduce the results of
microscopic calculations.
The single particle potential $U$ and the (Landau)
effective mass $M^*$ are in general density dependent.  Because
the functional dependence of the spectra on the momenta is similar to
that of the noninteracting case, the only modification to the free gas
results arises due to the nucleon effective mass. Thus, to obtain the
mean field or Hartree response we need only replace the free nucleon
mass $M$ in Eq. ~(\ref{impi}) by the in-medium mass $M^*$. The density
dependence of $M^*$ is model dependent and in nonrelativistic
approaches arises due to the presence of a momentum dependent
interaction. Modifications due to $M^*(n_B)$
are important and always act to decrease the scattering cross
sections. This is chiefly due to the fact that a lower $M^*$ decreases
the density of states $N_0=M^*k_F/\pi^2$ near the Fermi surface and
thereby the number of target neutrons available for scattering.
Kinematically, a lower $M^*$ favors larger energy transfers
by shifting the strength from low $q_0$ to regions of high $q_0$. For
thermal neutrinos, the dominant contribution to the total cross
section comes from the region $|q_0|\le \pi T$, since factors arising
from the principle of detailed balance and final state blocking for
the neutrinos exponentially suppress the phase space when $|q_0| \ge
\pi T$ .  Thus, for partially degenerate matter and for neutrino
energies of order $T$, this shift in the strength leads to significant
suppressions in the total cross sections.

The results in Fig.~\ref{sig1} show how $M^*(n_B)$, which depends
on the strong interaction model,
modifies the neutrino mean free path as a function of
density for different temperatures. For the illustrative
results presented here, we have used
${M^*}/{M}=[1+\alpha (n_B/n_0)]^{-1}$,
characteristic of a large class of Skyrme models for
$n_B \lsim  (3- 4) n_0$, where $n_0=0.16~{\rm fm}^{-3}$ is the nuclear
saturation density. Denoting the value of the nucleon mass at
$n_0$ by $M_0^*$, the factor
$\alpha=(M-M^*_0)/M^*_0$. For the results shown in Fig.~\ref{sig1}, we
have set $M_0^*=0.8$. Our findings indicate that the neutrino mean
free paths are roughly enhanced by a factor proportional to
$(M/M^*)^2$. With increasing temperature, this enhancement decreases
only moderately. We are thus led to the conclusion that for the
densities and temperatures of interest, effects due to a dropping
in-medium nucleon mass significantly increases the neutrino mean free
path.

We turn now to address the effects of strong interaction correlations.
We first note that the nucleon-nucleon interaction for densities of
relevance is not amenable to a perturbative treatment
\cite{BBN,GEB}. Thus, we are forced to adopt approximate schemes to incorporate
nonperturbative effects. The random phase approximation (RPA), also
known as the ring approximation, is particularly suited to describe
the response of matter at high densities.  In RPA, ring diagrams are
summed to all orders in order to incorporate p-h correlations. Single
pair excitations created by the neutrinos are now allowed to propagate
in the medium due to the presence of p-h interactions. The strength of
the interaction in a specific p-h channel will therefore play a
crucial role in determining the magnitude of the RPA corrections.
Deferring to later sections the discussion of the interaction
potentials, we set up here the basic formalism to calculate the cross
sections for a given interaction $V_{00}$ in the spin independent
channel and $V_{10}$ in the spin dependent channel.  The RPA
polarization, as a result of summing ring diagrams to all orders, is
\cite{FW}
\begin{eqnarray}
\Pi^{RPA}_{ij}=\frac{\Pi^0}{1-V_{ij}\Pi^0}\,.
\end{eqnarray}
The dynamic form factor which incorporates these correlations
is given by
\begin{eqnarray}
S_{ij}(q_0,q)&=&\left[ \frac{1}{1-\exp(q_0/T)}\right]
~\frac{Im~\Pi^{o}(q_0,q)}{\epsilon_{i,j}}\,,
\nonumber \\
\epsilon_{i,j}(q_0,q)&=&[1-V_{ij} Re~\Pi^{0}(q_0,q)]^2 +
[V_{ij} {\rm Im}~\Pi^{o}(q_0,q)]^2 \,.
\end{eqnarray}
In addition to the imaginary part of the polarization, which is
sufficient to evaluate the free gas response, the real part is also needed to
evaluate the corrections due to strong interaction correlations.  At
finite temperatures, the real part is easily evaluated numerically by
using the Kramers-Kronig relation
\begin{eqnarray}
{\rm Re}~\Pi^o(q_0,q)=-\frac{1}{\pi} {\mathcal P}
\int_{-\infty}^{\infty}\quad d\omega \quad
\frac{{\rm ~Im~}\Pi^o(\omega,q)}{\omega-q_0}\,.
\label{repi}
\end{eqnarray}
The dielectric screening function $\epsilon_{i,j}$ is the only
modification that arises in RPA. If the interaction is repulsive
$\epsilon_{i,j}(q_0=0,q)>1$, the medium response
is suppressed, while an attractive interaction will result in an
enhancement. If the repulsive interaction is strong, the RPA response
predicts the existence of collective excitations such as zero-sound
and spin zero-sound which could enhance the response when the
energy transfer is finite and equal to the energy of the collective
state.

\subsubsection{Particle-Hole Interactions}
The particle-hole (abbreviated as p-h) interaction, $V_{ij}$,  is in general a
function of density, temperature, $q_0$, and $q$. For the temperatures of
interest in the PNS  evolution, $q_0$ and $q$ are typically smaller than a few
tens of MeV's. Since the p-h interactions due to strong interactions are short
ranged ($\sim$~1/meson mass) compared with the typical wavelength of the
excitations probed by the neutrinos, the momentum dependence of the interaction
will not play a significant role.
It is well known that the effective interaction of meson exchange models
result in very large matrix elements due to the presence of a large
short-ranged component and cannot be used in perturbative expansion
\cite{BBN,GEB}. In nonrelativistic models, there are essentially two
complementary methods that have been used to estimate the p-h interactions in
nuclei and in bulk nuclear matter. We outline the basic features of these
methods and employ the corresponding p-h interaction potentials to compute the
neutrino scattering cross sections.

\subsubsection*{Schematic Potential Models}

Skyrme-type models start
from a zero-range, density, and momentum dependent potential from which
an energy density functional is constructed. The force parameters are
determined empirically by calculating the ground state in the
Hartree-Fock approximation and by fitting the observed ground state
properties of nuclei and nuclear matter.  Using Landau Fermi-liquid
theory, the effective p-h interaction is given by the second
functional derivative of the total energy density with respect to the
densities taken at the Hartree-Fock solution:
$V_{ij}=({\delta^2E(n)}/{\delta n_{i}\delta n_{j}})$,
and is usually expressed in terms of Fermi-liquid parameters. When the
interaction is short ranged compared to the wavelength of the
excitations of interest, justifying the assumption that only $l=0$ terms are
retained, $V_{ij}$ assumes a simple form.  For example,
in symmetric nuclear matter the p-h interaction is given by \cite{BBN}
\begin{eqnarray}
{\sc V}(k_1,k_2)=N_0^{-1}[F_0 +G_0~{\mathbf \sigma_1 \cdot \sigma_2}
+(F'_0 +G'_0~{\mathbf \sigma_1\cdot \sigma_2}) {\mathbf \tau_1 \cdot
\tau_2}] \,,
\label{fl}
\end{eqnarray}
where $\sigma$ and $\tau$ are spin and isospin Pauli matrices, which act on
the p-h states. To achieve full generality,  Eq.~(\ref{fl}) should be
supplemented with contributions from  tensor interactions.  The Fermi-liquid
parameters $F_0, G_0, F'_0$, and $G'_0$ are dimensionless numbers and
$N_0=2M^*k_F/\pi^2$ is the density of states at the Fermi surface.  In neutron
matter, since we have frozen the isospin degree of freedom, only the first two
terms contribute (with $N_0^{-1}=M^*k_F/\pi^2$), and the effective interaction
$V_{00}=N_0^{-1}F_0$ and $V_{10}=N_0^{-1}G_0$. The nucleon effective mass
$M^*$, which arises due to a momentum dependent interaction, is given by
$M^*/M=(1+F_1/3)$.

Skyrme models have been successful in describing nuclei and their
 excited states \cite{BT}. In addition, various authors have
explored its applicability to describe bulk matter at densities of
relevance to neutron stars.  The effective nucleon-nucleon interaction
of the standard Skyrme model is given by the
potential~\cite{VB}
\begin{eqnarray}
V_{NN}(r)&=&t_0(1+x_0P_{\sigma})\delta(r)
+ \frac{1}{6}t_3 n^{\gamma}(1+x_3P_{\sigma})\delta(r)\nonumber \\
&+& \frac{1}{2}t_1(1+x_1P_{\sigma})
[\hat{k}^2\delta(r)+\hat{k^\dagger}^2\delta(r)]
+ t_2(1+x_2P_{\sigma})\hat{k^\dagger}\delta(r)\hat{k} \,,
\label{nnint}
\end{eqnarray}
where, the force parameters $t_0,t_1,t_2,t_3,$ and the exchange force
parameters $x_0,x_1,x_2$, and $x_3$ are empirically
determined. The three-body interaction, which is written as a
density dependent two-body interaction (through the parameters $t_3$ and
$\gamma$) and the momentum dependent interactions arising due to $t_1$ and
$t_2$ terms dominate the high density behavior of these models.

The Fermi-liquid parameters for neutron matter, expressed explicitly
in terms of the Skyrme parameters, are collected in Appendix A.
Fig.~\ref{nfl} shows the results for four different sets of Skyrme
parameters for $n_B \lsim 4n_0$.  A striking feature is the large
density dependence of the Fermi-liquid parameters.  While the
qualitative behaviors of all four parameter sets are similar, significant
quantitative differences exist. The models SkM* and SGII~\cite{SG}
have been constrained by fitting the properties of systems with very
small isospin asymmetries, while the models SLy4 and SLy5 were further
constrained to reproduce the results of microscopic neutron matter
calculations
\cite{SLy}.  All four models become unstable to spin oscillations and
are driven towards a ferromagnetic ground state since
$G_0>-1$ for densities in the range $(2-4)n_0$.  For $0 > G_0 > -1$ , the RPA
corrections enhance the spin response approximately by a factor
$(1+G_0)^{-2}$ resulting in small neutrino mean free paths. For
$G_0 < -1$, the spin symmetric ground state is unstable.  At the phase
transition density, RPA response functions of the spin symmetric state
diverge, and the formalism developed thus far is inapplicable. The
response functions of the energetically favored, spin polarized ground
state needs to be calculated. We do not attempt to do this here; we
merely note that large enhancements in the RPA response functions are
precursors to a phase transition. We present results only for $n_B <
2n_0$, for which the ground state is spin symmetric. Spin instability
is a common feature associated with a large class of Skyrme models
\cite{spin_collapse}, but is not realized in more microscopic
calculations.  It must be emphasized that the interaction in the spin
dependent channel is a crucial ingredient in calculating the response
functions. A ``poorly'' behaved spin interaction
introduces a significantly large model
dependence in the neutrino cross sections.

$F_0$, on the other hand, is well behaved and is qualitatively
similar to the results of microscopic
calculations. This is expected, since interactions in this channel
are fit to the empirical value of the nuclear
compressibility $K$ at $n_0$. The models SLy4 and SLy5
($K_0=230$ MeV) were further constrained to reproduce the pressure
versus density curve for pure neutron matter of the
microscopic calculations of Wiringa et al.~\cite{WFF}. To
date, spin dependent interactions in the Skyrme models have not been directly
constrained. This accounts for the large differences between the
various models. Whether or not a simple Skyrme model
can mimic the bulk properties of a
microscopic calculation of both spin symmetric and spin asymmetric
systems remains an open question. For this reason, the use of a Skyrme
model to describe the spin response of neutron matter for densities of
interest in neutron stars does not appear to be promising until the
spin dependent interaction is constrained.  Nevertheless, the
energy density functional method provides a consistent framework to
study the response of hot and dense asymmetric nuclear matter. We hope
that further studies will provide a better Skyrme
energy density functional or an alternative parameterization that will be
useful for neutron star calculations.

\subsubsection*{Microscopic Potential Models}
In a microscopic approach, one starts with the bare interaction and
obtains an effective interaction by solving iteratively the
Bethe-Goldstone equation. Commonly known as the Brueckner G-matrix
method, this method provides a useful means of arriving at density dependent
effective p-h interactions. Further refinements include the use of correlated
basis states and inclusion of a larger class of diagrams~\cite{Andy}.  In this
article, we do not wish to address the merits and demerits of these
calculations; instead, we contrast the neutron matter results obtained by
Backmann et al.~\cite{backmann} with those of Jackson et al.~\cite{Andy} for
the Reid-$v_6$ and Bethe-Johnson-$v_6$ potentials (see Fig.~\ref{vfl}). In
sharp contrast to the Skyrme models, these calculations predict a large and
repulsive spin interaction ($G_0>1$). Although the qualitative behavior of
$F_0$ and $M^*$ are similar, important quantitative differences exist.  In view
of this, a knowledge of  at least the $\ell=0$ Fermi-liquid parameters of more
modern  calculations \cite{AP,APR} would be very useful.

\subsubsection{Neutral Current Neutrino Scattering Cross Sections}

In the density regions where spin stability is assured, the neutrino mean free
paths may be computed for the Skyrme models in degenerate neutron matter, since
$M^*$, $V_{00}$, and $V_{10}$ are easily expressible in terms of the Skyrme
parameters (see Appendix~A). The models SLy4 and SLy5 have been constrained to
better fit systems with large isospin asymmetries. For this reason, we choose
the model SLy4 to study the neutron matter response.  Fig.~\ref{dsig_SLy4}
shows the neutrino differential cross section for a neutrino energy $E_\nu=25$
MeV, at fixed momentum transfer $q=25$ MeV, as a function of $q_0/q$.   At
$n_B=0.04~{\rm fm}^{-3}$, the p-h interaction is attractive in the
density-density channel ($F_0=-0.7$) and repulsive in the spin-density channel
($G_0=0.6$). This accounts for the enhancement (over the Hartree result) in the
vector channel and suppression in the axial vector channel at small $q_0$. The
enhancement in the spin channel when $q_0/q \sim v_F$, where $v_F=k_F/M^*$ is
the velocity at the Fermi surface, is due to the presence of a collective spin
excitation that arises due to the large and repulsive spin interaction.  At
finite temperatures, the response is dominated by the kinematical region in
which $q_0< T$.

The bottom right panels of Fig.~\ref{dsig_SLy4} show that a strong and
attractive interaction in the vector channel gives rise to a large peak in the
vector response which dominates the total differential cross section. The weak
collective spin state seen at T=0 is almost completely damped at finite
temperatures and only modestly enhances the cross sections in this kinematical
region.

The top panels of Fig.~\ref{dsig_SLy4}  show results for $n_B=0.16$ fm$^{-3}$.
For densities close to $n_0$, the p-h interaction in both channels is
relatively weak. It remains repulsive in the spin channel ($G_0\cong 0.15$) and
attractive in the spin independent channel ($F_0\cong -0.28$). The RPA
correlations do not significantly alter the differential cross sections since
the modest suppression in the spin response is compensated by an enhancement in
the density response.  At higher densities, the Skyrme models investigated here
predict a phase transition to a ferromagnetic state and become acausal. In our
view, these are clear indicators of an inherent high density problem in the
Skyrme models due to the very strong momentum and density dependence of the
bare interaction.  The large differences in the density dependence of the
Fermi-liquid parameters for the different models suggest that this
is indeed the case.

In Fig.~\ref{dsigvar}, the neutrino differential cross sections for
the neutron matter EOS of Backmann et al. \cite{backmann} are
shown. At all relevant densities, the repulsive spin interaction
accounts for the suppression in the axial response and also for the
enhancement in the region where $q_0/q \sim v_F$ due to the presence
of a collective spin resonance. The vector part is enhanced at low
density and suppressed at higher densities as $F_0$ becomes positive.
At $2n_0$, the repulsion in this channel is sufficient to produce a
weakly damped zero-sound mode. The total differential cross sections
are significantly suppressed in all cases, and the suppression
increases with increasing density.

The collision mean free paths  for the
Skyrme model SLy4 and for the microscopic model of B\"{a}ckmann
et al. \cite{backmann} are compared in Fig.~\ref{nmnr_lambda}. The mean
free paths with and without RPA corrections are labelled
$\lambda_{RPA}$ and $\lambda_{M*}$, respectively.
The qualitative trends are similar to those observed at the level of
the differential cross sections. The ratio
$\lambda_{RPA}/\lambda_{M*}$ shown in the right panels indicate that
typical RPA corrections are large, but model dependent. Neutrino mean
free paths are suppressed in the model SLy4 due to an attractive p-h
interaction. When the Fermi-liquid parameters obtained by B\"{a}ckmann et
al. \cite{backmann} are employed, the large and repulsive spin interactions
enhance the mean free path by as much as a factor of four. The differences
between the Hartree results ($\lambda_{M^*}$) reflect the different density
dependencies of the nucleon effective mass predicted by these models (see
Fig.~\ref{nfl} and Fig.~\ref{vfl}). The density dependence of $M^*$ in the
Skyrme model SLy4 is very strong, but is relatively moderate in the microscopic
calculation due to B\"{a}ckmann et al. \cite{backmann}. The increase in the
Hartree mean free paths with increasing density is a result of the rapidly
decreasing nucleon effective mass and the use of a nonrelativistic dispersion
relation. At high density, a nonrelativistic description for the nucleon
kinematics breaks down and results in a spurious increase in the neutrino mean
free path. In the next section, we study the response of matter in relativistic
models in which some of these deficiencies are remedied.

\subsection{Relativistic Models}
In Walecka-type relativistic field-theoretical models, nucleons are
described as Dirac spinors which interact via the exchange of mesons.
In the simplest of these models, the $\sigma$- and $\omega$-meson
fields simulate the attractive and repulsive character of the
nucleon-nucleon interaction, respectively. The strengths of the
nucleon-meson couplings are fixed by the ground state properties of
nuclear matter at saturation density by solving the field theory at a
specified level of approximation~\cite{SW}.  Further, to account for
the isovector character of the nucleon-nucleon interaction, the
$\rho$-meson contribution is included and the $\rho NN$ coupling is
fixed by the empirical value of the nuclear symmetry energy. In this
section, we study the response of neutron matter whose ground state
properties are calculated in the mean field approximation to the
$\sigma\omega\rho$ model.

The differential cross section per unit volume $V$ for
$\nu$-neutron scattering is given by~\cite{WH}
 \begin{eqnarray}
\frac {1}{V} \frac {d^3\sigma}{d^2\Omega_3 dE_3}  =  -\frac {G_F^2}{32\pi^2}
\frac{E_3}{E_1}~
\left[1-\exp{\left(-\frac{q_0}{T}\right)}\right]^{-1}~
[1-f_3(E_3)]
\times{\rm Im}~(L^{\alpha \beta}\Pi^R_{\alpha \beta}) \,,
\label{dcross}
\end{eqnarray}
where $E_1$ and $E_3$ are the incoming and outgoing neutrino energies and
$q_0=E_1-E_3$ is the energy transfer in the reaction.
In terms of the incoming neutrino four momentum
$k=(E_1,\vec{k})$ and the four momentum transfer
$q_\mu=(q_0,\vec{q})$, the lepton tensor $L_{\alpha\beta}$ is
given by
\begin{eqnarray}
L^{\alpha\beta}= 8[2k^{\alpha}k^{\beta}+(k\cdot q)g^{\alpha\beta}
-(k^{\alpha}q^{\beta}+q^{\alpha}k^{\beta})\mp i\epsilon^{\alpha\beta\mu\nu}
k^{\mu}q^{\nu}] \,.
\end{eqnarray}
The target particle retarded polarization tensor, which is a function of the
neutron chemical potential, temperature,
the kinematical variables $q_0$, and $|\vec{q}|$, is
\begin{eqnarray}
{\rm Im} \Pi^R_{\alpha\beta} =\tanh{\left(\frac{q_0}{2T}\right)} {\rm
Im}~\Pi_{\alpha\beta}  \,,
\end{eqnarray}
where $\Pi_{\alpha\beta}$ is the time ordered or causal polarization and is
given by
\begin{eqnarray}
\Pi_{\alpha\beta}=-i \int
\frac{d^4p}{(2\pi)^4} {\rm Tr}~[T(G_2(p)J_{\alpha} G_4(p+q)J_{\beta})]\,.
\end{eqnarray}
The Green's functions $G_i(p)$ (the index $i$
labels particle species) describe the propagation of baryons at finite
density and temperature.  The current operator $J_{\mu}$ is $\gamma_{\mu}$
for the vector current and $\gamma_{\mu}\gamma_5$ for the axial current. Given
the structure of the particle currents, we have
\begin{eqnarray}
\Pi_{\alpha\beta}
&=&c_V^2\Pi_{\alpha\beta}^{V}+ c_A^2\Pi_{\alpha\beta}^{A}
-2 c_V c_A\Pi_{\alpha\beta}^{VA} \,.
\end{eqnarray}
For the vector polarization,
$\{J_\alpha,J_\beta\}::\{\gamma_\alpha,\gamma_\beta\}$,
for the axial polarization,  $\{J_\alpha,J_\beta\} ::
\{\gamma_\alpha\gamma_5,\gamma_\beta\gamma_5\}$, and for the mixed
part,  $\{J_\alpha,J_\beta\} ::\{\gamma_\alpha\gamma_5,
\gamma_\beta\}$. Using vector current conservation and translational
invariance, $\Pi_{\alpha\beta}^V$ may be written in terms of two independent
components. In a frame where $q_{\mu}=(q_0,|q|,0,0)$, we have
\begin{eqnarray*}
\Pi_T = \Pi^V_{22} \qquad {\rm and} \qquad
\Pi_L = -\frac{q_{\mu}^2}{|q|^2}\Pi^V_{00} \,.
\end{eqnarray*}
The axial current correlation function can be written as a vector
piece plus a correction term:
\begin{eqnarray}
\Pi_{\mu \nu}^A=\Pi^V_{\mu \nu}+g_{\mu \nu}\Pi^A \,.
\end{eqnarray}
The mixed axial vector current correlation function is
\begin{eqnarray}
\Pi_{\mu \nu}^{VA}= i\epsilon_{\mu, \nu,\alpha,0}q^{\alpha}\Pi^{VA}\,.
\end{eqnarray}
The above mean field or Hartree polarizations, which characterize the
medium response to the neutrino, have been explicitly evaluated in
previous works \cite{LH,SMS} and are collected in Appendix~B.
At the Hartree level, the only modification that
arises is due to the density dependent nucleon effective mass.

In the $\sigma\omega\rho$ model, p-h excitations propagate
in the medium via the interactions arising due to these mesons. The RPA
response includes a subclass of these excitations; namely, the ring
diagrams are summed to all orders by solving the Dyson equation~\cite{HW}.
To calculate the required polarizations, we begin with
the Lagrangian density
\begin{eqnarray*}
L &=& \sum_{B} \overline{B}(-i\gamma^{\mu}\partial_{\mu}-g_{\omega B}
\gamma^{\mu}\omega_\mu
-g_{\rho B}\gamma^{\mu}{\bf{b}}_{\mu}\cdot{\bf t}-M_B+g_{\sigma B}\sigma)B \\
&-& \frac{1}{4}W_{\mu\nu}W^{\mu\nu}+\frac{1}{2}m_{\omega}^2\omega_{\mu}\omega^
{\mu} - \frac{1}{4}{\bf B_{\mu\nu}}{\bf
B^{\mu\nu}}+\frac{1}{2}m_{\rho}^2 b_{\mu}b^{\mu}
+\frac{1}{2}\partial_{\mu}\sigma\partial^{\mu}\sigma -\frac{1}{2}
m_{\sigma}^2\sigma^2-U(\sigma)\\
&+& \sum_{l}\overline{l}(-i\gamma^{\mu}\partial_{\mu}-m_l)l \,.
\end{eqnarray*}
Here, $B$ are the Dirac spinors for baryons and $\bf t$ is the isospin
operator. The field strength tensors for the $\omega$- and
$\rho$-mesons are $W_{\mu\nu} = \partial_\mu\omega_\nu-\partial_\nu\omega_\mu$
and ${\bf B}_{\mu\nu} =  \partial_\mu{\bf b}_\nu-\partial_\nu{\bf b}_\mu$,
respectively, and $U(\sigma)$ represents the scalar self-interactions
and is taken to be of the form
$U(\sigma) = (b/3)M_n(g_{\sigma N}\sigma)^3 +(c/4)(g_{\sigma
N}\sigma)^4$.
The model may then be solved either in the mean field approximation
(MFT) or in the relativistic Hartree approximation (RHA). In either
case, the nucleon propagator $G(p_\mu)$ retains the same structure as
in the noninteracting case, but with constant density dependent shifts
to the mass and energy due the presence of scalar and vector mean
fields. Explicitly, $M \rightarrow M^*=M-g_{\sigma}\sigma$ and $p_0
\rightarrow p^*_0=p_0-g_\omega \omega + \frac{1}{2}g_{\rho}b $, where
$\sigma,\omega,$ and $b$ are the zeroth components of the $\sigma-,
\omega-, {\rm and}~\rho$-meson mean fields.  The Hartree polarization
functions (see Appendix~B) are only sensitive to
mass shifts since the momentum independent energy shifts exactly
cancel.  The Dyson equations for the various polarizations are
\begin{eqnarray}
\tilde{\Pi}^{\mu,\nu}= \Pi^{\mu,\nu} + \tilde{\Pi}^{\mu,\tau} D_{\tau,\gamma}
\Pi^{\gamma,\nu} \,,
\end{eqnarray}
where $\tilde{\Pi}^{\mu,\nu}$ is the RPA polarization function and
$D_{\tau,\gamma}$ is the interaction  due to the exchange of
scalar and vector mesons. The RPA response in these field-theoretical
models have been studied previously \cite{WH}.
Here, we gather the important steps leading to the
evaluation of the RPA corrections. The correction to the vector part
of the response is given by
\begin{eqnarray}
\delta{\Pi^{\mu,\nu}_V}=\tilde{\Pi}^{\mu,\nu}-\Pi^{\mu,\nu} \,.
\end{eqnarray}
In order to account for the mixing between the attractive scalar and the
repulsive vector interactions,  the Dyson equation in the isoscalar
longitudinal channel becomes
\begin{eqnarray}
\left[\begin{array}{cc}\tilde{\Pi}^{S} & \tilde{\Pi}^{M} \\
\tilde{\Pi}^{M} & \tilde{\Pi}^{00} \end{array}\right] =
\left[\begin{array}{cc}\Pi^{S} & \Pi^{M} \\
\Pi^{M} & \Pi^{00}\end{array}\right] +
\left[\begin{array}{cc}\tilde{\Pi}^{S} & \tilde{\Pi}^{M} \\
\tilde{\Pi}^{M} & \tilde{\Pi}^{00} \end{array}\right]
\left[\begin{array}{cc}-\chi_S & 0 \\ 0 & -\tilde{\chi}_V \end{array}\right]
\left[\begin{array}{cc}\Pi^{S} & \Pi^{M} \\
\Pi^{M} & \Pi^{00}\end{array}\right]
\end{eqnarray}
The interaction matrix is written in terms of the variables
\begin{eqnarray}
\chi_S=\frac{(g_\sigma/m_\sigma)^2}{1-q^2_\mu/m_\sigma^2} \,, \qquad
 \chi_\omega=\frac{(g_\omega/m_\omega)^2}{1-q^2_\mu/m_\omega^2} \,, \qquad
\chi_\rho=\frac{(g_\rho/m_\rho)^2}{1-q^2_\mu/m_\rho^2} \,,
\end{eqnarray}
and $\tilde{\chi}_V=q_\mu^2 \chi_V /q^2$, where
$\chi_V=\chi_\omega+\chi_\rho/4$ is the sum of $\omega$- and
$\rho$-meson contributions to the vector interaction. Note that the
$\rho$-meson contribution to the interaction is significantly smaller
than that of the $\omega$-meson and provides a very small additional
repulsion. The dominant contributions arise due to $\sigma$- and
$\omega$-mesons. The solution to the above matrix equation yields
\begin{eqnarray}
\delta \Pi_V^{00}&=&-\left[\frac{\tilde{\chi}_V \Pi_{00}^2 +
\chi_S \Pi_M^2 -
\tilde{\chi}_V \chi_S (\Pi_M^2-\Pi_{00}\Pi_S)}{\epsilon_L}\right]
\nonumber \\
\epsilon_L&=& 1 + \chi_S \Pi_S + \tilde{\chi}_V \Pi_{00}
+ \tilde{\chi}_V \chi_S (\Pi_M^2 - \Pi_{00}\Pi_S) \,.
\end{eqnarray}
Only the vector mesons contribute to the transverse parts and the RPA
corrections in this case take  the simple form
\begin{eqnarray}
\delta\Pi_T = \delta\Pi_{22}=\frac{\Pi_T\chi_V \Pi_T}{\epsilon_T}; \quad
\epsilon_T  = 1+\chi_V \Pi_T \,.
\end{eqnarray}
The axial vector polarization is not significantly modified in the
$\sigma\omega\rho$ model, since the vector mesons modify the axial response
only through the vector-axial vector mixing. The corrections to the
axial vector response due to the correlations induced by the
vector mesons are given by the  relations
\begin{eqnarray}
\tilde{\Pi}^{\mu \nu}_A = \Pi^{\mu \nu}_A +\Pi_{V}\chi_V \tilde{\Pi}_{VA}
\quad {\rm and} \quad
\tilde{\Pi}_{VA}= \Pi_{VA} + \tilde{\Pi_{VA}}\chi_V\Pi_T .
\end{eqnarray}
To directly modify the axial response, we require a meson with a
pseudo-vector coupling to the nucleons.  The lightest meson which may
be coupled to the nucleons in this way is the pion.  However, the
direct part of the pion contribution in this channel will be
suppressed in the long wavelength limit, since the derivative $\pi NN$
coupling vanishes in the limit $q_\mu \rightarrow 0$. In addition, it
is well known from nuclear phenomenology that there exists a large
repulsive component in the spin-isospin channel. To account for this
large repulsion, various authors (see, for example, Ref. \cite{BBN})
have suggested the use of
a very short range repulsive interaction parametrized through the
so-called Migdal parameter $g'$. This is usually accomplished by
modifying the pion propagator according to \cite{KPH}
\begin{eqnarray}
\frac{q_\mu q_\nu}{q_\mu^2-m_\pi^2} \rightarrow
\frac{q_\mu q_\nu}{q_\mu^2-m_\pi^2}-~g'~g_{\mu \nu} \,.
\label{pionp}
\end{eqnarray}
The value of $g'$ is expected to be in the range (0.5-0.7)
in nonrelativistic models.  However, a detailed study to empirically
determine this value in relativistic models does not exist.  For this
reason, we introduce a point-like interaction with a pseudo-vector
coupling to the nucleons and parameterize its strength through the
Migdal parameter $g'$. The strength of the coupling is given by
$\chi_A=g'f^2_\pi/m_\pi^2$. The pion contribution, which is expected
to be attractive albeit small, is not explicitly taken into account,
since we are treating $g'$ as a parameter that characterizes the total
strength in the pseudo-vector channel.  With this choice, the RPA
corrections to the axial vector polarization function is given by
\begin{eqnarray}
\delta\Pi^{\mu \nu}_A = \Pi_A^{\mu \tau}
\tilde{D}^{\tau \gamma}_A \Pi^{\gamma \nu}_A \quad {\rm with} \quad
\tilde{D}_A = (1-D_A)^{-1}D_A \qquad {\rm and } \quad D_A=\chi_A g_{\mu \nu}
\,.
\label{axe}
\end{eqnarray}
Having determined the correlated polarization functions, we may
calculate the differential cross section in a relativistic model. For
this purpose, we have chosen a mean field model labelled GM3 and a
model which includes the vacuum contributions incorporated through the
relativistic Hartree approximation (RHA). The various coupling
constants for the model GM3 and a detailed discussion of these field
theoretical models can be found in Ref.~\cite{RPL}.

\subsubsection{Neutrino Scattering Cross Sections}

We begin by discussing the results of neutrino mean free paths
obtained when RPA correlations are
ignored (Fig.~\ref{sighartree}).
In this case, the only medium dependence arises through the
density dependent $M^*$.  With increasing density, the decreasing $M^*$ nearly
compensates for the increase in particle density so as to make the
neutrino mean free path approach a constant value at high
density. This feature, seen in both field-theoretical models, is not
seen in the free gas case, in which the neutrino mean free paths
decrease far more rapidly.  The nonrelativistic cross
sections, which are roughly proportional to $M^{*^2}$, are reduced
significantly with decreasing $M^*$. In the relativistic case, the
cross sections are proportional to $E_F^2=k_F^2+M^{*^2}$, resulting in
a relatively weaker density dependence. Thus, at high densities of
interest in neutron stars, the use of nonrelativistic reaction
kinematics underestimates the cross sections by approximately a factor
($M^{*^2}/E_F^2$) due to the decreasing behavior of $M^*$ with
density.  This factor could be as small as $0.5$ at high density.

RPA corrections in models GM3 and RHA due to the $\sigma,\omega,\rho$,
and $g'$ correlations suppress the cross sections at high density owing 
to the repulsive character of these interactions. At lower density,
when $M^*\gg k_F$, the mixing between the scalar and vector mesons is
large and the interaction in the longitudinal vector channel is attractive
due to the dominant contribution of the scalar meson. At higher
densities, mixing becomes unimportant and vector meson contributions
dominate leading to significant suppression. The transverse components
of the response in the vector channel are reduced at all densities,
since only the vector mesons contribute to this correlation
function. In Fig.~\ref{dsigprh} and Fig.~\ref{dsiggm3}, the
differential cross sections with and without correlations are shown
for the RHA and GM3 models. In the absence of pseudo-vector
correlations arising due to $g'$, only the vector part is modified.

The correction to the axial vector response functions are negligibly small when
$g'=0$. The corrections to the vector part are significant at high density
(more than a factor of two). However, since the axial current contributions
dominate the differential cross sections, even large suppressions to the vector
part of the response will translate to only modest changes in
the neutrino scattering rates.  To explore the sensitivity to the magnitude of
$g'$, we study the response for $g'=0.3$ and $g'=0.6$. Effects due to
repulsive correlations are quite significant even for $g'=0.3$ (see
Fig.~\ref{dsigprh} and Fig.~\ref{dsiggm3}).

Fig.~\ref{sigrel} shows the mean
free paths for the models investigated in this work. The left panels show the
Hartree cross sections and the right panels show the extent to which RPA
correlations modify the Hartree results as a function of both density and
temperature.
RPA correlations due to $\sigma-,\omega-,~{\rm and}~\rho$-mesons result
in $(20-30)\%$ increase in the neutrino mean free paths over the
Hartree results at high density. These modifications are modest and
show little density dependence, since they predominantly act only on
the vector part of the neutron matter response. The pseudo-vector
corrections modify the dominant axial vector response and
significantly increase the mean free path. With increasing density,
these corrections first increase and then decrease due to the small
nucleon effective mass at high density.

\section{NEUTRAL CURRENT NEUTRINO CROSS SECTIONS IN MULTI-COMPONENT MATTER}
The response of multi-component matter differs in several respects from
that of a single component system. At the free gas level, the neutrino
cross sections are influenced by compositional effects arising due to
species specific neutrino-matter couplings, and more importantly due
to a reduction in the Pauli-blocking. At fixed baryon density and
temperature, additional baryonic components will significantly decrease
the degree of particle degeneracy and enhance the scattering cross
sections. In charge neutral matter containing nucleons and leptons in
$\beta-$equilibrium, the proton fraction depends very sensitively on the
dense matter EOS through the density dependent nuclear symmetry energy.
The role of the nuclear symmetry energy in determining the neutrino
scattering and absorption mean free paths calculated in the Hartree
approximation was investigated in Ref.~\cite{RPL}. Here, we
investigate the effects arising of RPA correlations in a
multi-component system. As in Sec. II, we study the response in both
nonrelativistic and relativistic modes  of dense matter.
\subsection{Nonrelativistic Models}
We begin by considering a two-component system comprised of neutrons
and protons. At first, we ignore the electromagnetic correlations
between the protons and develop the basic formalism required to
calculate the correlated response of the strongly interacting system
to neutrino scattering. For a two-component system, the Hartree
polarization function may be written as a $2 \times 2 $ matrix,
diagonal$(\Pi_p^0,\Pi_n^0$),
where $\Pi^0_p$ and $\Pi^0_n$ are the proton and neutron
polarizations, whose real and imaginary parts may be explicitly
evaluated using Eq. (\ref{impi}) and Eq. (\ref{repi}), respectively. To
compute the $2 \times 2$ RPA polarization function, we parameterize the
p-h interactions in the spin independent and spin dependent
channels. The spin independent interaction is given by $f_{nn},f_{pp}$,
and $f_{np}$, and the spin dependent interaction by $g_{nn},g_{pp}$,
and $g_{np}$. It must be borne in mind that these parameters are
associated with in-medium p-h interaction and not the
particle-particle interactions. For example, the parameter $f_{nn}$
($g_{nn}$) is a measure of the matrix element between two $nn^{-1}$
states with identical (opposite) spin projections, and $f_{np}$
($g_{np}$) measures the matrix element between a $nn^{-1}$ and
$pp^{-1}$ states with identical (opposite) spin projections
\cite{Migdal}. In matrix form, the interactions in the spin
independent and spin dependent channels are given by
\begin{eqnarray}
D_V=
\left [
\begin {array}{cc} f_{pp}
&  f_{pn}
\\\noalign {\medskip}
f_{pn}
& f_{nn}
\end {array}
\right ] \quad {\rm and} \quad
D_A=
\left [
\begin {array}{cc} g_{pp}
&  g_{pn}
\\\noalign {\medskip}
g_{pn}
& g_{nn}
\end {array}
\right ] \,,
\end{eqnarray}
respectively. The Dyson equation for the vector polarization is
\begin{equation}
\Pi^{RPA}_V=\Pi^0+\Pi^{RPA}~D_V~\Pi^0 \,.
\end{equation}
The axial polarization is obtained by replacing $D_V$ in the above equation
by $D_A$. Solving the matrix equations, we obtain
\begin{eqnarray}
\Pi_V^{RPA}&=&\frac{1}{\Delta_V}
\left [
\begin {array}{cc}
\left (1-f_{nn}\,\Pi^0_n\right )\Pi^0_p  &
f_{np}\,\Pi^0_p\,\Pi^0_n
\\\noalign{\medskip}
f_{pn} \,\Pi^0_p\,\Pi^0_n &
\left (1-f_{pp}\,\Pi^0_p\right ) \Pi^0_n
\end{array}
\right]\\
\Delta_V &=& \left[ 1 - f_{nn}\, \Pi^0_n - f_{pp}\,\Pi^0_p +
f_{pp}\,\Pi^0_p\,f_{nn} \,\Pi^0_n - {f_{np}}^{2} \Pi^0_n\,\Pi^0_p
\right]
\end{eqnarray}
for the vector part, and
\begin{eqnarray}
\Pi_A^{RPA}&=&\frac{1}{\Delta_A}
\left [
\begin {array}{cc}
\left (1-g_{nn}\,\Pi^0_n\right )\Pi^0_p  &
g_{np}\,\Pi^0_p\,\Pi^0_n
\\\noalign{\medskip}
g_{pn} \,\Pi^0_p\,\Pi^0_n &
\left (1-g_{pp}\,\Pi^0_p\right ) \Pi^0_n
\end{array}
\right]\\
\Delta_A &=& \left[ 1 - g_{nn}\, \Pi^0_n - g_{pp}\,\Pi^0_p +
g_{pp}\,\Pi^0_p\,g_{nn} \,\Pi^0_n - {g_{np}}^{2} \Pi^0_n\,\Pi^0_p
\right]
\end{eqnarray}
for the axial part. Multiplying the above polarization matrices by
the appropriate vector and axial vector couplings for the neutrons and
protons, we arrive at the vector and axial vector response functions:
\begin{eqnarray}
S_V(q_0,q)&=&\frac{1}{1-\exp{(-q_0/T)}} {\rm Im}~\Pi^{RPA}_V \nonumber \\
\Pi^{RPA}_V&=&
\left[\frac{(c_V^p)^2(1-f_{nn}\Pi^0_n)\Pi^0_p +
(c_V^n)^2(1-f_{pp}\Pi^0_p)\Pi^0_n +
2c_V^p c_V^n f_{np} \Pi^0_n\Pi^0_p} {\Delta_V}\right]
\,,\nonumber \\
S_A(q_0,q)&=&\frac{1}{1-\exp{(-q_0/T)}}{\rm Im}~\Pi^{RPA}_A \nonumber \\
\Pi^{RPA}_A&=&
\left[\frac{(c_A^p)^2(1-g_{nn}\Pi^0_n)\Pi^0_p +
(c_A^n)^2(1-g_{pp}\Pi^0_p)\Pi^0_n +
2c_A^p c_A^n g_{np} \Pi^0_n\Pi^0_p}{ \Delta_A }\right]
\,,
\end{eqnarray}
respectively. In terms of these response functions, the differential
cross sections per unit volume is given by
\begin{eqnarray}
\frac{1}{V}\frac{d^3\sigma(E_1)}{d\Omega^2~dq_0}&=& \frac{G_F^2}{8\pi^3}
\quad E_3^2\quad [1-f_{\nu}(E_3)]~ \nonumber \\
&\times& \left[(1+\cos{\theta})S_{V}(q_0,q) +
(3-\cos{\theta})S_{A}(q_0,q)\right] \,,
\label{nrdsig2}
\end{eqnarray}
where the various kinematical labels appearing above are as in Eq.
(\ref{nrdsig}). Using Eq.  (\ref{nrdsig2}), the neutrino cross sections may be
computed if the p-h interaction is specified. From the discussion of the
previous section, it is clear that these interactions  are model dependent and
that  uncertainties are large. Therefore, we explore  different dense matter
models to identify the generic trends.

\subsubsection{Particle-Hole Interactions in Nuclear Matter}
In contrast to neutron-rich matter, for which we have little empirical
information, the p-h interaction in nuclear matter may be related to
empirical values of the Fermi-liquid parameters at nuclear saturation
density. The p-h interaction is written in terms of the Fermi-liquid
parameters in Eq. (\ref{fl}). In isospin-symmetric matter,
$f_{nn}=f_{pp}$ and $g_{nn}=g_{pp}$. This allows us to directly relate
the interaction strengths to Fermi-liquid parameters \cite{Migdal}:
\begin{eqnarray}
f_{nn}=\frac{F_0+F_0'}{N_0} \,,\quad f_{np}=\frac{F_0-F_0'}{N_0} \,,\quad
g_{nn}=\frac{G_0+G_0'}{N_0} \,,\quad g_{np}=\frac{G_0-G_0'}{N_0}\,,
\label{ftof}
\end{eqnarray}
where $N_0=2M^*k_f/\pi^2$ is the density of states at the Fermi
surface.  Note that we have retained only the
the $l=0$ (s-wave) terms, since we expect
the momentum transfer in the p-h channel to be small for low
energy neutrino scattering. However, the momentum dependent part of
the interaction, which gives rise to the $l=1$ term, $F_1=3(1-M^*/M)$
is still important since it determines the nucleon effective mass and
hence $N_0$.

From FLT, we know that the $l=0$ Fermi-liquid parameters are directly
related to macroscopic observables. $F_0$ is related to the isoscalar
incompressibility $K=9\partial P / \partial n_B $, and $F_0'$ is related
to the nuclear symmetry energy $a_4=(n_B/2) \partial^2\epsilon /
\partial n_3^2$ , where $n_3=n_n-n_p$ is the isospin density. Explicitly,
\ber
F_0 = ({K}/{6E_{F}})-1 \, \quad {\rm and} \quad
F_0'= ({3a_4}/{E_F})-1    \,,
\label{emprel1}
\eer
with $E_F=(k_F^2/2M^*)$ being the Fermi energy.  Note that in extracting the
p-h interaction potential from experimental observables such as $K$ and $a_4$,
a consistent value of $M^*$ must be employed, since both $F_0$ and $F_0'$
depend on $M^*$.

At nuclear saturation density, investigation of the monopole
resonances in nuclei suggest that the isoscalar compressibility $K
\cong 240 \pm 40$ MeV \cite{BLZ}. Information from neutron-rich nuclei
and observed isovector giant dipole resonances in nuclei require that
$a_4 \cong 32 \pm 5$ MeV, and empirical determinations of the nucleon
effective mass from level density measurements in nuclei favor $M^*/M
\cong 0.7 \pm 0.1$.  For typical values of $K=240$ MeV, $a_4=30$ MeV,
and $M^*/M=0.75$, the Fermi-liquid
parameters are: $F_0=-0.18$, $F_0'=0.83$, and $F_1=0.75$.  The spin
dependent parameter $G_0$, which is related to the experimentally
observed isoscalar spin-flip resonances is estimated to be small,
$G_0=0.1 \pm 0.1$ \cite{IST,LS}.  In contrast, the parameter $G_0'$,
which is related to the isovector spin-flip (Giant Gamow-Teller)
resonances in nuclei is empirically estimated to be large,
$G_0'\cong 1.5 \pm 0.2 $ \cite{BBN,IST}.

Using the empirical values for the Fermi-liquid parameters:
$F_0=-0.18,F_0'=0.83,G_0=0,G_0'=1.7$, and $M^*/M=0.75$, we find
that the numerical values for the p-h interaction strengths
in the various channels are:
\ber
f_{nn}&=&f_{pp}=1.7 \times 10^{-5}~ {\rm MeV}^{-2} \,,\quad
f_{np}=-2.7\times 10^{-5}~ {\rm MeV}^{-2} \,,\nonumber \\
g_{nn}&=&g_{pp}=4.5 \times 10^{-5} ~{\rm MeV}^{-2} \,,\quad
g_{np}=-4.5\times 10^{-5}~ {\rm MeV}^{-2} \,.
\label{flemp}
\eer
Using these values, the neutrino
scattering differential cross sections in nuclear matter are shown
in Fig.~\ref{nucdsig}. The upper panels are
for matter at zero temperature, where only the positive $q_0$ response
exists. The suppression due to $M^*$
effects are important amounting to $\sim 50\%$ reduction.
Correlations, which are predominantly repulsive
in nature due to the large empirical values for $G_0'$ and $F_0'$,
result in significant further reductions. The RPA result
indicates a factor of two reduction in the differential
cross sections in the region $q_0/q \ll v_F$, where $v_F$ is the Fermi
velocity. The presence of a collective state in the region $q_0/q \sim
v_F$ enhances the cross sections in this region. This enhancement, 
however, is not significant enough to override the large suppression seen
in the region where $q_0/q$ is small.

Integrating over the $q_0-q$ space, we obtain the total cross section
per unit volume or equivalently the inverse collision mean free
path. This is shown in Fig.~\ref{nucsig}. The left panels show the
cross sections calculated by taking into account only effects due to
$M^*$. The results shown are for different temperatures and for a
neutrino energy $E_\nu=\pi T$. Right panels show the ratio
$\lambda_{RPA}/\lambda_{M*}$. The resulting increase in $\lambda$ due
to the presence of a repulsive p-h interaction is approximately a
factor of 2.5 at low temperature and decreases with increasing
temperature.

\subsubsection*{Density Dependence}

The density dependence of the Fermi-liquid parameters is poorly
constrained by data. Although numerous theoretical models have been
constructed to gain insight into their high density behavior, there
appears to be no general consensus at the present time. Microscopic
calculations of neutron matter
differ quantitatively depending on their underlying assumptions. These
model dependencies are so large that no generic qualitative trends may
be identified. The exception is the isoscalar parameter $F_0$,
which becomes positive and increases with increasing density, a
feature which may be expected on general grounds as the repulsive
vector meson contributions dominate. The uncertainties
associated with $F_0'$ are related to the model
dependence of the nuclear symmetry energy. In models that favor a less
than linear increase of $a_4$ with density, $F_0'$ is expected to
decrease with increasing density (see Eq. (\ref{emprel1})).
The state of the art microscopic many-body
calculations favor a modest increase in the nuclear symmetry energy at
intermediate densities \cite{WFF,APR}; thus, we may expect
that $F_0'$ will generally decrease. The parameter $G_0'$ is related
to pion condensation, since it is a measure the spin-isospin
susceptibility of nuclear matter. The large repulsive character of
$G_0'$ strongly inhibits  s-wave pion condensation in the vicinity
of the nuclear saturation density. However, at higher densities pion
condensation cannot be ruled out a priori \cite{AP,APR}. Thus, while we may
expect $G_0'$ to decrease somewhat with increasing density, quantitatively it
remains very sensitive to the underlying model. The density dependence of the
isoscalar spin parameter $G_0$, which is not well constrained even at nuclear
density, is largely unknown.

Faced with these uncertainties, we begin by assuming that the spin
dependent parameters are fixed at their empirical values (determined
at saturation density), and use schematic models to explore the
influence of the density dependence of $F_0$ and $F_0'$. For this
purpose, we employ a simple parametric form for the EOS \cite{PIPELR}
(see Appendix~A).  This model does not explicitly address the role of
spin dependent interactions and assumes that the favored ground state
is spin symmetric. In particular, we choose the Skyrme-like models
labelled `SLn2' with a linear increase in the nuclear symmetry energy.
The index `n' in 'SLn2' takes on the values $n=1,2$ and $3$ for which
$K=120,180$ and $240$ MeV, respectively. The magnitudes of the
RPA corrections to the neutrino mean free paths for these different
EOS models are shown in Fig. \ref{nucSLn2}. Since the dominant
contribution to the scattering cross section arises from the axial
vector response function, the magnitude of the RPA corrections are
mostly sensitive to the spin dependent parameters. Thus, although the
vector response of the nuclear medium is modified by about 50-80 \%
at high density due to
RPA effects, the changes due to the varying stiffness of the dense
matter EOS are small.  This suggests that the neutrino mean free paths
will not be significantly altered due to variations in the nuclear
compressibility ($F_0$) or due to variations in the nuclear symmetry
energy ($F_0'$) as long as the axial contributions are not drastically
reduced.

Fig.~\ref{pnucsig} shows the behavior of the neutrino mean free paths in
symmetric nuclear matter for the EOS labelled SL22 as a function of density for
the temperatures of interest.  A comparison of the upper and lower panels show
that there is significant temperature dependence in the Hartree and RPA
results, and that RPA corrections decrease with increasing temperature.

To explore the sensitivity to density dependent spin interactions, we
employ the Skyrme models discussed earlier. In Fig.~\ref{fliquid} the
Fermi-liquid parameters in symmetric matter are shown. As in the case of pure
neutron matter, the results show a large model dependence.  The top panels in
Fig.~\ref{skynucsig} contain the neutrino cross sections for the model SGII,
which has been constrained to fit the isovector giant dipole and Gamow-Teller
resonances observed in nuclei.  The neutrino cross sections are enhanced at
high density due to an attractive interaction in the spin channels.  Similar
trends may also be expected for the model SkM*, since the Fermi-liquid
parameters show similar qualitative trends. The models SLy4 and SLy5 show
distinctly different behavior, especially in the $G_0$ and $G_0'$ channels, in
addition to favoring a larger symmetry energy.  Neutrino mean free paths for
the model SLy4 are shown in the lower panels of Fig.~\ref{skynucsig}.  This
model is poorly behaved in the spin-isospin channel, the negative value for
$G_0'$ at nuclear density testifying to this fact. Unlike the model SGII, which
is stable up to $\sim 3n_0$ (for higher densities $G_0 < -1$), the model SLy4
is stable in the spin channel for all densities. However, it becomes unstable
to long wavelength spin-isospin oscillations (pion condensation) for  $n>2n_0$.
It is this large attraction that is responsible for the significant reductions
in the mean free path observed (see lower right panel Fig.~\ref{skynucsig}).

We wish to emphasize that the above  model predictions are strictly valid only
for those densities at which the stability conditions, which require all the
$\ell$ =0 Fermi-liquid parameters be greater $-1$, are satisfied. In view of
this, the range of applicability of these models is limited to a small window
in density centered around the saturation density (see Fig.~\ref{fliquid}). We
have investigated a large class of Skyrme models and find this to be a generic
feature. It is not clear at the present time if the choice of a better
parameter set will help alleviate this feature. More likely, additional
momentum and density dependent terms are required. Despite these shortcomings,
the Skyrme energy density functional method provides a very simple and
transparent scheme to calculate both the EOS and consistent opacities required
for astrophysical problems. We hope that, in the future, a Skyrme
parameterization that mimics the results of microscopic many-body calculations
will become available so as to facilitate finite temperature calculations of
dense matter at arbitrary asymmetry.

\subsubsection{$\beta-$Stable Asymmetric Matter}
We now study the response of charge neutral matter in $\beta-$equilibrium using
the schematic model labelled SL22, and the Skyrme models SGII and SLy4. The
composition of matter with and without neutrino trapping, and at finite
temperature is calculated as outlined in Ref.~\cite{RPL}.  The linear increase
in the nuclear symmetry energy in the model SL22 favors relatively large proton
fractions compared to those predicted by the SGII and SLy4. Using the neutron
and proton chemical potentials obtained by solving the EOS, neutrino mean free
paths are calculated as described earlier.  The electrons, which are
relativistic for all densities of interest, cannot be consistently accounted
for in a nonrelativistic treatment. Therefore, we defer the discussion of the
response which includes electrons to the next section.

At first, we ignore effects due to electromagnetic correlations
between the charged components and consider only the response of the
strongly interacting baryonic components. The differential cross
sections are computed using Eq. (\ref{nrdsig2}) with the appropriate
baryon chemical potentials and effective masses. The p-h interaction
potentials in the spin independent channel are computed consistently
with the schematic EOS as described earlier, i.e. by taking functional
derivatives of the energy density with respect to neutron and proton
number densities. In general, for asymmetric matter $f_{nn} \ne
f_{pp}$. Therefore, we require three independent quantities to
characterize the p-h interaction in the spin independent
channel (unlike only two that were required for nuclear matter). The
three relevant quantities may be computed directly, if the form of the
potential part of the single particle spectrum is known. For the model
SL22 the single particle spectrum obtained by a functional
differentiation of the potential energy density is given
in Eq. (\ref{upot}) of Appendix~A.  The functional derivatives
of the single particle potential energy are then related to the
p-h interaction parameters as follows:
\be
f_{nn} = \frac{\delta U_n}{\delta n_n} \,, \quad
f_{pp} = \frac{\delta U_p}{\delta n_p} \,, \quad {\rm and} \qquad
f_{np} = \frac{\delta U_n}{\delta n_p}=\frac{\delta U_n}{\delta n_n} \,.
\ee
For the spin dependent part, we use empirical values since our
schematic EOS does not explicitly contain spin dependent
interactions. These empirical values are determined at saturation
density and for symmetric nuclear matter.  There is a priori no reason
to expect that they will remain unchanged with increasing density and
asymmetry. Our choice is therefore primarily motivated by its
simplicity, and to a lesser extent by microscopic calculations that
indicate that these parameters do not change appreciably with density
\cite{Andy}. Fig.~\ref{paldsig} shows the differential cross
sections in stellar matter with zero neutrino chemical potential.
Results for $n_B=n_0$ (bottom panels) and for $n_B=3n_0$ (top panels),
with proton fractions $Y_p=0.049$ and $Y_p=0.22$, respectively, are
shown.  The structure seen in the Hartree results (dashed curves) at
$T=0$ is associated with the fact that the single pair component of the
the proton response is available only in the kinematical region where
$q_0 \le |\vec{q}|v_F$, where $v_F=k_F^p/M^*_p$ is the velocity at the
Fermi surface. For $n_B=n_0$, the RPA cross section is suppressed by
roughly a factor of two. The choice of a large and repulsive spin
interaction is somewhat compensated by the attractive isoscalar component of
the p-h interaction. For $n_B=3n_0$, p-h interactions are repulsive in
all channels and account for the larger (factor of four) suppression
seen in the upper panels of Fig.~\ref{paldsig}.

In the Skyrme models, the p-h interaction in all the relevant channels is
explicitly calculable as functions of density and proton fraction, and
are given in Appendix A. The differential cross sections are
computed using Eq. (\ref{nrdsig2}) with the appropriate model dependent
inputs, i.e. the p-h interaction, the neutron and proton chemical potentials and
the effective masses. Results for the models SGII and SLy4 are shown
in Fig~\ref{betadsig2}.
As a result of the large density and asymmetry dependence of the p-h
interaction predicted by Skyrme models, the RPA corrections vary
significantly with changing density and proton fraction. The top
panels show results for $n_B=0.16$ fm$^{-3}$ and proton fraction
$Y_p=0.05$. The differences seen in the Hartree results are due to the
different effective masses and proton fractions predicted by these
models. The RPA results differ qualitatively between these two models. At
nuclear saturation density (left panels), both models are stable with
respect to spin-isospin fluctuations and contain nearly equal proton
fractions. The RPA response differs mainly due to the different
spin-dependent interactions: SLy4 favors a large and attractive spin-isospin
interaction, which strongly correlates the neutron and proton spins
and accounts for the enhancement in the kinematical region where
the protons contribute (region where $q_0/q \le v_F^p$). In model
SGII, the situation is reversed since the p-h interaction is weakly
repulsive in the spin channels and accounts for the modest suppression
seen in the left-lower panel of Fig.~\ref{betadsig2}. At $2n_0$, the
p-h interaction is large and attractive in both the isoscalar and
isovector spin channels for both models. For model SLy4, this turns out
to be so large that the spin-symmetric $\beta-$equilibrium ground state
is unstable to long wavelength spin-isospin fluctuations, a feature
which was already encountered in the case of symmetric matter. Although
the spin forces are large and attractive even for the model SGII, they
are not sufficiently large to energetically favor a spin-asymmetric state. The
resulting enhancement in the RPA response is about a factor of 4 at
small $q_0$ (note that for T=10 MeV, the RPA response plotted has been
multiplied by $0.25$) and decreases with increasing $q_0$.

The collision mean free paths are shown in Fig.~\ref{betasig}.  The
Hartree mean free paths are shown in the left panels and the ratio of
the RPA mean free paths to the Hartree results are shown in the right
panels.  The results show trends similar to those observed earlier for
neutron matter and for symmetric nuclear matter.  Results for the Skyrme
models are not shown, since the RPA modes become unstable with
increasing density and asymmetry. At low density
and for small asymmetries, the trends are similar to those shown in
Fig.~\ref{skynucsig}.

\subsection{Relativistic Models}
The basic formalism required to calculate the response of a system
containing nucleons and electrons within the framework of a
relativistic mean field theory may be found in Ref.\cite{HW}. Here, we
summarize the main results of their work and extend it to incorporate
spin dependent correlations.

Neutron and proton single pair excitations are coupled due to strong
interactions. In addition, the proton and electron p-h excitations are
coupled due to electromagnetic interactions. To incorporate these
correlations consistently, we write the various polarizations functions
in matrix form. The vector part of the polarization has
transverse and longitudinal components. The transverse part is
affected only due to the vector mesons. For a system comprised of
neutrons, protons, and electrons, the RPA polarization is given by
the Dyson equation
\ber
\Pi_T^{RPA}& =& \Pi_T + \Pi_T^{RPA}~D_V~\Pi_T \,,
\eer
where the three-component transverse polarization matrix is
diagonal$(\Pi^e_T,\Pi^p_T,\Pi^n_T)$
and the interaction matrix, defined in terms of
$\chi_\gamma=-e^2/q_\mu^2$, $\chi_V=\chi_\omega+\chi_\rho $, and $
\chi_I = \chi_\omega - \chi_\rho $, is given by
\ber
D_V=
\left [
\begin {array}{ccc} -\chi_\gamma &  \chi_\gamma & 0
\\\noalign{\medskip} \chi_\gamma & -\chi_\gamma - \chi_V & -\chi_I
\\\noalign{\medskip} 0 & -\chi_I  & -\chi_V
\end {array}
\right ]\,.
\eer
The above equation accounts for electromagnetic, isoscalar, and
isovector strong interaction correlations. Since the scalar $\sigma$-
and the vector $\omega$-mesons mix strongly in the longitudinal channel, the
polarization matrix is written as a $4 \times 4$ matrix
equation \cite{chin}. Explicitly, the
Dyson equation for the longitudinal part is
\ber
\Pi_L^{RPA}&=&\Pi_L + \Pi_L^{RPA}~D_L~\Pi_L \,,
\eer
with the longitudinal polarization for the $n,p$, and $e$ system given by
\ber
\Pi_L&=&
\left [
\begin {array}{cccc} \Pi^e_{00}& 0 & 0 & 0
\\\noalign{\medskip} 0 & \Pi^p_S+\Pi^n_S & \Pi^p_M & \Pi^n_M
\\\noalign{\medskip} 0 & \Pi^p_M & \Pi^p_{00} & 0
\\\noalign{\medskip} 0 & \Pi^p_M & 0  & \Pi^p_{00}
\end {array}
\right ] \,,
\eer
and the interaction matrix
\ber
D_L&=&
\left [
\begin {array}{cccc} -\tilde{\chi}_\gamma & 0 &  \tilde{\chi}_\gamma & 0
\\\noalign{\medskip} 0 & -\chi_S & 0 & 0
\\\noalign{\medskip} \tilde{\chi}_\gamma & 0 &
-\tilde{\chi}_\gamma - \tilde{\chi}_V & -\tilde{\chi}_I
\\\noalign{\medskip} 0 & 0 & -\tilde{\chi}_I  & -\tilde{\chi}_V
\end {array}
\right ]\,.
\eer
The interaction parameters are:
$\tilde{\chi}_V= q_\mu^2 \chi_V / q^2$, $\tilde{\chi}_I= q_\mu^2
\chi_I/ q^2$, and $\tilde{\chi}_\gamma=q_\mu^2\chi_\gamma/q^2$.
The dominant modification to the axial polarizations at small momentum
transfers arises due to the pseudo-vector contact interaction discussed
earlier and is parameterized by the Migdal parameter $g'$. Further, since
the electromagnetic contributions to spin correlations are negligible,
we account only for the correlations due to strong
interactions between neutrons and protons. It is conventional to
introduce the short range correlations by modifying the pion propagator in
the pseudo-vector form of the pion-nucleon interaction described by
the Lagrangian density
\be
L= -({f_\pi}/{m_\pi})\bar{\psi}\gamma^5\gamma^\mu {\bom{\tau}}\psi
\partial_\mu {\bom{\pi}}
\,.
\ee
The Migdal parameter $g'$ enters through the modified pion propagator
\be
\frac{q_\mu q_\nu}{q_mu^2-m_\pi^2+i\epsilon} \rightarrow
\frac{q_\mu q_\nu}{q_mu^2-m_\pi^2+i\epsilon} - g' g_{\mu \nu} \,,
\ee
and the $NN \pi$ vertex becomes
\be
\Gamma^\mu_{NN \pi}= \sqrt{2} ({f_\pi}/{m_\pi}) \gamma_5 \gamma^\mu  \,.
\ee
At small momentum transfers, the repulsive contact term overwhelms the
attraction arising due to one-pion exchange. For this reason, we retain
only the contribution from the $g'$ term in the RPA
equations.  For a two-component system, this requires us to extend
the Dyson equation for the axial vector part in Eq. (\ref{axe})
to an $8\times 8$ matrix. However, it can be explicitly shown that for
$c^p_A=-c^n_A$, the cross terms cancel and the RPA axial polarization
for the n-p system closely resembles that for a one-component system, but
with $\Pi_A=\Pi_A^n + \Pi_A^p$. This greatly simplifies the formalism
required to incorporate spin dependent correlations. The RPA axial
polarization is therefore once again given in terms of Eq. (\ref{axe}).

We have ignored effects due to mixing of the vector and axial
correlations, since they are proportional to $\Pi_{VA}$, which is
negligible at small momentum and energy transfers.

Substituting the vector and axial vector RPA polarizations for the
correlated 3 component system into Eq. (\ref{dcross}),
the differential cross section for the $npe$ system is
\begin{eqnarray}
\frac{1}{V}\frac{d^3\sigma}{d^2\Omega dE_{\nu}} &=& -\frac{G_F^2}{32
\pi^3}~ \frac {E_3}{E_1} ~[1-f_3(E_3)]~
{\rm Im} \left[
\frac{L_{\alpha \beta} \Pi_{RPA}^{\alpha \beta}}
{1-\exp(-q_0/T)}
\right]\,.
\end{eqnarray}
Utilizing the particle fractions, nucleon effective masses, and particle
chemical potentials in $\beta-$equilibrated stellar matter,  the neutrino cross
sections and scattering mean free paths  are straightforwardly computed  using
the above equation.

In the model labelled RHA, the effects arising due to vacuum loops are
explicitly taken into account both at the level of the EOS and the
response functions. The differential cross sections for this model are
shown in Fig.~\ref{rdsig} for the model GM3,
wherein nonlinear scalar self-interactions play an important role
\cite{GM} (it acts to lower the compression modulus, increase the
nucleon effective mass and make the ``effective'' $\sigma$-meson mass
density dependent).  Since the
Migdal parameter $g'$ plays a dominant role in determining the RPA
response, results for different values of $g'$ are shown. The
modification to the vector response due to scalar, vector, and
isovector mesons is large, but, it does not
translate to large suppressions in the differential cross sections
(top left panel of Fig.~\ref{rdsig}), since the response is
dominated by the axial current coupling of the neutrinos. The total
suppression remains very sensitive to $g'$ for which
empirical values are expected to be in the range $0.5-0.7$. RPA
correlations suppress the neutrino scattering cross sections by
typically a factor of $2-3$.

The RPA and Hartree  mean free paths of thermal neutrinos for the model GM3 are
compared in Fig.~\ref{rlam}.  Effects due to correlations increase with
density and decrease with temperature. The decrease in the ratio
$\lambda_{RPA}/\lambda_H$ (lower panels) at high density is due to the
decreasing $M^*$.  The pseudovector point-like coupling introduced through $g'$
is sensitive to $M^*$ \cite{KPH} and its effects decrease with increasing
density. The results in the lower left panels indicate that the RPA
correlations, even for a large $g'$, will typically increase the mean free path
by a factor $2-2.5$.

\section{CHARGED CURRENT NEUTRINO CROSS SECTIONS IN MULTI-COMPONENT MATTER}

The discussion thus far has focussed upon the role of correlation
effects on the neutral current scattering reactions. While these
reactions are the only source of opacity for the $\mu$ and $\tau$
neutrinos, the electron neutrino mean free path is for the most part
dominated by the charged current reaction $\nu+n \rightarrow e^-+p$.
The absorption reaction is kinematically different from scattering,
since the energy and momentum transfers are not limited by the
matter's temperature alone. The energy transfer is typically of order
$\hat{\mu}=\mu_n-\mu_p$. In the extremely degenerate case, for which
$T/\mu <<1$, and in neutrino-poor matter, final state Pauli-blocking
due to electrons and momentum conservation restricts the available
phase space. During the deleptonization epoch, the typical neutrino
momenta are large ($\sim 100-200$ MeV) and despite the mismatch of
the neutron, proton, and electron Fermi momenta, momentum conservation
is easily satisfied with the aid of the neutrino's momentum. This is
no longer true in the late stages of the cooling phase, in which the
neutrino energies are of order $k_BT$.  Further, a finite neutrino
chemical potential affects the relative displacement between the
neutron and proton chemical potentials according to
$\hat{\mu}=\mu_e-\mu_\nu$.

\subsection{Non-Relativistic Models}
We begin by extending the formalism developed in Sec. III to
incorporate effects due to correlations.  The differential cross
section per unit volume, for the absorption reaction in matter
containing neutrons, protons, and electrons follows from Eq.
(\ref{nrdsig}), and is given by
\begin{eqnarray}
\frac{1}{V}\frac{d^3\sigma(E_1)}{d\Omega^2~dq_0}&=& \frac{G_F^2}{8\pi^3}
\quad E_3^2\quad [1-f_{e}(E_3)] \nonumber \\
&\times& \left[g_V^2(1+\cos{\theta})S^\tau_V(q_0,q) +
g_A^2(3-\cos{\theta})S^\tau_A(q_0,q)\right] \,,
\label{anrdsig}
\end{eqnarray}
where $g_V=1$ and $g_A=1.26$ are the charged current vector and axial
vector coupling constants. The particle labels 2 and 4 correspond to
the neutron and proton, respectively, and 3 to final state electron.
The other kinematical variables are similar to the case of scattering.
The isospin and spin-isospin density response functions are defined in
terms of the asymmetric polarization $\Pi^\tau$. Due to the difference
between neutron and proton chemical potentials, and of the single
particle energies, $\Pi^\tau$ differs significantly from $\Pi^0$. For
the simple form of the single particle energy given by Eq.
(\ref{nrspec}), an analytic expression for the imaginary part may be
derived (see Ref.~\cite{RPL} for details), \ber {\rm Im}\Pi^\tau&=&
\frac{M_2^*M_4^*T}{\pi q}~\left[\xi_--\xi_+\right] \,\\ 
\xi_\pm &=&
\ln\left[\frac{1+\exp[(e_\pm-\mu_2+U_2)/T]}
  {1+\exp[(e_\pm+q_0-\mu_4+U_2)/T]}\right] \nonumber \\
e_\pm&=&\frac{p_\pm^2}{2M_2^*} \quad p_{\pm}^2
=\frac{2q^2}{\chi^2}\left [\left(1+\frac{\chi M_4^*c}{q^2}\right) \pm
  \sqrt{1+\frac{2\chi M_4^*c}{q^2}}~\right]\,.
\label{pic}
\end{eqnarray}
The factors \be \chi = 1- \frac{M_4^*}{M_2^*} \quad {\rm and} \quad c
= q_0 + U_2 - U_4 - \frac{q^2}{2M_4^*} \,, \ee arise due to the
isospin dependence of the single particle potential energies. The real
part is calculated numerically by using the Kramers-Kronig relation
Eq. (\ref{repi}). The RPA response functions are related to the
Hartree polarization $\Pi^\tau$ by the Dyson equation. Explicitly,
\ber S^\tau_V(q_0,q)&=&
\frac{1}{1-\exp{[-(q_0+(\mu_2-\mu_4))/T]}}~{\rm Im}~
\left[\frac{\Pi^\tau}{1-f_c\Pi^\tau}\right] \,, \\ S^c_A(q_0,q)&=&
\frac{1}{1-\exp{[-(q_0+(\mu_2-\mu_4))/T]}}~{\rm Im}~
\left[\frac{\Pi^\tau}{1-g_c\Pi^\tau}\right] \,, \eer where the p-h
interaction in the vector and axial vector channels are given by $f_c$
and $g_c$, respectively.

\subsection*{Symmetric Nuclear Matter}

To elucidate the role of correlations, we begin with a brief
discussion of the neutrino absorption reactions in symmetric nuclear
matter. Here, the kinematical complications arising due to dissimilar
neutron and proton Fermi surfaces and electron blocking are absent. To
begin, we once again need the strength of the p-h interaction for the
spin independent and the spin dependent channels. In contrast to
scattering, both charge or isospin are transferred along the p-h
channel in the charged current absorption reaction. Thus, only the
isospin changing part of the p-h interaction is relevant for the
charged current reaction. In terms of the Fermi-liquid parameters,
this implies that only $F'_0$ and $G'_0$ contribute. For zero range
interactions, the potentials required to calculate the RPA response
are given by $f_c=2F'_0/N_0$ and $g_c=2G'_0/N_0$, where the factor two
arises due to isospin considerations. In Fig.~\ref{absnuc}, the
differential cross section for the charged current neutrino absorption
reaction in symmetric nuclear matter is shown. As discussed earlier,
large and positive empirical Fermi-liquid parameters indicate a
correspondingly large and repulsive p-h interaction. This accounts for
the large suppression seen at small $q_0$. The suppression is roughly
proportional to $(1+F'_0)^{-2}$ in the vector response and
$(1+G'_0)^{-2}$ in the axial response. At high $q_0$ and for small
temperature, well defined collective modes in the isospin-density and
spin-isospin density channels appear.  These correspond to the well
known Giant Dipole (labelled as GD in Fig.~\ref{absnuc}) and Giant
Gamow-Teller (labelled as GT) resonances found in nuclei. With
increasing temperature, these collective states are significantly
broadened due to Landau damping (see the right panel of
Fig.~\ref{absnuc}).

\subsection*{Asymmetric Nuclear Matter}

In asymmetric nuclear matter, strong interactions affect the charged
current opacity both because of correlations as discussed above, and
because of the important role they play in determining the quantity
$\hat{\mu}=\mu_n-\mu_p$. The latter, was discussed earlier in
Ref.~\cite{RPL}.  In the following discussion, we show that the p-h
interaction in the vector channel and $\hat{\mu}$ are directly
related.  It was noted in Eq. (\ref{emprel1}) that $F'_0$ was related
to the nuclear symmetry energy. The nuclear symmetry energy in turn
determines $\hat{\mu}$, and thereby the proton fraction for stellar
matter in $\beta-$equilibrium \cite{PIPELR}. Thus, a particular choice
of the p-h interaction in the isospin-density channel uniquely
determines both $\hat{\mu}$ and the extent to which correlations
suppress the cross sections. Relative to the case in which no
interactions are considered, these effects oppose each other. From Eq.
(\ref{emprel1}), it may be verified that $\hat{\mu}$ is linearly
proportional to $(1+F'_0)$. Correlations suppress the isospin density
fluctuations by a factor roughly proportional to $(1+F'_0)^{-2}$.  The
cumulative effect on the total absorption cross section is easily
deduced by noting that the cross section is a linear function of
$\hat{\mu}$ (see Ref.~\cite{RPL}), and is inversely proportional to
$(1+F'_0)^{2}$. This underscores the need to employ a p-h interaction
that is consistent with the EOS employed to compute the composition of
charge neutral $\beta-$equilibrated matter.  Here, we employ the
schematic potential model SL22 to compute the composition and the p-h
interaction in the spin independent channel.

The p-h interaction in the spin-isospin channel has been studied
extensively in the context of charge-exchange nuclear reactions
\cite{IST}, muon capture rates on nuclei \cite{CO}, and pion
condensation \cite{MSTV}. In these studies, a p-h interaction arising
due to the exchange of $\pi$- and $\rho$-mesons are included to account
for the momentum dependence, in addition to the short range repulsion
parametrized through the Migdal parameter $g'$. The isovector
interaction in the longitudinal channel arises due to $\pi$ exchange
and in the transverse channel due to $\rho$-meson exchange. We employ
this widely-used form for the p-h interaction, which has been
successful in describing a variety of nuclear phenomena \cite{OTW}.
The longitudinal and transverse potentials in the $\pi + \rho + g'$
model are given by
\begin{eqnarray}
V_{L}(q_0,q)&=&\frac{f^2_{\pi}}{m^2_{\pi}}\left
(\frac{{\bf q}^2}{q^2_0-{\bf q}^2-m^2_{\pi}}F^2_{\pi}(q)+ g'\right)\\
V_{T}(q_0,q)&=&\frac{f^2_{\pi}}{m^2_{\pi}}\left
(\frac{{\bf q}^2~C_\rho}{q^2_0-{\bf q}^2-m^2_{\rho}}F^2_{\rho}(q)+ g'\right)\,,
\end{eqnarray}
where $C_\rho=2$, $F_{\pi}=(\Lambda^2-m^2_{\pi})/(\Lambda^2-q^2)$ and
$F_{\rho}=(\Lambda_{\rho}^2-m^2_{\rho})/(\Lambda_{\rho}^2-q^2)$ are
the $\pi NN$ and $\rho NN$ form factors \cite{OTW}, and the numerical
value of $g'$ used was $0.6$. With this choice for the p-h
interaction, the axial response function is given by
\begin{eqnarray}
S^c_{A}(q_0,q)&=&\left[ \frac{1}{1-\exp[-(q_0+\hat{\mu})/T]}\right]
~{\rm Im}~\Pi^{\tau}(q_o,q)
\left(\frac{1}{3 \epsilon_{L}} + \frac{2}{3 \epsilon_{T}}\right)\\
\epsilon_{L}&=&[1-2 V_{L} {\rm Re}~\Pi^{0}(q_o,q)]^2 +
[2 V_{L} {\rm Im}~\Pi^{0}(q_o,q)]^2 \,, \\
\epsilon_{T}&=&[1- 2 V_{L} {\rm Re}~\Pi^{0}(q_o,q)]^2 +
[2 V_{T} {\rm Im}~\Pi^{0}(q_o,q)]^2 \,.
\end{eqnarray}
Using the axial response function given above and a p-h interaction in
the isospin-density channel, which is directly related to the nuclear
symmetry energy predicted by the EOS, the neutrino absorption cross
sections are computed.

In Fig.~\ref{dabs_y4}, the Hartree and RPA results for the
differential cross section for the absorption reaction are shown for
$n_B=0.16$ \fm3 and $n_B=0.48$ \fm3. The neutrino chemical potential
and proton fraction determined by the finite temperature EOS are also
shown in the figure. Since only those neutrinos close to the Fermi
surface participate in transport, we have chosen the neutrino energy
$E_\nu=\mu_\nu$. The response peaks in the region where $q_0\sim
-\hat{\mu}$. The RPA suppression is roughly a factor of two for the
kinematics shown here. The relatively smaller suppression seen here is
due to the large energy and momentum transferred to the baryons. Since
the charged current probes small distances, the effects of many-body
correlations are somewhat suppressed.  Further, the pion exchange
contribution to the p-h interaction is attractive at large $\vec{q}$
and acts to decrease the short range repulsion due to $g'$.

The neutrino absorption mean free path is shown in Fig.~\ref{asig_yl}
for neutrino trapped matter at $Y_L=0.4$ for different temperatures
and densities of relevance. The trends are very similar to the case of
scattering. The left panel shows the Hartree results, wherein effects
arising due to the density dependent nucleon effective masses and
single particle potentials are included. The increase in the mean free
path at high density arises due to the rapidly dropping $M^*(n_B)$ and
the use of a nonrelativistic form for the single particle energy. The
magnitude of the RPA corrections are shown in the right panels. On
average, the enhancement is about a factor $2-3$. At intermediate
baryon density, this enhancement increases with density and decreases
with temperature. At higher density, the rapid decrease in $M^*$
decreases the density of p-h states at the Fermi surface and thereby
reduces the magnitude of the RPA corrections.

During the cooling epoch of a protoneutron star, the neutrino chemical
potential becomes progressively smaller. For typical electron neutrino
energies of order $T$, the kinematical restriction on the charged
current process becomes important at low temperature.  The density
dependence of the symmetry energy essentially determines the density
and temperature at which these kinematical restrictions severely
inhibit the charged current rates. In the model SL22, which is
characterized by a linear symmetry energy, the kinematical
restrictions are unimportant for $n_B
\ge 2$. At lower densities, and for $T \ll \mu_p$, there is
significant suppression.

In Fig.~\ref{dabs_y0}, the Hartree and RPA differential cross sections
are compared. The momentum transfer was held constant at
$|\vec{q}|=\mu_e+\pi T$, to ensure momentum conservation. The response
peaks in the region $q_0 \sim (-\mu_e+E_\nu)$ and exponentially
decreases for larger $q_0$ due to Pauli-blocking of the degenerate
electrons. On the other hand, the final state Pauli-blocking due to
protons is negligible, since protons are fairly nondegenerate. This,
combined with the large negative energy transfers, results in the real
part $\Pi^\tau$ becoming large and negative, leading to significant
reductions (factor $\sim 4$) in the RPA cross sections.  For
$n_B=0.48$ \fm3, a larger proton fraction and a lower nucleon
effective mass decreases the magnitude of the real part $\Pi^\tau$. In
this case, the suppression due to correlations is only about a factor
of $1.5-2$.

In Fig.~\ref{asig_y0}, neutrino absorption mean free paths for matter
in which $\mu_\nu=0$ are shown. The trend identified in the preceding
discussion, by which we may expect to find a large enhancement at
intermediate densities and low temperatures, is clearly seen. For $n_B
\ge 2 n_0$, correlations enhance the mean free paths by a factor
$\sim3-4$. For intermediate densities, the enhancement is
significantly larger and very sensitive to the ambient temperature.

\subsection{Relativistic Models}

In field theoretical models, isovector correlations are induced by
interactions mediated by the $\rho$-meson.  Axial vector
correlations, in the spin-isospin channel, arise due to the presence
of a strong short ranged component, which is commonly parameterized as
a contact interaction of strength $g^\prime$.  The calculation of
these vector and axial vector responses is similar to that discussed
in Sec. IIB, but with two important differences.  First, the $\sigma$-
and $\omega$-mesons do not contribute when isospin is transferred
along the p-h channel. Second, the polarization functions now describe
the propagation of neutron-hole and proton-particle states, which
depend on $\hat{\mu}=\mu_n-\mu_p$ and the isovector part of the single
particle potential energy. The relevant polarizations are collected in
Appendix B.

The differential cross section for the charged current absorption
reaction is the same as that for neutral current scattering,
Eq.~(\ref{dcross}), with the modification that $f_3(E_3)$ is replaced
by $f_e(E_e)$.  The vector part of the correlation function
$\Pi^R_{\alpha\beta}$ is modified due to the $\rho$-meson
contribution. Explicitly, the correction to the longitudinal part is
\begin{eqnarray}
  \delta \Pi_{00}^V=\frac{-2\tilde{\chi}_I \Pi_{00}^2}
    {1+2\tilde{\chi}_I \Pi_{00}}
\end{eqnarray}
and to the transverse part is
\begin{eqnarray}
\delta\Pi_T = \delta\Pi_{22}=\frac{2\Pi_T\chi_I\Pi_T}{1+2\chi_I \Pi_T} \,.
\end{eqnarray}
For momentum transfers of interest, the axial vector current
correlation function is modified mainly by the large and repulsive
contact interaction of strength g$^\prime$. As discussed earlier, this
is conventionally introduced as a correction to the pion propogator
(see Eq. (\ref{pionp})). In principle, the contact term will lead to
$\rho-\pi$ mixing via in-medium particle-hole states.  These mixing
effects are ignored in this work. This is justified both because
$g^\prime$ is a phenomenological parameter, which is determined by
ignoring these mixing effects, and because the mixing, which is
proportional to the vector-axial vector polarization function, is
negligible at small momentum transfers. The RPA equation for the axial
vector polarization is as in Eq. (\ref{axe}), but with $\chi_A$
replaced by $2\chi_A$, where the factor 2 arises from isospin
considerations.

In Fig. \ref{ardsig_y4}, the RPA and Hartree
differential cross sections for neutrino
absorption are shown for $n_B=n_0$ and $n_B=3n_0$, respectively.
The results are presented for lepton rich, neutrino trapped matter
with $Y_L=0.4$. The electron neutrino energy
and the momentum transfer are both set equal to the neutrino chemical
potential. The composition, individual chemical potentials, and the
density dependent baryon effective masses are computed using the field
theoretical model GM3. The RPA differential cross sections are
significantly suppressed relative to the Hartree results, particularly
around the peak position i.e., when $q_0\approx
-(\mu_n-\mu_p)$. The suppression decreases with increasing density.

The absolute magnitude of the differential cross section depends
sensitively on the neutrino energy, the electron chemical potential,
composition, and temperature.  In order to assess the overall effects
of RPA correlations, both as a function of density and temperature, we
integrate over $q_0$ and $q$ and compare the RPA and Hartree results
for the neutrino mean free paths in neutrino-trapped matter in
Fig. \ref{arsig_y4}.  As expected, RPA correlations increase the
neutrino mean free paths.  The trends are very similar to those seen
in the nonrelativistic models.  This is mainly due to the fact that
$g^\prime$ plays a dominant role in suppressing the axial vector
response in both cases. At low density, the minor differences seen
between the two approaches may be attributed to the different p-h
interactions employed.  Relativistic effects become important with
increasing density and this accounts for the differences seen at
higher density.

The RPA and Hartree differential cross sections in neutrino poor
matter ($\mu_\nu=0$) are shown in Fig. \ref{ardsig_y0}.  Due to
important compositional and kinematical differences, the neutrino
absorption cross sections differ from those of nonrelativistic models.
At low density, Pauli blocking due to protons is small due to their
low concentrations, which acts to enhance the polarization of the
medium in the isospin channel. The suppression at $n_B=n_0$ (left
panel) is approximately a factor of 4.  The suppression is reduced at
high density; at $n_B=3n_0$ (right panel), the suppression is approximately a factor of 2.

The RPA and Hartree absorption mean free paths are compared in
Fig. \ref{arsig_y0} as a function of density and temperature.  The
qualitative trends are again very similar to those encountered in the
nonrelativistic case (see Fig. \ref{asig_y0}). The quantitative
differences are due to the different compositions arising due to the
different isovector interactions employed. It is clear that
relativistic effects play an increasingly important role with
increasing density, decreasing the level of suppression compared with
the non-relativistic case.

\section{EFFECTS ON PROTONEUTRON STAR COOLING}
Within a tenth of a second of the core bounce which precedes a
supernova explosion in the death of a massive star, the newly-formed
neutron star, or protoneutron star (PNS), becomes quasistatic since
its structure further evolves only in response to thermal and
compositional changes in its interior.  These changes are, for the
most part, controlled by the diffusion of neutrinos from the interior.
At high density and for temperatures above several MeV, the
neutrino-matter reaction rates are sufficiently high to ensure that
neutrinos are in thermal and chemical equilibrium.  Thus, in the
equilibrium diffusion aproximation (EDA), the neutrino distribution
function in these regions is both nearly Fermi-Dirac and
isotropic. These assumptions have been used to study the evolution of
the PNS \cite{BL,KJ,JRPLM}.

In the EDA, the distribution function in a spherically symmetric background
can be approximated by an expansion in terms of Legendre polynomials
to $O(\mu)$, with $\mu$ being the angle between the neutrino momentum
and the radial direction. Explicitly,
\begin{equation}
f(\omega,\mu)= f_0(\omega) +  \mu f_1(\omega) \,, \quad
f_0 = (1+exp[(\omega-\mu_\nu)/T])^{-1} \,,
\end{equation}
where $f_0$ is the Fermi--Dirac distribution function at equilibrium
($T=T_{mat}$, $\mu_{\nu}=\mu_{\nu}^{eq}$), and
$\omega$ and $\mu_\nu$ are the neutrino energy and chemical potential,
respectively. This allows
the energy-integrated fluxes of lepton number and energy to be expressed
as a linear combination of gradients of the degeneracy parameter 
$\eta=\mu_\nu/T$ and the temperature~\cite{JRPLM}:
\begin{eqnarray}
\label{nflux}
F_n&=&- \, \frac{e^{-\Lambda} e^{-\phi}T^2}{6 \pi^2}
\left[ D_3 \frac{\partial (T e^{\phi})}{\partial r} +
(T e^{\phi}) D_2 \frac{\partial \eta}{\partial r}  \right] \\
\label{eflux}
F_e&=&- \, \frac{e^{-\Lambda} e^{-\phi}T^3}{6 \pi^2}
\left[ D_4 \frac{\partial (T e^{\phi})}{\partial r} +
(T e^{\phi}) D_3 \frac{\partial \eta}{\partial r}  \right] \,.
\label{fluxes}
\end{eqnarray}
The diffusion coefficients $D_2$, $D_3$, and $D_4$ arise
naturally from the transport equations and contain all the
microphysics of the neutrino-matter interactions. They
are defined by
\be
D_n = \int_0^\infty dx~x^n D(\omega)f_0(\omega)(1-f_0(\omega))~,
\label{d2d3}
\ee where $x=\omega/T$ and
$D(\omega) = {\left( \kappa_a+\kappa_s \right)}^{-1}$.  The latter term
includes the contribution of both absorption-emission opacities
($\kappa_a$) and scattering opacities ($\kappa_s$),
which are related to the relevant cross sections by
\begin{equation}
\kappa_{a,s}(\omega) = \frac{1}{1-f_0(\omega)} \frac{\sigma_{a,s}}{V}\,.
\end{equation}
The fluxes in Eqs. (65) and (66) are technically for one
particle species. To include all six neutrino types, we redefine them in an
obvious notation as
\begin{eqnarray}
D_2=D_2^{\nu_e}+D_2^{\bar{\nu}_e}\,, \quad
D_3=D_3^{\nu_e}-D_3^{\bar{\nu}_e}\,, \quad
D_4=D_4^{\nu_e}+D_4^{\bar{\nu}_e}+4 D_4^{\nu_\mu}\,.
\end{eqnarray}
Effects on PNS evolution produced by changes
in the neutrino opacities can thus be understood in terms of
changes to the diffusion coefficients.

In Fig.~\ref{difc} we show the ratio of each of the diffusion
coefficients calculated in the RPA approximation to those calculated
in the Hartree approximation. The left panels display the results for
neutrino-free matter and the right panels show the results for
neutrino-rich matter ($Y_{\nu}=0.06$). In the region of interest
during the first, or deleptonization, phase of the PNS evolution, in
which the trapped neutrino abundances are high and 10 MeV $< T <$ 50
MeV, the RPA corrections to the diffusion coefficients are in the
range 1.3--1.75 for matter above nuclear density. This correction, at
first glance, seems smaller than expected; a correction factor
$\approx2$ was obtained for the mean free paths at $E_\nu=\pi$T (see
Fig 21).  However, the RPA correction redistributes strength in the
spectrum (see Fig. 20), decreasing the cross section at energies near
the peak while increasing the cross section for larger energy
exchange.  The diffusion coefficients result from an energy
integration over the entire spectrum. Another important feature is
that in the region below about half nuclear density, and at low
temperatures, the RPA corrections are diminished since the composition
of matter becomes dominated by nuclei instead of the $npe$ liquid.
Therefore, the diffusion coefficients in the outer regions of the
PNS will not be strongly altered by RPA corrections.

During the second, or cooling, phase of PNS evolution, in which the
excess neutrinos have been largely lost, the neutrino chemical
potential is relatively low and nearly constant throughout the star.
The evolution is then governed by the neutrino energy flux,
Eq. (\ref{eflux}), in which the electron neutrino chemical potential
gradient can be ignored.  Thus, $D_4$ will become the dominant
diffusion coefficient. The temperatures are now smaller than in the
earlier, deleptonization phase, being in the range $1 < T/\rm{MeV} <
20$.  Fig.~\ref{difc} shows that the RPA corrections, at three times
nuclear density (0.5 fm$^{-3}$), are larger than in neutrino-rich
matter, and can exceed a factor of 2.5.  Nevertheless, at
and below nuclear density, the corrections are considerably smaller, as
observed in the previous case.  The effect of RPA corrections
may be expected to be more evident in the cooling phase than in the
deleptonization phase.

To study the effect in a realistic PNS simulation, we approximated the
functions $D_i^{RPA}/D_i^{H}$, for $i=2-4$, by a fit in terms of
$n_B$, $T$ and $Y_\nu$. Fig. \ref{factor} shows the function assumed
for $Y_\nu = 0$; the fits for other values of $Y_\nu$ are similar.
Notice the strong density dependence in the region $n_B < n_0$.  We
used these fits to correct the diffusion coefficients calculated in
Ref.~\cite{JRPLM}, and performed a new PNS cooling simulation with the
same code.  We also wished to compare these results to those of
Burrows and Sawyer~\cite{BS}, who also studied the effects of
correlations in the RPA approximation but simply assumed that the
total opacities were decreased by a factor of 3.33 above either 1/2.7
or 1/5.4 times nuclear density, respectively.  Therefore, we performed
two additional exploratory simulations with the identical assumptions.  The
results of all three simulations are displayed in Figs.~\ref{cquant} --
\ref{lumslong}.

In Fig. \ref{cquant}, the evolution of the central values of the important
thermodynamic quantities is shown.  The solid line corresponds to the baseline
model, the dashed and dot-dashed lines to the exploratory models A and B, and
the dotted line to our RPA calculation.  The main effects of the larger mean
free paths produced by the exploratory models and by our RPA corrections are
that the inner core deleptonizes more quickly.  In turn, the maxima in central
temperature and entropy are reached on shorter timescales. In addition,
the faster increase in thermal pressure in the core slows the compression
associated with the deleptonization stage, although after 10 s the net
compressions of all models converge.

The relatively large, early, changes in the central thermodynamic
variables do not, however, translate into similarly large effects on
observables such as the total neutrino luminosity and the average
radiated neutrino energy, relative to the baseline simulation.  The
luminosities for the different models are shown as a function of time
in Fig. \ref{lums} and Fig. \ref{lumslong}.  Figure \ref{lums} shows the
early time development in detail.  The exploratory models agree with
the results reported by Burrows and Sawyer~\cite{BS}.  However, the
magnitude of the effects when our full RPA corrections are applied is
somewhat reduced compared to the exploratory models.  It is
especially important that at and below nuclear density, the
corrections due to correlations are relatively small.  Since
information from the inner core is transmitted only by the neutrinos,
the time scale to propagate any high density effect to the
neutrinosphere is the neutrino diffusion time scale. Since the
neutrinosphere is at a density approximately 1/100 of nuclear density,
and large correlation corrections occur only above 1/3 nuclear density
where nuclei disappear, we find that correlation corrections
calculated here have an effect at the neutrinosphere only after 1.5 s.
Moreover, the RPA suppresion we have calculated is considerably
smaller than those reported in Ref.~\cite{BS}, reaching a maximum of
about 30\% after 5 s, compared to a luminosity increase of 50\% after
only 2 s.  However, the corrections are still very important during
the longer-term cooling stage (see Fig.~\ref{lumslong}), and result in
a more rapid onset of neutrino transparency compared to the Hartree results.

\section{CONCLUSIONS AND OUTLOOK} The effects due to correlations are
important, albeit model dependent. Due to the large neutrino-baryon axial
vector coupling, the residual interaction in the spin dependent channels plays
a crucial role in determining the magnitude of the RPA
corrections. In pure neutron matter, these model dependencies were
carefully studied. In the Skyrme model,
the p-h interaction in the spin channel becomes large and attractive
with increasing density. This drastically reduces the neutrino mean
free path by enhancing the spin response. The opposite behavior is
suggested by variational calculations of the Fermi-liquid parameters
in neutron matter. In this case, the residual interaction at high
density is strongly repulsive in the spin channels. A repulsive p-h
interaction generically shifts the strength from regions of small
energy transfer $q_0$ to large energy transfer. In the case of strong
repulsion, collective spin excitations arise and the strength at small
$q_0$ is highly suppressed.

Not unexpectedly, the Skyrme models, in which the force parameters are
calibrated only at nuclear density,  either become unstable to spin
fluctuations or violate causality at high density.  Schematic models such as
SL22, which are partly based on the Skyrme approach, are somewhat better
behaved at high density, but they do not address the issue of spin dependence
in the nucleon-nucleon interaction. In these models, the RPA corrections to the
vector response may be calculated consistently with the EOS.   For the axial
part,  an empirical spin-isospin interaction determined by the value of the
Fermi-liquid parameter G$_0^\prime$ in symmetric nuclear matter may be used.

The situation is quite similar in field-theoretical models,
wherein the full relativistic structure of the baryon currents and
more importantly, relativistic kinematics of the scattering
reactions are naturally incorporated. In these models, the
scalar-attractive contribution dominates at low density leading to
enhancement in the vector part of the scattering response. With
increasing density, the scalar decouples and the repulsive vector
mesons suppress the cross sections. This is a generic feature. The RPA
corrections in the vector channels generally
suppress the neutrino cross sections at high density and
moderately enhance it at low density.  In addition, the RPA corrections
in the vector channel and the EOS properties such as the compressibility
and nuclear symmetry energy are intimately related. While
this stresses the need for consistency between the EOS and neutrino
opacities, there appears to be some degree of decoupling between the
strong interaction effects that influence the EOS and those that
influence the neutrino opacities. This is almost entirely due to the
different roles played by a repulsive spin dependent interaction in EOS
and opacity calculations. They directly influence the dominant part of
the neutrino opacities but play little role in determining
the properties of spin symmetric matter as long as
the spin interaction is repulsive and the spin correlation energy
is negligible compared to the other interaction energies.
This is true for the field-theoretical model, wherein the
introduction of a strong repulsive pseudovector-isovector interaction
through the Migdal parameter does not contribute to the EOS at the
mean field level, but plays an important role in determining the neutrino
mean free path.

The RPA correlations in multi-component matter are qualitatively different from
those in single-component system. The presence of even a small proton fraction
affects the shape of the matters response function. The generic trend is that
it increases the strength at small energy transfer. The inclusion of
electromagnetic correlations does  not significantly affect the neutrino mean
free paths. The mixing of the photon with the vector mesons via the p-h states
could result in the presence of plasmon like collective states in the system.
Although this might play an important role in  determining the average energy
transfer in neutrino matter interactions, it does not affect the overall
neutrino mean free paths.

The dominant role played by spin dependent nucleon-nucleon interactions
in the medium and its present poorly constrained status does not allow
us to draw definite conclusions. For the vector part, we showed that
RPA opacities are consistent with the underlying EOS. In
subsequent works, we hope that better constraints on the spin dependent
interactions in dense matter will allow calculations which will pin down its
effects on both the EOS and on the opacities. At the present time,
however, the RPA correlations remain sensitive to this poorly known
density and isospin dependence of the p-h spin interaction.

Notwithstanding the above caveats, and the differences between
nonrelativistic and relativistic approaches such as the spin- and
isospin-dependent interactions and the nucleon effective masses,
suppressions of order 2--3, relative to the case in which correlations
are ignored, are obtained.  It is also satisfying that the independent
investigations of Burrows and Sawyer~\cite{BS} have led to a similar
conclusion.

We studied the effects of RPA correlations on the evolution of
protoneutron stars.  Our calculations indicate that while RPA
correlations suppress the neutrino cross section by a factor
$\approx3$ at thermal neutrino energies, the net effect on
energy-integrated quantities such as neutrino diffusion coefficients
is more moderate.  Moreover, the effects of correlations are
diminished below nuclear densities, and disappear in low-density
matter that is dominated by nuclei.  Thus, we observe
significant effects in observable quantities such as neutrino
luminosities only after 1 second, and it is only after several
seconds that the neutrino luminosities are increased by as much as
30\%.  Larger effects occur during the late-time thermal
cooling phase, and the protoneutron star will become transparent on
shorter timescales when RPA corrections are included.

\section*{ACKNOWLEDGEMENTS}
We thank Adam Burrows, Chuck Horowitz, and Ray Sawyer for many beneficial
discussions.  This work was supported in part by the U.S. Department of Energy
under Contracts DOE/DE-FG02-88ER-40388 and DOE/DE-FG02-87ER-40317 and by the
NASA grant NAG52863. J. Pons gratefully acknowledges research support from the
Spanish DGCYT grant PB97-1432.

\subsection*{APPENDIX A: PARTICLE-HOLE INTERACTIONS}
\label{appx_phi}

The p-h interaction in the Skyrme and Skyrme-like schematic
models may be directly obtained by double functional differentiation
of the potential energy density. The potential energy density for the
density and temperatures of interest is mainly a function of baryon
density and proton fraction. The temperature dependence, which enters
only through the explicitly momentum dependent interactions, is weak. In what
follows, we provide analytical expressions for the p-h interaction at
zero temperature.

The effective nucleon-nucleon interaction in the standard
Skyrme model \cite{VB} is given in Eq.~(\ref{nnint}).
The potential energy density in the Hartree-Fock approximation for the above
interaction may be computed using the standard method by employing plane wave
states for the nucleons \cite{VB}. In
terms of the neutron and proton Fermi momenta $k^n_F$ and $k^p_F$, the neutron
and proton densities $n_n$ and $n_p$, and the total baryon density
$n=n_n+n_p$,  the potential energy density is
given by
\ber
\frac{1}{\Omega} \langle V \rangle &=& \frac{1}{2} t_0 \left(1 +
\frac{1}{2} x_0\right) n^2 + \frac{1}{12} t_3 n^{\gamma +
2} \left(1 + \frac{1}{2} x_3\right)
 \nonumber \\
&+& \frac{3}{20} \left[ t_1 \left(1 + \frac{1}{2} x_1\right) +
t_2 \left(1 +\frac{1}{2} x_2\right) \right] n^2 (3\pi^2 n/2)^{4/3}
\nonumber \\
&-& \frac{1}{2} t_0 \left(\frac{1}{2} + x_0\right) \left(n_p^2
+ n_n^2\right) -
\frac{1}{12} t_3 n^{\gamma} \left(\frac{1}{2} + x_3\right)
\left(n_p^2 + n_n^2\right)
 \nonumber \\
&+& \frac{3}{20} \left[ - t_1 \left(\frac{1}{2} + x_1\right) +  t_2
\left(\frac{1}{2} + x_2\right) \right] 
 \left[ n_p^2\,k_F^2(p) + n_n^2\, k_F^2(n)  \right] \,.
\label{upot}
\eer
The single particle potential energy for a given nucleon with isospin index
$\tau$ and with momentum $k$ is
\be
U_{\tau}(k) = \frac{\displaystyle \delta \langle V
\rangle}{\displaystyle \delta n_{kk}} \,,
\ee
where $n_{k k}$ is the $(\tau \tau)$-diagonal  element of the
occupation number matrix
\be
n_{ij} \equiv \langle \Psi \vert a_j^+ a_i \vert \Psi \rangle \,,
\ee
each label $i,j$ denoting momentum, spin, and isospin. For Skyrme
interactions, one gets
\ber
U_{\tau}(k) &=& t_0 \left( 1 + \frac{1}{2} x_0\right) n
+ \frac{1}{6} t_3 n^{\gamma} \left(1 + \frac{1}{2} x_3\right)
 n
\nonumber \\
&& + \frac{1}{12} \gamma t_3 n^{\gamma -1} \left[
\left(1+\frac{1}{2}x_3\right) n^2 - \left(\frac{1}{2} + x_3\right)
\left(n_Z^2 +n_N^2\right) \right]
\nonumber \\
&& + \frac{1}{4} \left[ t_1 \left( 1 + \frac{1}{2} x_1\right) +
t_2 \left( 1 + \frac{1}{2} x_2\right) \right] n \left( k^2
+ \frac{3}{5} k_F^2\right)
\nonumber \\
&& - t_0 \left(\frac{1}{2} + x_0\right) n_{\tau} - \frac{1}{6} t_3
n^{\gamma} \left(\frac{1}{2} + x_3\right) n_{\tau}
  \nonumber \\
&& - \frac{1}{4} \left[ t_1 \left(\frac{1}{2} + x_1\right)
- t_2 \left(\frac{1}{2} + x_2\right) \right] n_{\tau} \left[ k^2
+ \frac{3}{5} k_F^2(\tau) \right]\,.
\eer
The p-h interaction is obtained by functional
differentiation of the single particle potential energy or equivalently
the double functional
differentiation of the total potential energy, namely
\be
\langle k_1 k_3^{-1} \vert V_{ph} \vert k_4 k_2^{-1}\rangle =
\frac{\displaystyle \delta^2 \langle V \rangle}{
\displaystyle \delta n_{k_3 k_1} \delta n_{k_4 k_2}}.
\ee
Hereafter, we will employ the standard notation for
the participating momenta, namely  ${\bf k}_1 = {\bf q} +
{\bf q}_1$, ${\bf k}_2 = {\bf q}_2$, ${\bf k}_3 = {\bf q}_1$,
${\bf k}_4 = {\bf q} +{\bf q}_2$, which indicates that ${\bf q}$
is the transferred momentum. The p-h interaction can be expressed
as
\be
V_{ph}({\bf q}_1,{\bf q}_2,{\bf q}) = \sum_{(\tau \tau',S)}
V^{(\tau \tau',S)}({\bf q}_1,{\bf q}_2,{\bf q})\, P^{(S)}
\ee
with the p-h spin projectors $P^{(0)}=1/2$ and $P^{(1)}=\vec{\sigma}
\cdot \vec{\sigma}'/2 $. The symbol $(\tau \tau')$ indicates the
four isospin combinations
\ber
(pp^{-1};pp^{-1})
\quad & {\rm and} & \quad (nn^{-1};
nn^{-1})\,\,\,\,{\rm with}\,\,\,\,
\tau' = \tau,
\label{tt} \\
(pp^{-1};nn^{-1})
\quad & {\rm and} & \quad (nn^{-1};pp^{-1})\,\,\,\,{\rm with} \,\,\,\,
\tau' = - \tau
\label{tmt}
\eer
Two other isospin combinations, namely $(pn^{-1};~pn^{-1})$ and
$(np^{-1};~np^{-1})$, are needed to describe the  p-h interaction
relevant for the charge exchange processes. These are, however, not
independent, and may be related to  interactions in the
$(pp^{-1};~pp^{-1})$, $(nn^{-1};~nn^{-1})$, and $(pp^{-1};~nn^{-1})$ by
isospin considerations. Denoting the p-h interaction in the spin
independent channels as $f_{pp}$, $f_{nn}$, and $f_{np}$ for the
particle hole states $(pp^{-1};pp^{-1})$, $(nn^{-1};nn^{-1})$, and
$(pp^{-1};nn^{-1})$, respectively, we arrive at the relations \cite{HNP}
\ber
f_{nn} &=& t_0 (1-x_0) + \frac{1}{6} t_3 n^{\gamma} (1-x_3)
+ \frac{2}{3} \gamma t_3 n^{\gamma -1} \left[ (1 +
\frac{1}{2} x_3 ) n - (\frac{1}{2} + x_3) n_{n} \right]
\nonumber \\
&& + \frac{1}{6} \gamma (\gamma -1)  t_3 n^{\gamma -2} \left[ (1 +
\frac{1}{2} x_3 ) n^2 - (\frac{1}{2} + x_3) (n_p^2 +
n_n^2) \right] \nonumber \\
&& + \frac{1}{4} \left[ t_1 (1-x_1) - 3 t_2 (1+x_2) \right] q^2
+
\frac{1}{4} \left[ t_1 (1-x_1) +3 t_2 (1+x_2) \right] k_F(n)^2\,, \\
f_{np} &=& \frac{1}{2} t_0 (x_0-1) + \frac{1}{12} t_3
n^{\gamma} (x_3-1)
+ \frac{1}{8} \left[ t_1 (x_1-1) -  t_2 (1+x_2) \right] q^2 \nonumber \\
&-& \frac{1}{8} \left[ t_1 (x_1-1) + t_2 (1+x_2) \right]
(k_F(n)^2+k_F(p)^2)\,.
\eer
The interaction between p-h proton states $f_{pp}$ is related to
$f_{nn}$ evaluated at proton fraction $(1-x)$ by isospin symmetry. The
spin dependent p-h interactions characterized by $g_{nn}$, $g_{pp}$,
and $g_{np}$ are obtained by taking functional derivatives of the
energy density with arbitrary spin excess and are given by
\ber
g_{nn} &=& t_0 (x_0-1) + \frac{1}{6} t_3 n^{\gamma} (x_3-1)
+ \frac{1}{4} \left[ t_1 (x_1-1) -  t_2 (1+x_2) \right] q^2 \nonumber \\
&+& \frac{1}{4} \left[ t_1 (x_1-1) + t_2 (1+x_2) \right]~k^2_{F_n}\,, \\
g_{np} &=& \frac{1}{2} t_0 x_0 + \frac{1}{12} t_3 n^{\gamma} x_3
+ \frac{1}{8} \left[ t_1 x_1 - t_2 x_2 \right] q^2 \nonumber \\
&+& \frac{1}{8} \left[ t_1 x_1 + t_2 x_2 \right](k_F(n)^2+k_F(p)^2)\,.
\eer
As noted earlier, the quantity $g_{pp}$ is equal to $g_{nn}$ evaluated at
proton fraction $1-x$.

In the schematic models of Ref.~{\cite{PIPELR},
one begins with a
parametric form for the energy density for spin symmetric, but 
arbitrary isospin asymmetric, matter.
The single particle potential, obtained by functional differentiation
of the potential energy density, is given by
\ber
U_i(n,x,k;T)  &=& \frac{1}{5} u \left[ \sum_{i=1,2} \left\{
5C_i\pm (C_i-8Z_i)(1-2x) \right\} \right] g(k,\Lambda_i)
\nonumber \\
&+& Au \left[ 1 \mp \frac{2}{3} \left( \frac{1}{2} +x_0 \right) (1-2x) \right]
\nonumber \\
&+& Bu^\sigma
\left[1 \mp \frac{4}{3} \frac {1}{\sigma + 1} (1-2x)
- \frac{2}{3} \frac {(\sigma-1)}{(\sigma+1)} \left(\frac{1}{2}
+ x_3\right) (1-2x)^2 \right]
\nonumber \\
&+& \frac{2}{5} \frac{1}{n_0} \sum_{i=1,2} \left\{ (2C_i + 4Z_i) 2
\int \frac {d^3k}{(2\pi)^3} \, g(k,\Lambda_i) f_i(k) \right. \nonumber \\
&+&\left. (3C_i-4Z_i)2
\int \frac {d^3k}{(2\pi)^3} \, g(k,\Lambda_i) f_j(k) \right\} \,,
\end{eqnarray}
where the upper (lower) sign in $\mp$ is for neutrons (protons) and $i \neq j$.
Taking functional derivatives of the single particle potential energy
density, we arrive at the p-h interaction parameters:
\be
f_{nn} = \frac{\delta U_n}{\delta n_n} \,, \quad
f_{pp} = \frac{\delta U_p}{\delta n_p} \,, \quad
f_{np} = \frac{\delta U_n}{\delta n_p}=\frac{\delta U_n}{\delta n_n} \,.
\ee
Their explicit algebraic forms are:
\ber
f_{nn}&=& A\left(1-\frac{2}{3}(x_0+\frac{1}{2})\right) + C \nonumber \\
&+& \left[ 1-\frac{4}{3}\frac{1}{\sigma+1}(\sigma(1-2x)-2x)\right]
 B u^{\sigma-1} \nonumber \\
&-&\left[\frac{2}{3}\frac{\sigma-1}{\sigma+1}(x_3+\frac{1}{2})\left(
\sigma(1-2x)^2-4x(1-2x)\right)\right] Bu^{\sigma-1} \nonumber \\
&+& \frac{2}{5}\left[(3C-4Z)x+(2C+4Z)(1-x)\right] \nonumber \\
&-& \frac{2}{3} \alpha \left[(3\bar{C}-4\bar{Z})x^{2/3}+(2\bar{C}+4\bar{Z})
(1-x)^{2/3}\right] u^{2/3}\nonumber \\
&+& \frac{2}{5} \alpha \left[(3\bar{C}-4\bar{Z})x^{5/3}-(2\bar{C}+4\bar{Z})
(1-x)^{5/3}\right] u^{2/3}
+\frac{2}{5}(\tilde{C}-8\tilde{Z})x~{k_{F_n}}^2\,, \\
f_{np}&=& A\left(1+\frac{2}{3}(x_0+\frac{1}{2})\right) + C \nonumber \\
&+& \left[ 1-\frac{4}{3}\frac{1}{\sigma+1}(\sigma(1-2x)-2(1-x))\right]
B~u^{\sigma-1} \nonumber \\
&-&\left[\frac{2}{3}\frac{\sigma-1}{\sigma+1}(x_3+\frac{1}{2})\left(
\sigma(1-2x)^2-4(1-x)(1-2x)\right)\right] Bu^{\sigma-1} \nonumber \\
&+& \frac{2}{5}\left[(3C-4Z)x+(2C+4Z)(1-x)\right] \nonumber \\
&-& \frac{2}{3} \alpha \left[(3\bar{C}-4\bar{Z})x^{2/3}+(2\bar{C}+4\bar{Z})
(1-x)^{2/3}\right]~ u^{2/3}\nonumber \\
&+& \frac{2}{5} \alpha \left[(3\bar{C}-4\bar{Z})x^{5/3}-(2\bar{C}+4\bar{Z})
(1-x)^{5/3}\right]~ u^{2/3}
+\frac{2}{5}(\tilde{C}-8\tilde{Z}){k_{F_n}}^2 \,,\\
&{\rm with}& \nonumber \\
C&=&C_1+C_2 \,, \quad \tilde{C} = \left(\frac{C_1}{\Lambda_1}
+\frac{C_2}{\Lambda_2} \right)\,,
\quad Z=Z_1+Z_2 \,, \quad \tilde{Z} = \left(\frac{Z_1}{\Lambda_1}
+\frac{Z_2}{\Lambda_2} \right) \nonumber \,. \\
\eer
The form of the energy density employed in these schematic models does
not explicitly account for the explicit spin dependence of the nucleon-nucleon
interaction. The potential energy density is independent of any spin
excess indicating that $g_{nn}$ and $g_{np}$ are zero.

The p-h interactions parameters discussed in this appendix are related
to the Fermi-liquid parameters. In symmetric nuclear matter the
appropriate relations are
\begin{eqnarray}
f_{nn}=\frac{F_0+F_0'}{N_0} \,,\quad f_{np}=\frac{F_0-F_0'}{N_0} \,,\quad
g_{nn}=\frac{G_0+G_0'}{N_0} \,,\quad g_{np}=\frac{G_0-G_0'}{N_0}\,,
\end{eqnarray}
where the normalization factor $N_0=2M^*k_F/\pi^2$ is the density of
states at the Fermi surface. For neutron matter $F_0=N_0 f_{nn}$ and
$G_0=g_{nn} N_0$, with $N_0=M^*k_F/\pi^2$.

\subsection*{APPENDIX B: POLARIZATION FUNCTIONS}
\label{appx_pol}
The various polarization functions required to evaluate the Hartree
and RPA response functions are presented in this appendix. The zero
temperature polarization functions may be found in Ref. \cite{LH} and
for finite temperatures in Ref. \cite{SMS}. Here, we collect the
present extensions of these results to asymmetric matter, and, in
particular, to unlike p-h excitations. For space like
excitations, $q_{\mu}^2\le0$, they are given by
\begin{eqnarray}
{\rm Im}~ \Pi_L(q_0,\vec{q})
&=&2 \pi\int\frac{d^3p}{(2\pi)^3}
~~\frac{E_p^{*2}-|p|^2\cos^2\theta}{E_p^*E_{p+q}^*}~~\Theta\,,\\
{\rm Im}~ \Pi_T(q_0,\vec{q})
&=&\pi\int\frac{d^3p}{(2\pi)^3}
~~\frac{q_{\mu}^2/2-|p|^2(1-\cos^2\theta)}{E_p^*E_{p+q}^*}~~\Theta\,,\\
{\rm Im}~ \Pi_A(q_0,\vec{q})
&=&2\pi\int\frac{d^3p}{(2\pi)^3}
~~\frac{M_2^{*^2}}{E_p^*E_{p+q}^*}~~\Theta\,,\\
{\rm Im}~ \Pi_{VA}(q_0,\vec{q})
&=&2\pi\int\frac{d^3p}{(2\pi)^3}
~~\frac{q_\mu^2M_2^*}{|q^2|E_p^*E_{p+q}^*}~~\Theta \,.
\end{eqnarray}
In the above,
\begin{eqnarray}
&&\Theta=F(E_p^*,E_{p+q}^*)[\delta(q_0-(E_{p+q}-E_p))+
\delta(q_0-(E_p-E_{p+q})]\,, \\
&&F(E_p^*,E_{p+q}^*)=f_2(E^*_p)(1-f_4(E_{p+q}^*))\,,\\
&&E_{p}^*=\sqrt{|p|^2+M_2^{*2}}\,,  ~~\quad
E_{p} = E_{p}^*+U\,.
\end{eqnarray}
The particle distribution functions $f_i(E)$ are given by the
Fermi-Dirac distribution functions
\begin{eqnarray}
f_i(E_p^*)=(1+\exp[(E_p^*-\nu_i)/kT])^{-1} \,,
\label{FD}
\end{eqnarray}
where $\nu $ is the effective chemical potential defined by
\begin{equation}
\nu_i=\mu_i - U_i =\mu_i- (g_{\omega B_i}\omega_0 +
t_{3 B_i}g_{\rho B_i}b_0)\,,
\end{equation}
and the particle labels $2$ and $4$ correspond to the initial and
final baryons.

The angular integrals are performed by exploiting the delta functions.
The three dimensional integrals can be reduced to the following one
dimensional integrals:
\begin{eqnarray}
{\rm Im}~ \Pi_L(q_0,{q})
&=& \frac{q_{\mu}^2}{2\pi |q|^3}
\int_{e_-}^{\infty}dE~~[(E+q_0/2)^2-|q|^2/4]\nonumber\\
&\times&[F(E,E+q_0)+F(E+q_0,E)]\,,\\
{\rm Im}~ \Pi_T(q_0,{q})
&=& \frac{q_{\mu}^2}{4\pi |q|^3}
\int_{e_-}^{\infty}dE~~[(E^*+q_0/2)^2+|q|^2/4+|q|^2M_2^{*^2}/q_{\mu}^2)]
\nonumber\\&\times & [F(E,E+q_0)+F(E+q_0,E)]\,,\\
{\rm Im}~ \Pi_A(q_0,{q})
&=&\frac{M_2^{*^2}}{2\pi|q|}\int_{e_-}^{\infty}dE~~[F(E,E+q_0)+F(E+q_0,E)]\,,\\
{\rm Im}~ \Pi_{VA}(q_0,{q})
&=&\frac{q_\mu^2}{8\pi |q|^3}
\int_{e_-}^{\infty}dE~~[2E+q_0] \nonumber \\
&\times&[F(E,E+q_0)+F(E+q_0,E)]\,.
\end{eqnarray}
The lower cut-off $e_-$ arises due to kinematical restrictions
and is given by
\begin{eqnarray}
e_-&=&-\beta \frac{\tilde{q_0}}{2}
+\frac{q}{2}\sqrt{\beta^2-4\frac{M_2^{*2}}{q^2-\tilde{q_0^2}}}\,,
\end{eqnarray}
where
\begin{equation}
\tilde{q_0}=q_0+U_2-U_4 \,,\quad
\beta = 1+\frac{M^{*^2}_4-M^{*^2}_2} {q^2-\tilde{q_0^2}}\,.
\end{equation}
It is convenient to re-express the polarization functions as follows:
\begin{eqnarray}
{\rm Im}~ \Pi^R_L(q_0,q)&=&\frac{q_{\mu}^2}{2\pi |q|^3}
\left[I_2 + q_0I_1+\frac{q_{\mu}^2}{4}I_0\right]\,, \\
{\rm Im}~ \Pi^R_T(q_0,q)&=&\frac{q_{\mu}^2}{4\pi |q|^3}
\left[I_2 + q_0I_1+
\left(\frac{q_{\mu}^2}{4} +
\frac{q^2}{2}+M_2^{*^2}\frac{q^2}{q_{\mu}^2}\right) I_0\right]\,, \\
{\rm Im}~ \Pi^R_A(q_0,q)&=&\frac{M_2^{*^2}}{2\pi |q|}I_0\,,\\
{\rm Im}~ \Pi^R_{VA}(q_0,q)&=&\frac{q_{\mu}^2}{8\pi |q|^3}[q_0I_0+2I_1] \,,
\label{appolar}
\end{eqnarray}
where we used the one-dimensional integrals
\begin{equation}
I_n=\tanh{\left(\frac{q_0+(\mu_2-\mu_4)}{2T}\right )}
\int_{e_-}^{\infty}dE~E^n~[F(E,E+q_0)+F(E+q_0,E)]\,.
\end{equation}
These integrals may be explicitly expressed in terms of the Polylogarithmic
functions
\begin{equation}
Li_n(z) = \int_0^z \frac {Li_{n-1}(x)}{x} \,dx \,,
\qquad Li_1(x) = \ln (1-x)
\end{equation}
which are defined to conform to the definitions of Lewin~\cite{LL}.
This Polylogarithm representation is particularly useful and compact:
\begin{eqnarray}
I_0 &=& T~z\left(1+\frac{\xi_1}{z}\right)\,, \\
\label{iis0}
I_1 &=& T^2~z\left(\frac{\mu_2-U_2}{T}  -
\frac{z}{2}+\frac{\xi_2}{z}+\frac{e_-\xi_1}{zT}\right) \,,\\
\label{iis1}
I_2 &=& T^3~z \left(\frac{(\mu_2-U_2)^2}{T^2} -
z\frac{\mu_2-U_2}{T} +
\frac{\pi^2}{3}+\frac{z^2}{3}\right) \nonumber \\
&-&T^3~z
\left(2\frac{\xi_3}{z}-2\frac{e_-\xi_2}{Tz}
-\frac{e_-^2\xi_1}{T^2z}\right) \,,
\label{iis2}
\end{eqnarray}
where $z=(q_0+(\mu_2-\mu_4))/T$ and the factors $\xi_n$ are given by
\begin{equation}
\xi_n = Li_n(-\alpha_1) - Li_n(-\alpha_2)\,,
\label{dlis}
\end{equation}
with
\begin{equation}
\alpha_1 = \exp\left((e_--\mu_2+U_2)/T\right)\,,\quad
\alpha_2 = \exp\left((e_-+q_0-\mu_4+U_2)/T\right) \,.
\end{equation}
\noindent We note that the nonrelativistic structure function for
neutral current scattering, Eq.~(\ref{impi}) is, aside from the
factor $M_2^{*^2}T/\pi q$, equal to $I_0$ since
$\xi_1\equiv\xi_-$.
In the case of neutral currents, some of the terms in the above are
simplify,
\begin{eqnarray}
z = \frac{q_0}{T} \,, \quad \mu_2 = \mu_4\,, \quad
e_- = -\frac{q_0}{2}
+\frac{q}{2}\sqrt{1-4\frac{M_2^{*2}} {q_{\mu}^2}}\,,
\end{eqnarray}
and we recover the results obtained earlier in Ref. \cite{SMS}.

For the real part, analytic forms do not exist. In the following we
present, in integral form, the required polarization functions. The
results presented below generalize the results of Ref. \cite{SMS} to
the case when the particle and hole states correspond to different
baryons.
\ber
&{\rm Re}~\Pi_L& = -\frac{4}{\pi^3} \int d^3p
\frac{{\sc F}(E^*_p)}{2E_p^*}
\left[\frac{q_{\mu}^2(E_p^{*2}-|\vec{p}|^2 \cos^2{\theta})}
{q_\mu^4-4(p\cdot q)^2} \right]_{p_0=E^*_p} \nonumber \\
&=&\frac{-q_\mu^2}{\pi^2 |\vec{q}|^3}
\int^\infty_0 dp~\frac{p}{E_p^*} {\sc F}(E^*_p) \nonumber \\
&\times&\left[|{\vec{p}}||\vec{q}|-\frac{1}{2}\left(
(E_p^{*2}-q_0E_p^*+q_\mu^2/4)L_- - (E_p^{*2}+q_0E_p^*+q_\mu^2/4)L_+\right)
\right] \nonumber \,,\\
\\
&{\rm Re}~\Pi_T& = -\frac{4}{\pi^3} \int d^3p
\frac{{\sc F}(E^*_p)}{2E_p^*}
\left[\frac{ (p\cdot q)^2-q_\mu^2*|\vec{p}|^2(1- \cos^2{\theta})/2)}
{q_\mu^4-4(p\cdot q)^2} \right]_{p_0=E^*_p} \nonumber \\
&=&\frac{1}{\pi^2 |\vec{q}|^3}
\int^\infty_0 dp~\frac{p}{E_p^*} {\sc F}(E^*_p)~
\left[|\vec{q}|^3|\vec{p}|+|\vec{q}||\vec{p}|q^2_\mu/2 \right.\nonumber \\
&-& \left.\frac{q_\mu^2}{4}\left((E_-^2+M^{*2}\frac{|\vec{q}|^2}{q_\mu^2}+
\frac{|\vec{q}|^2}{4})L_- + (E_+^2+M^{*2}\frac{|\vec{q}|^2}{q_\mu^2}+
\frac{|\vec{q}|^2}{4})L_+\right)\right] \nonumber \,,\\
\\
&{\rm Re}~\Pi_S& = -\frac{4}{\pi^3} \int d^3p
\frac{{\sc F}(E^*_p)}{2E_p^*}
\left[\frac{ q_\mu^2M^{*2}-(p\cdot q)^2}
{q_\mu^4-4(p\cdot q)^2} \right]_{p_0=E^*_p} \nonumber \\
&=&-\frac{1}{\pi^2|\vec{q}|}
\int^\infty_0 dp~\frac{p}{E_p^*} {\sc F}(E^*_p)~
\left[|\vec{q}||\vec{p}|-(q_\mu^2-4M^{*2})(L_-+L_+)/8 \right] \nonumber \,,\\
\\
&{\rm Re}~\Pi_A& = -\frac{4}{\pi^3} \int d^3p
\frac{{\sc F}(E^*_p)}{2E_p^*}
\left[\frac{ q_\mu^2M^{*2}}
{q_\mu^4-4(p\cdot q)^2} \right]_{p_0=E^*_p} \nonumber \\
&=&-\frac{M^{*2}}{2\pi^2|\vec{q}|}
\int^\infty_0 dp~\frac{p}{E_p^*} {\sc F}_+(E^*_p)~(L_-+L_+)\,,\\
&{\rm Re}~\Pi_M& = -\frac{4M^*}{\pi^3} \int d^3p
\frac{{\sc F}(E^*_p)}{2E_p^*}
\left[\frac{ q_\mu^2E_p^*-q_)(p\cdot q)}
{q_\mu^4-4(p\cdot q)^2} \right]_{p_0=E^*_p} \nonumber \\
&=&-\frac{M^*}{2\pi^2|\vec{q}|}
\int^\infty_0 dp~\frac{p}{E_p^*} {\sc F}(E^*_p)~
\left[E_-L_-+E_+L_+\right] \,,
\eer
where
\ber
{\sc F}(E)&=& \frac{f_i(E)+f_j(E)}{2}\,, \quad E_\pm=E_p^*\pm q_0/2\,,\\
L_+&=&\ln\left[1+\frac{4|\vec{q}||\vec{p}|}
{2E_p^*q_0-q_\mu^2+2|\vec{q}||\vec{p}|}\right]
\frac{f_i(E_p^*)}{{\sc F}(E)} \,, \\
L_-&=&\ln\left[1-\frac{4|\vec{q}||\vec{p}|}
{2E_p^*q_0+q_\mu^2-2|\vec{q}||\vec{p}|}\right]
\frac{f_j(E_p^*)}{{\sc F}(E)} \,.
\eer
The particle labels $i$ and $j$ refer to the initial and final state
baryon. Anti-particle contributions have been neglected in the above
formulae as they provide negligible contributions for the temperatures
and densities of interest. Since the typical temperatures never exceed
a few tens of MeV's they will be only important for electrons and only
when $\mu_e \ll T$.

The vacuum contributions, which only play a role in the relativistic
Hartree approximation, are not explicitly given here, but may be found in
Ref. \cite{HW}.

\newpage
\section*{REFERENCES}
\newcommand{\IJMPA}[3]{{ Int.~J.~Mod.~Phys.} {\bf A#1}, (#2) #3}
\newcommand{\JPG}[3]{{ J.~Phys. G} {\bf {#1}}, (#2) #3}
\newcommand{\AP}[3]{{ Ann.~Phys. (NY)} {\bf {#1}}, (#2) #3}
\newcommand{\NPA}[3]{{ Nucl.~Phys.} {\bf A{#1}}, (#2) #3 }
\newcommand{\NPB}[3]{{ Nucl.~Phys.} {\bf B{#1}}, (#2)  #3 }
\newcommand{\PLB}[3]{{ Phys.~Lett.} {\bf {#1}B}, (#2) #3 }
\newcommand{\PRv}[3]{{ Phys.~Rev.} {\bf {#1}}, (#2) #3}
\newcommand{\PRC}[3]{{ Phys.~Rev. C} {\bf {#1}}, (#2) #3}
\newcommand{\PRD}[3]{{ Phys.~Rev. D} {\bf {#1}}, (#2) #3}
\newcommand{\PRL}[3]{{ Phys.~Rev.~Lett.} {\bf {#1}}, (#2) #3}
\newcommand{\PR}[3]{{ Phys.~Rep.} {\bf {#1}}, (#2) #3}
\newcommand{\ZPC}[3]{{ Z.~Phys. C} {\bf {#1}}, (#2) #3}
\newcommand{\ZPA}[3]{{ Z.~Phys. A} {\bf {#1}}, (#2) #3}
\newcommand{\JCP}[3]{{ J.~Comput.~Phys.} {\bf {#1}}, (#2) #3}
\newcommand{\HIP}[3]{{ Heavy Ion Physics} {\bf {#1}}, (#2) #3}
\newcommand{\RMP}[3]{{ Rev. Mod. Phys.} {\bf {#1}}, (#2) #3}
\newcommand{\APJ}[3]{{Astrophys. Jl.} {\bf {#1}}, (#2) #3}

{}

\newpage
\section*{FIGURE CAPTIONS}
\vskip 10pt
FIG. 1:  The ratio of the neutrino mean free paths computed with ($M^*$) and
without ($M$) effective mass corrections as a function of nucleon density $n_B$
in units of the nuclear equilibrium density $n_0=0.16~{\rm fm}^{-3}$.  Results
are shown for pure neutron matter at $T=5$ and 30 MeV.  The dash-dot curve
shows the density dependence of the neutron effective mass.

\vskip 10pt
FIG. 2:  The density dependence of the Fermi-liquid parameters and the neutron
effective mass for neutron matter for  Skyrme models studied in this work.

\vskip 10pt
FIG. 3: The density dependence of  Fermi-liquid parameters of microscopic
calculations from  Refs.~\cite{Andy}  and \cite{backmann} (see text for
details).

\vskip 10pt
FIG. 4: The neutrino differential cross sections in neutron matter  as a
function of the ratio of energy to momentum transfer for $T=0$ MeV (left
panels) and $T=10$ MeV (right panels) at $n_0$ (top panels) and $n_0/4$ (bottom
panels).  Results are for the Skyrme model SLy4 for $q=E_\nu=25$ MeV.  The
symbols $V$ and $A$ refer to the vector and axial vector contributions,
respectively.  Thin lines refer to the Hartree results and thick lines
to the RPA results.

\vskip 10pt
FIG. 5: Same as Fig. 4, except that results are for the model of
Backmann et al.~\cite{backmann} and are shown for three densities: $2n_0$ (top
panels), $n_0$ (center panels), and $n_0/4$ (lower panels).

\vskip 10pt
FIG. 6: The density dependence of the neutrino mean free paths at energy
$E_\nu=\pi T$ in neutron
matter for $T=10, 20$ and 30 MeV.  Results for the microscopic calculations of
B\"{a}ckmann et al. are in the top panels and those for the Skyrme model SLy4
are in the bottom panels. The right panels show the extent to which RPA
corrections modify the Hartree results.

\vskip 10pt
FIG. 7: The density dependence of the neutrino mean free paths in the
relativistic models RHA and GM3 in the Hartree approximation (i.e., RPA
correlations were ignored) at energy $E_\nu=\pi T$  for $T=0, 10$ and 20 MeV.
For reference, the results for the case in which effective mass corrections are
ignored are also shown as dashed lines.

\vskip 10pt
FIG. 8: The neutrino differential cross sections in neutron matter in the model
RHA for $q=E_\nu=25$ MeV.  Dotted lines show the Hartree results, while solid,
dashed and dot-dashed curves show the results including RPA correlations with
$g^\prime=0, 0.3$ and 0.6, respectively.  The left panels have $T=0$ MeV and
right panels have $T=10$ MeV; the top panels are for $4n_0$ and bottom panels
are for $n_0$.

\vskip 10pt
FIG. 9: Same as in Fig.~\ref{dsigprh}, but for the model GM3.

\vskip 10pt
FIG. 10: The density and temperature dependences of neutrino mean free  paths
at energy $E_\nu=\pi T$  in neutron matter.  The top panels are for the model
GM3; the bottom panels are for RHA.  The left panels show the results in the
Hartree approximation.  In the right panels, effects due to  spin dependent
correlations introduced through the Migdal parameter $g'$ are shown.  The
different curves correspond to the same temperatures as in the left panels.

\vskip 10pt
FIG. 11: The neutrino differential cross sections in symmetric nuclear matter
for $q=E_\nu=30$ MeV.  The Fermi-liquid parameters employed are given in Eq.
(\ref{flemp}).  Results for the free gas, the Hartree approximation, and with
RPA correlations are compared for $T=0$ and 10 MeV.

\vskip 10pt
FIG. 12: The temperature dependence of the neutrino scattering mean free path
in symmetric nuclear matter at density $n_0$ for the Fermi-liquid parameters in
Eq. (\ref{flemp}).  The left panel shows results for thermal neutrinos
($E_\nu=\pi T$) calculated in the Hartree approximation, and the right panel
shows the effect of RPA correlations.

\vskip 10pt
Fig. 13: The density dependence of the effect of RPA correlations on
the neutral current neutrino mean free paths in the nonrelativistic
SLn2 models with different compressibilities.

\vskip 10pt
FIG. 14: The density dependence of the neutrino scattering mean free path  in
symmetric nuclear matter for the EOS SL22 at $T=10, 20,$ and 30 MeV.
The upper panel shows results for thermal neutrinos ($E_\nu=\pi T$)
calculated in the Hartree approximation, and the lower panel shows the effect
of RPA correlations.

\vskip 10pt
FIG. 15: The density dependence of the Fermi-liquid and effective nucleon mass
in symmetric nuclear matter for Skyrme models used in this paper.

\vskip 10pt
FIG. 16: The neutrino scattering mean free paths in symmetric nuclear matter
for the Skyrme models SGII (top panels) and SLy4 (bottom panels), at a neutrino
energy $E_\nu=\pi T$.  Left panels show  the Hartree approximation and right
panels show the effect of RPA correlations.

\vskip 10pt
FIG. 17: Differential scattering cross sections in $\beta-$stable
neutrino-free matter for the schematic model SL22. The proton fraction $x_p=0.049$
for $n_B=n_0$ (lower panels) and $x_p=0.22$ at $n_B=3n_0$ (upper panels).  The
left panels show results for $T=0$ while right panels are for $T=10$ MeV.
$E_\nu=q=30$ MeV are assumed.

\vskip 10pt
FIG. 18: Differential scattering cross sections in $\beta-$stable neutrino-free
matter for the Skyrme models SLy4 (top panels) and SGII (bottom panels), for
the indicated proton fractions and temperatures.  Results for both the Hartree
approximation (dashed lines) and with RPA correlations (solid lines) are
displayed.

\vskip 10pt
FIG. 19: The density dependence of the
scattering mean free paths in $\beta-$stable asymmetric matter for
the schematic model SL22 at the indicated temperatures, in the Hartree
approximation (left panel) and with RPA correlations included (right panel).
In the right panel, thick lines show results obtained by employing empirical
values for the spin parameters.  Thin lines correspond to results obtained
without any spin-dependent p-h interaction.

\vskip 10pt
FIG. 20: Differential cross sections at $n_B=0.32$ fm$^{-3}$ in $\beta-$stable
neutrino-free matter for the field-theoretical models RHA (left panels) and GM3
(right panels) for $T=0$ (lower panels) and $T=10$ MeV (upper panels).  Results
for the Hartree approximation alone and also including RPA corrections with
$g^\prime=0, 0.1$ and 0.6 are compared.

\vskip 10pt
FIG. 21: The density and temperature dependences of the neutral current mean
free paths for $\beta-$stable neutrino free matter in the field-theoretical
model GM3. The upper left panel shows the Hartree results for the case
$E_\nu=3T$.  The influence of the spin correlations introduced via the Migdal
parameter $g^\prime$ is strong, as can be deduced from the results shown in
the upper right and bottom panels.

\vskip 10pt
FIG. 22: Charged current diferential cross sections in symmetric nuclear
matter at nuclear saturation density, for the case $E_\nu=q=50$ MeV.
Empirical values of the Fermi-liquid parameters are employed, and results
for the Hartree approximation and with RPA correlations are compared.   For
$T=0$ (left panel), the peaks labelled `GD' and `GT' correspond to the giant
dipole and Gamow-Teller resonances, respectively. For $T=10$ MeV (right panel),
the collective states are  significantly broadened and overlap.

\vskip 10pt
FIG. 23: Charged current differential cross sections in $\beta-$stable
matter for $Y_L=0.4$ and $T=10$ MeV at $n_B=n_0$ (left panel) and
$n_B=3n_0$ (right panel).  The schematic model SL22 was employed.
Results for the Hartree approximation and with RPA correlations are compared
for the case $q=E_\nu=\mu_\nu$.

\vskip 10pt
FIG. 24: The density and temperature dependences of the charged current
neutrino  mean free path in $\beta-$stable matter for the SL22 model
assuming $Y_L=0.4$. Results for the Hartree approximation (left panel) are
compared with those including RPA corrections (right panel) with
$g^\prime=0.6$.

\vskip 10pt
FIG. 25: Same as Fig. 21, except for neutrino-free matter and for
$q=E_\nu+\mu_e$ and $E_\nu=\pi T$.

\vskip 10pt
FIG. 26: Same as Fig. 22, except for neutrino-free matter and for
$q=E_\nu+\mu_e$ and $E_\nu=\pi T$.

\vskip 10pt
FIG. 27: Charged current differential cross sections in $\beta-$stable
matter for $Y_L=0.4$ and $T=10$ MeV at $n_B=n_0$ (left panel) and
$n_B=3n_0$ (right panel).  The relativistic model GM3 was employed.
Results for the Hartree approximation and with RPA correlations are compared
for the case $q=E_\nu=\mu_\nu$.

\vskip 10pt
FIG. 28: The density and temperature dependences of the charged current
neutrino  mean free path in $\beta-$stable matter for the GM3 model
assuming $Y_L=0.4$. Results for the Hartree approximation (left panel) are
compared with those including RPA corrections (right panel) with
$g^\prime=0.6$.

\vskip 10pt
FIG. 29: Same as Fig. 27, except for neutrino-free matter and for
$q=E_\nu+\mu_e$ and $E_\nu=\pi T$.

\vskip 10pt
FIG. 30: Same as Fig. 28, except for neutrino-free matter and for
$q=E_\nu+\mu_e$ and $E_\nu=\pi T$.

\vskip 10pt
FIG. 31: RPA corrections to the diffusion coefficients in the
density-temperature plane for the field-theoretical model GM3.  Results are
shown for the three diffusion coefficients $D_2$, $D_3$, and $D_4$ (upper,
middle, and lower panels, respectively) for neutrino-free matter (left panels)
and neutrino-rich matter (right panels).

\vskip 10pt
FIG: 32: The density and temerature dependence of the ratio of the  RPA and
Hartree diffusion coefficients ($D_i^{RPA}/D_i^{H}$) in neutrino-free matter.

\vskip 10pt
FIG: 33: Central values of thermodynamic quantities (entropy $s$, temperature
$T$, baryon density $n_B$, $\nu_e$ concentration $Y_\nu$, $\nu_e$ chemical
potential $\mu_\nu$, electron concentration $Y_e$) during the evolution of
a baryon mass 1.4 M$_\odot$ protoneutron star.  The relativistic model GM3 for
nucleons-only matter was employed.  Results are compared for the
Hartree approximation (baseline case), the inclusion of RPA correlations, and a
constant factor of three reduction of the opacities for densities greater than
$u=n_B/n_0=1/2$ and 1/4.  The latter case corresponds to that studied in Ref.
\cite{BS}.

\vskip 10pt
FIG 34: The upper panel shows the total emitted neutrino luminosity for the
protoneutron star evolutions described in Fig. 33.  The curves are labelled
as in Fig. 32.  The lower panel shows the ratio of the luminosities
obtained in the three  models which contain corrections to the baseline
(Hartree approximation) model.

\vskip 10pt
FIG 35: The total emitted neutrino luminosity for long-term
protoneutron star evolutions described in Fig. 34.  The curves are labelled
as in Fig. 34.

\newpage
\begin{figure}
\begin{center}
\epsfxsize=6.0in
\epsfysize=7.0in
\epsffile{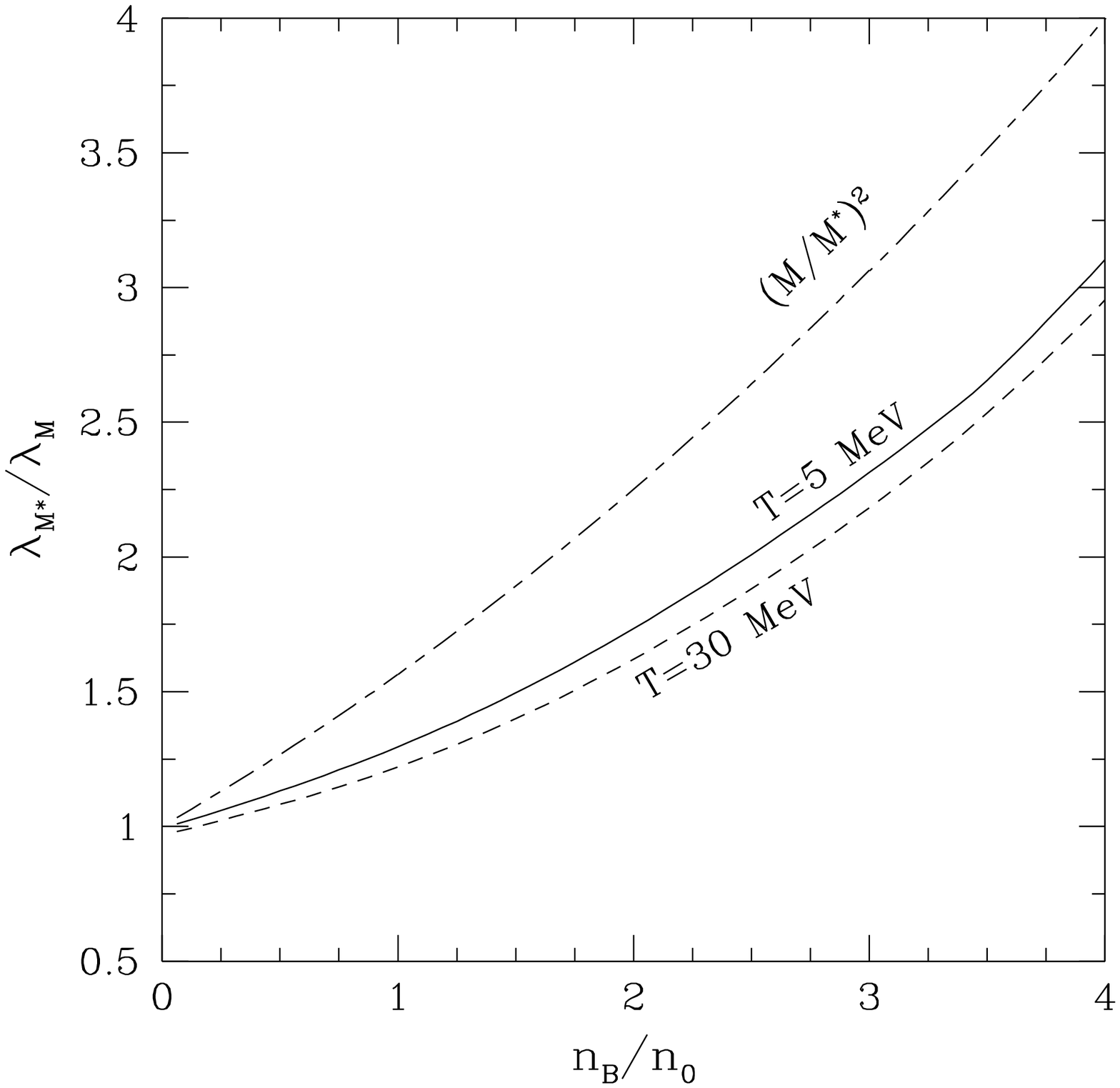}
\caption[]{}
\label{sig1}
\end{center}
\end{figure}

\newpage
\begin{figure}
\begin{center}
\epsfxsize=6.0in
\epsfysize=7.0in
\epsffile{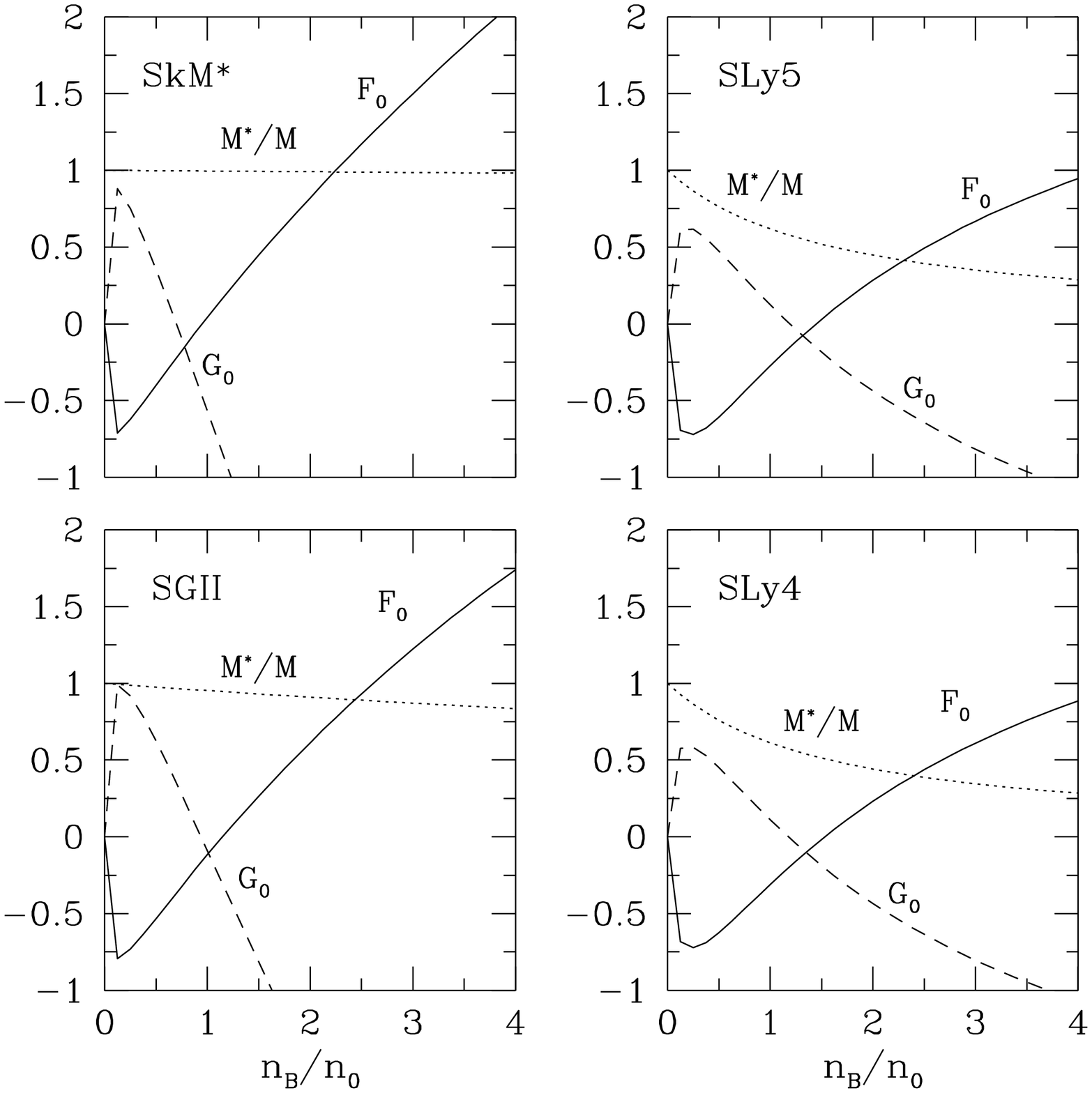}
\caption[]{}
\label{nfl}
\end{center}
\end{figure}

\newpage
\begin{figure}
\begin{center}
\epsfxsize=6.0in
\epsfysize=7.0in
\epsffile{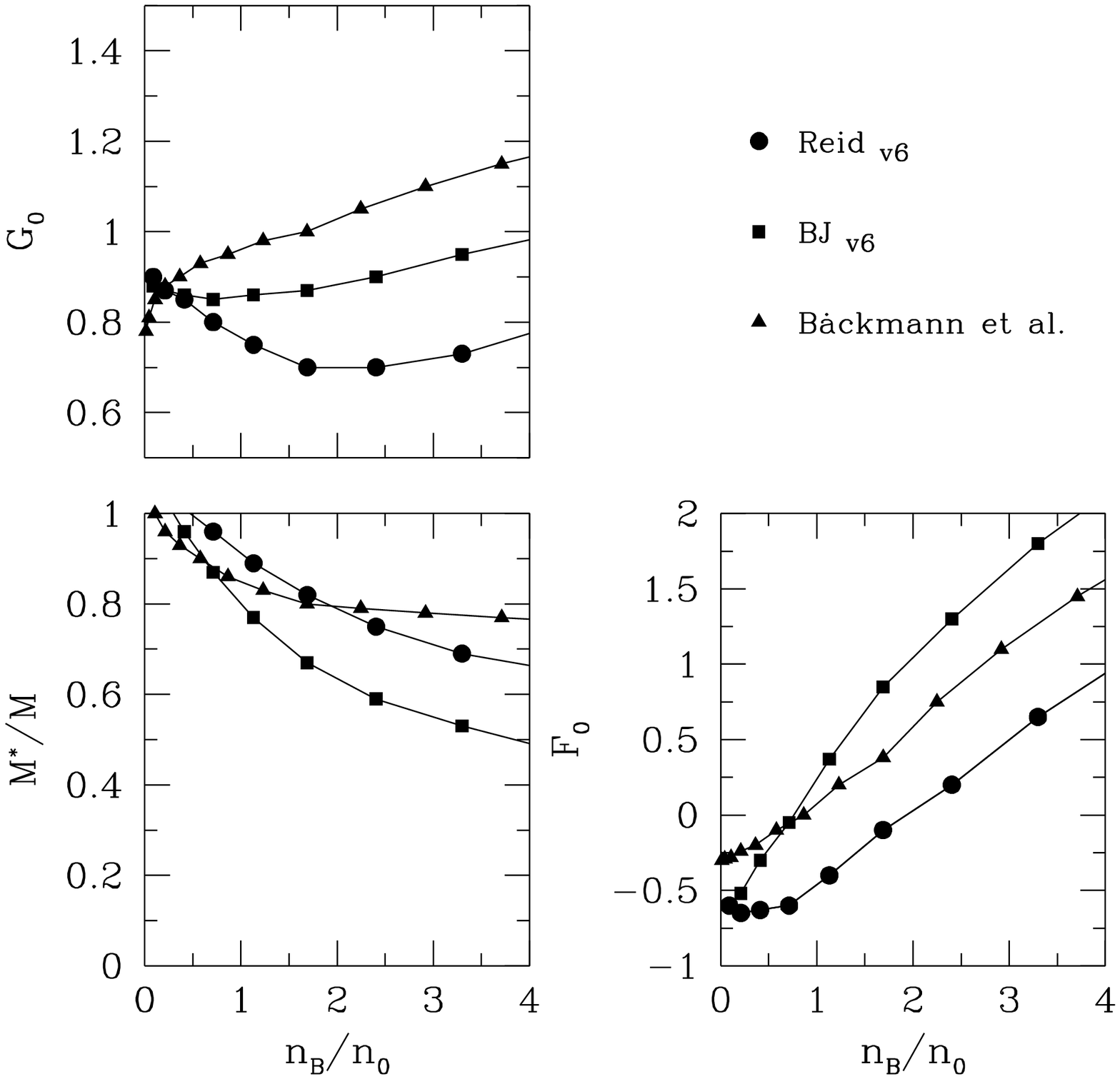}
\caption[]{}
\label{vfl}
\end{center}
\end{figure}

\newpage
\begin{figure}
\begin{center}
\epsfxsize=6.0in
\epsfysize=7.0in
\epsffile{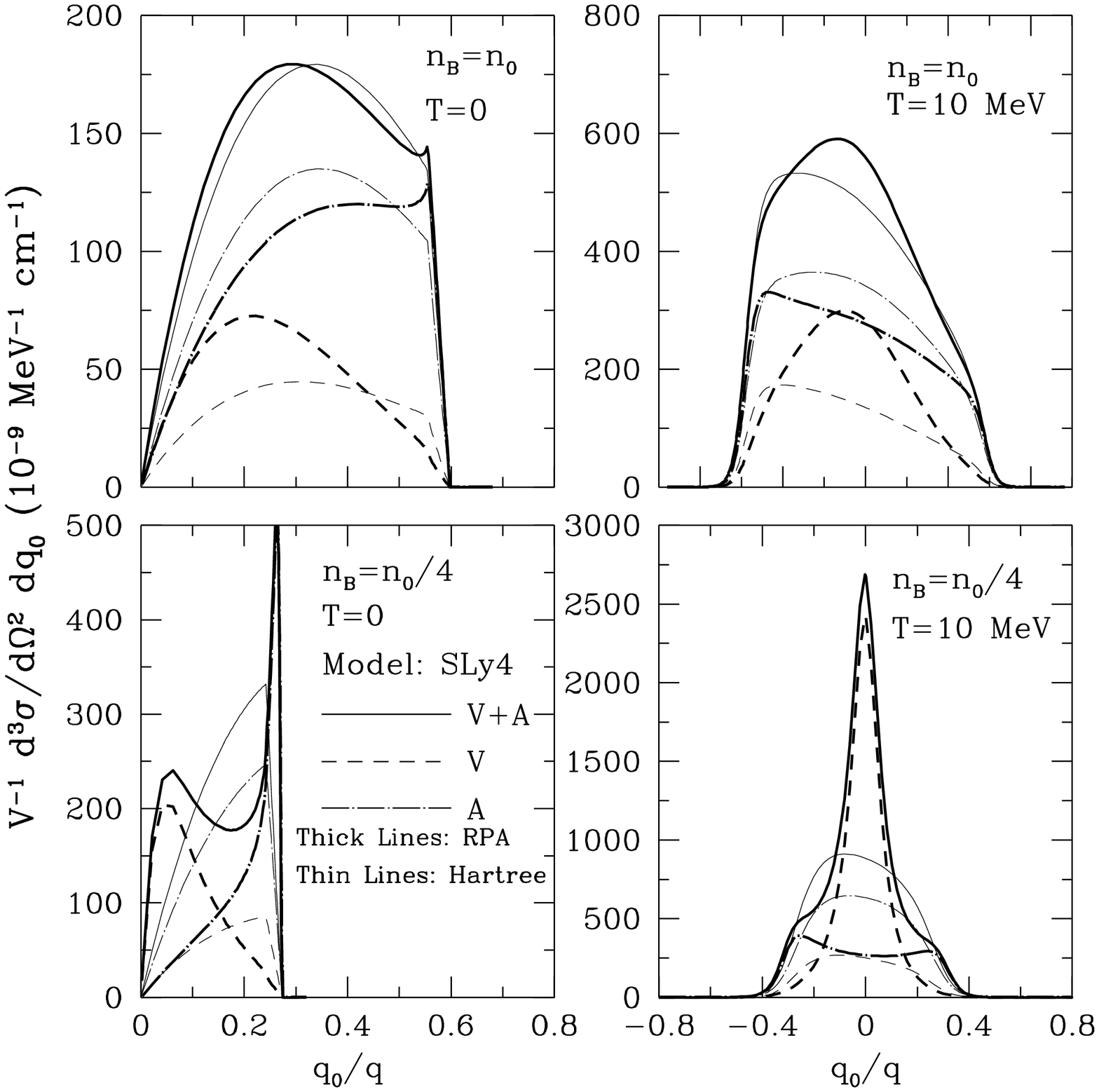}
\caption[]{}
\label{dsig_SLy4}
\end{center}
\end{figure}

\newpage
\begin{figure}
\begin{center}
\epsfxsize=6.0in
\epsfysize=7.0in
\epsffile{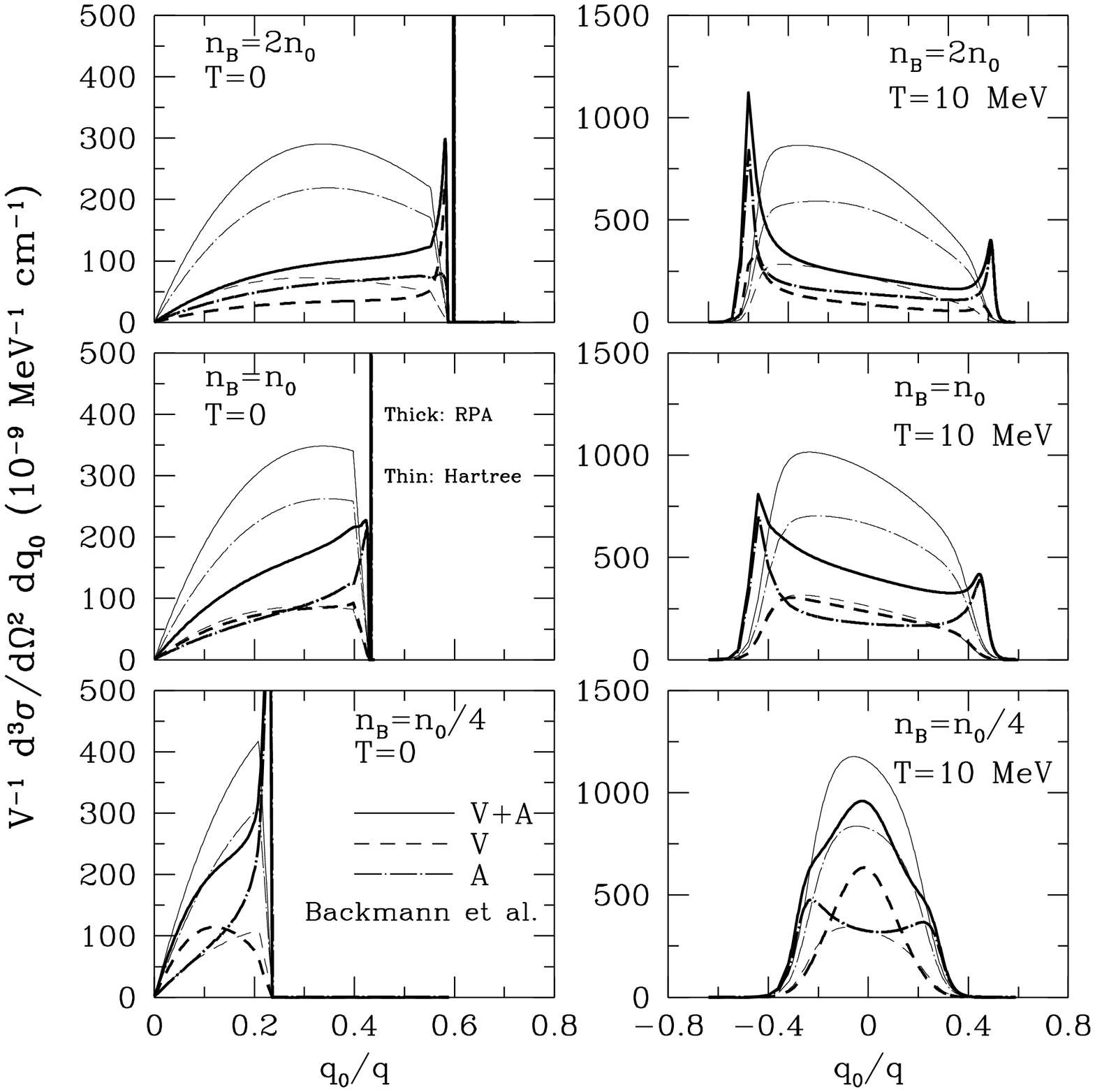}
\caption[]{}
\label{dsigvar}
\end{center}
\end{figure}

\newpage
\begin{figure}
\begin{center}
\epsfxsize=6.0in
\epsfysize=7.0in
\epsffile{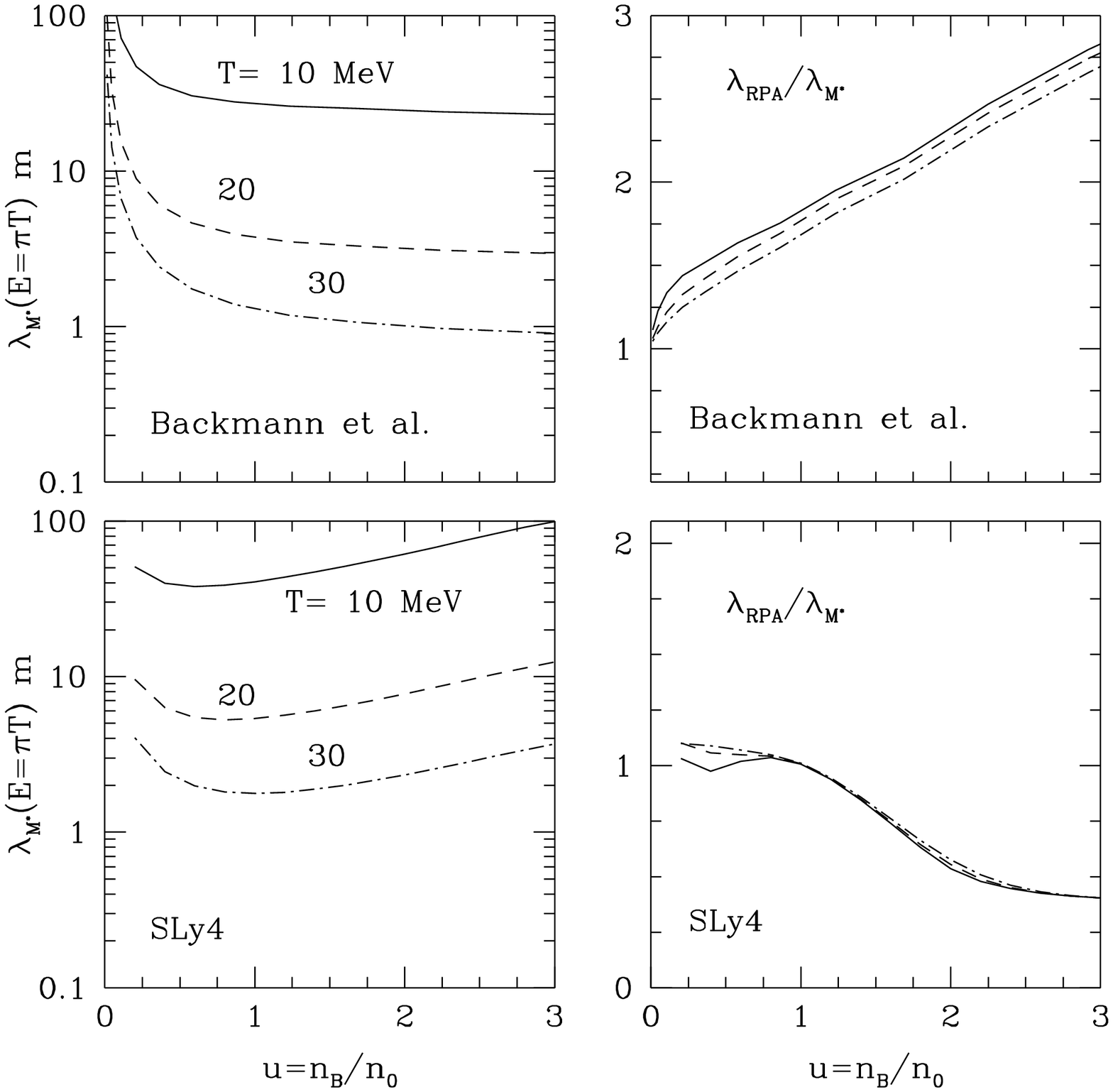}
\caption[]{}
\label{nmnr_lambda}
\end{center}
\end{figure}

\newpage
\begin{figure}
\begin{center}
\epsfxsize=6.0in
\epsfysize=7.0in
\epsffile{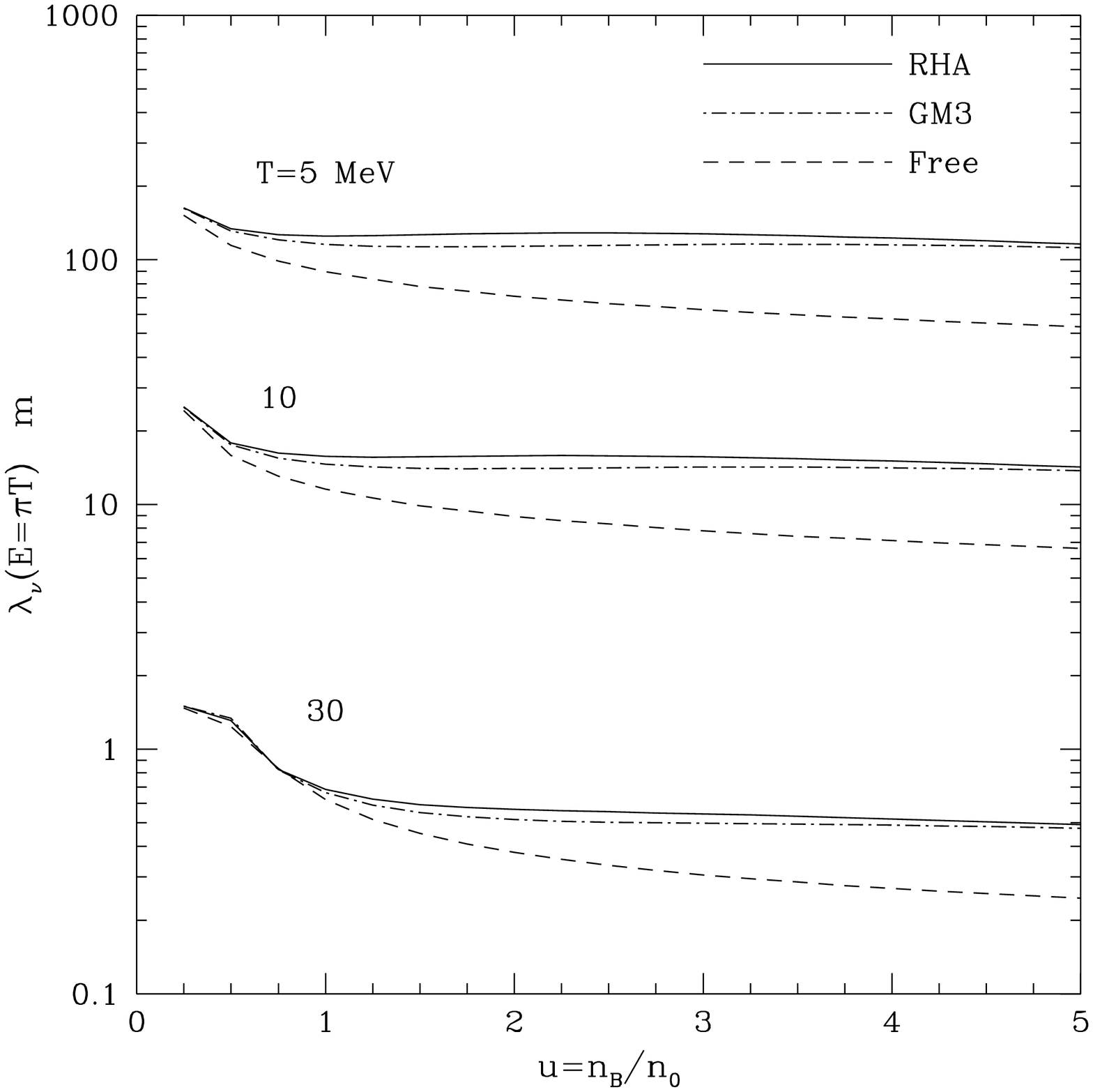}
\caption[]{}
\label{sighartree}
\end{center}
\end{figure}

\newpage
\begin{figure}
\begin{center}
\epsfxsize=6.0in
\epsfysize=7.0in
\epsffile{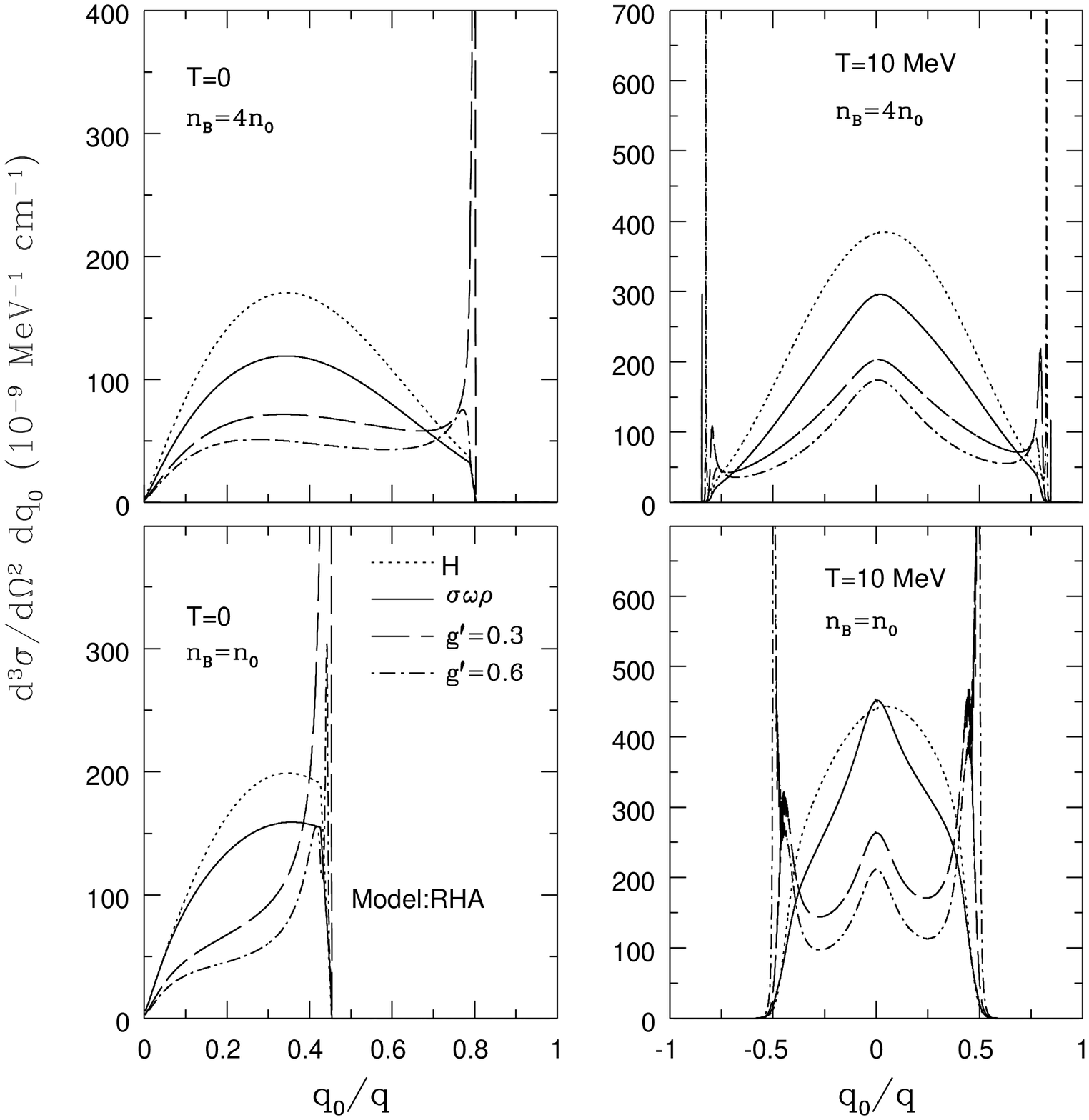}
\caption[]{}
\label{dsigprh}
\end{center}
\end{figure}

\newpage
\begin{figure}
\begin{center}
\epsfxsize=6.0in
\epsfysize=7.0in
\epsffile{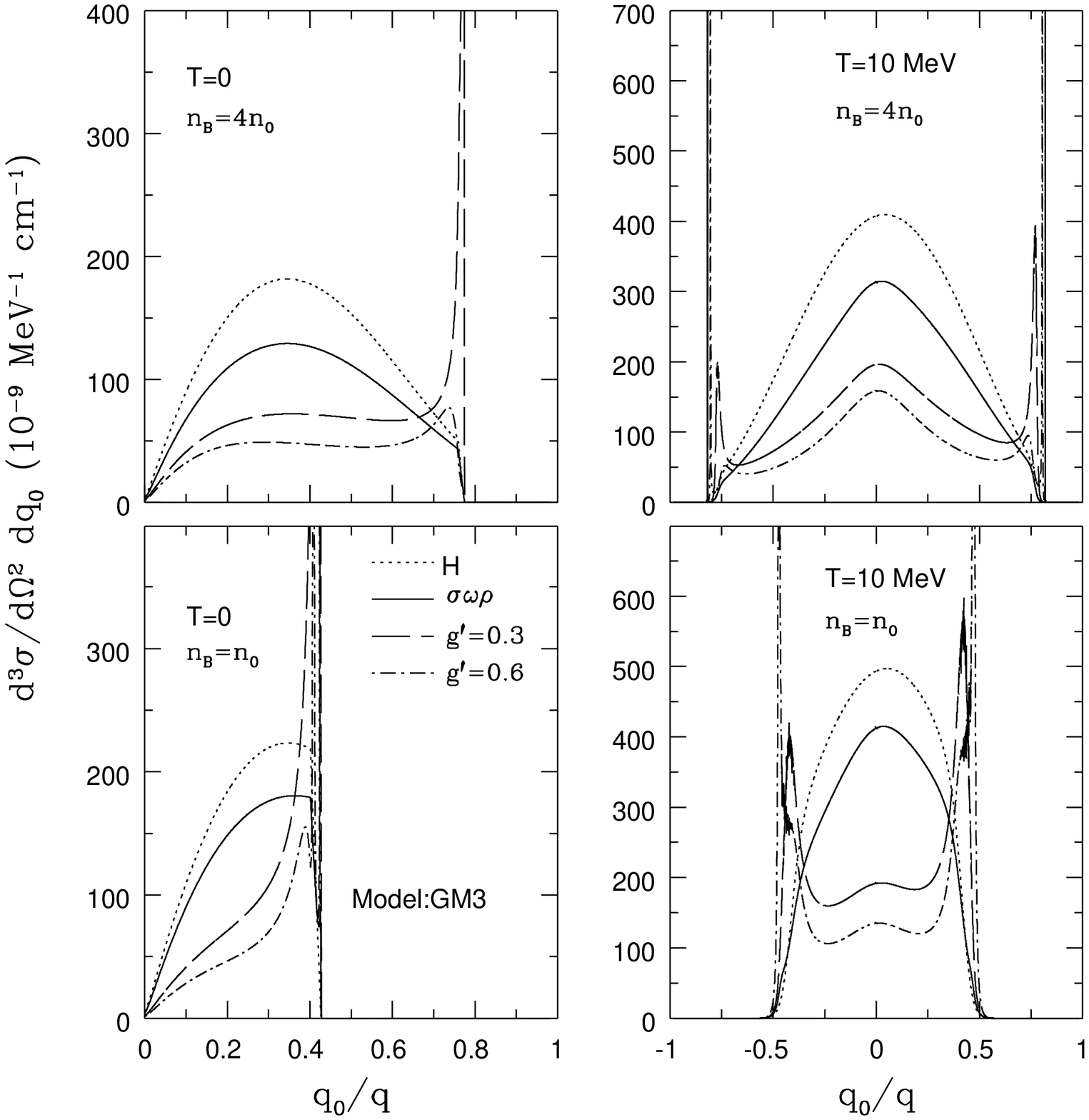}
\caption[]{}
\label{dsiggm3}
\end{center}
\end{figure}

\newpage
\begin{figure}
\begin{center}
\epsfxsize=6.0in
\epsfysize=7.0in
\epsffile{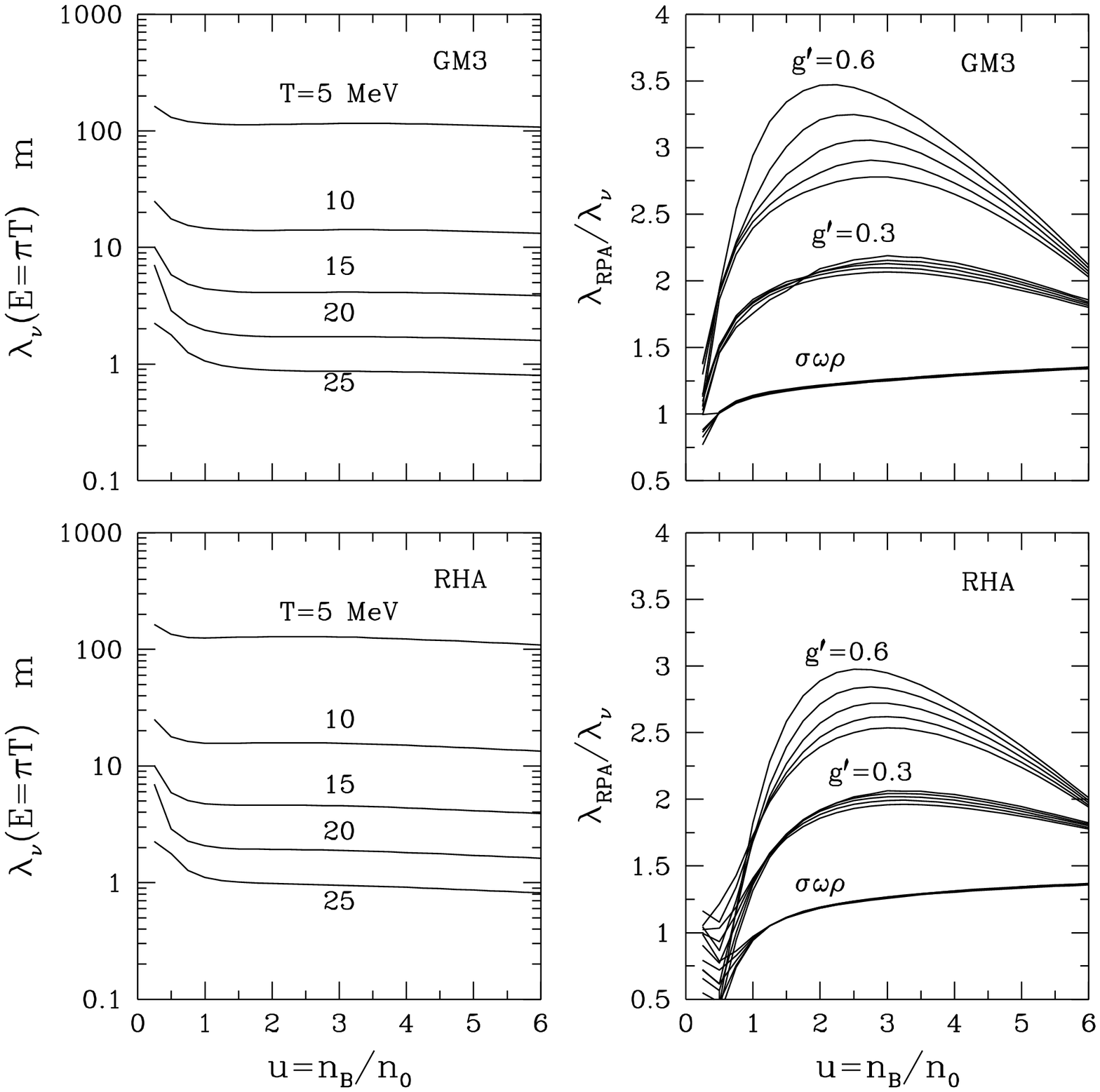}
\caption[]{}
\label{sigrel}
\end{center}
\end{figure}

\newpage
\begin{figure}
\begin{center}
\epsfxsize=6.0in
\epsfysize=7.0in
\epsffile{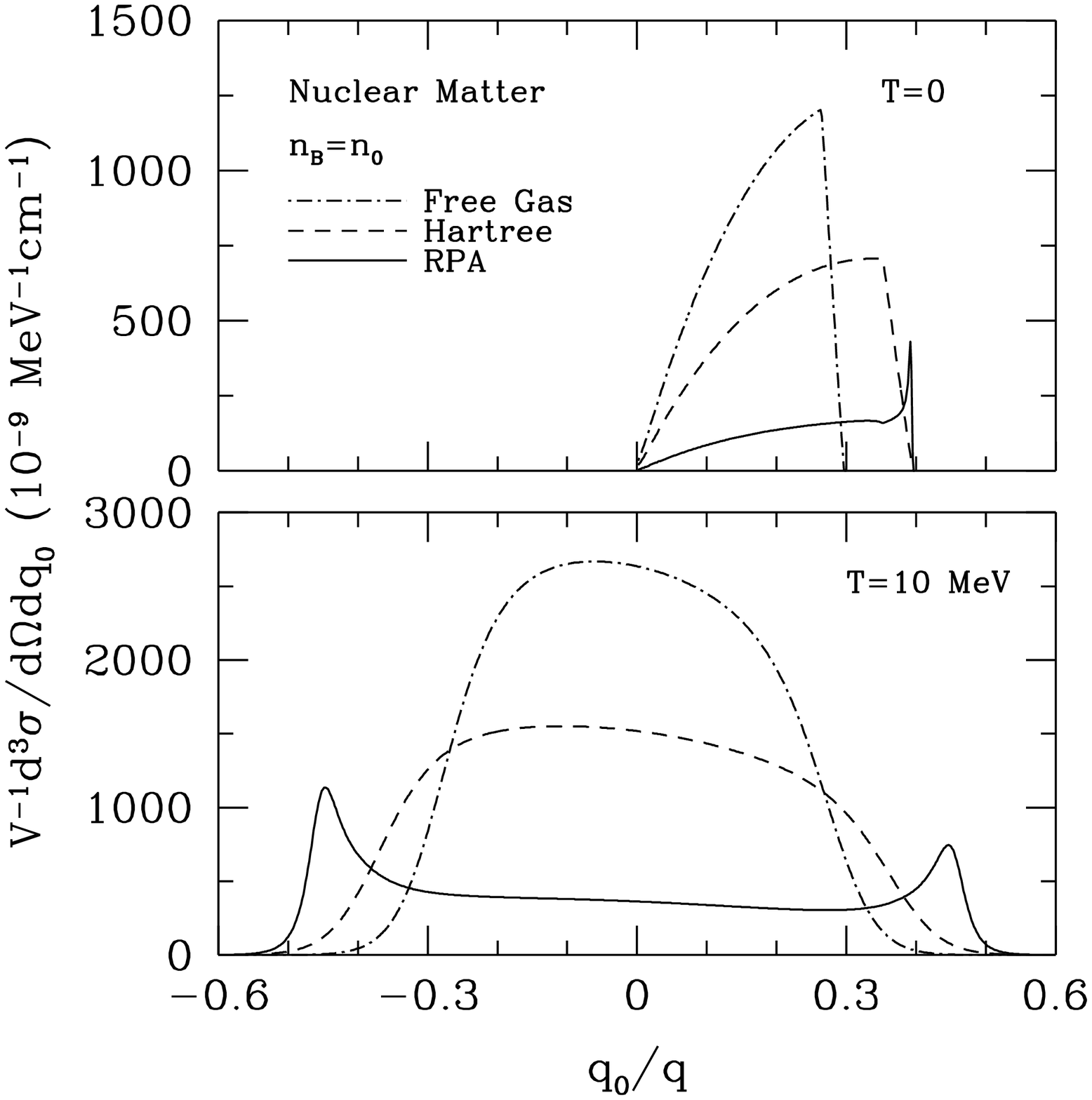}
\caption[]{}
\label{nucdsig}
\end{center}
\end{figure}

\newpage
\begin{figure}
\begin{center}
\epsfxsize=6.0in
\epsfysize=7.0in
\epsffile{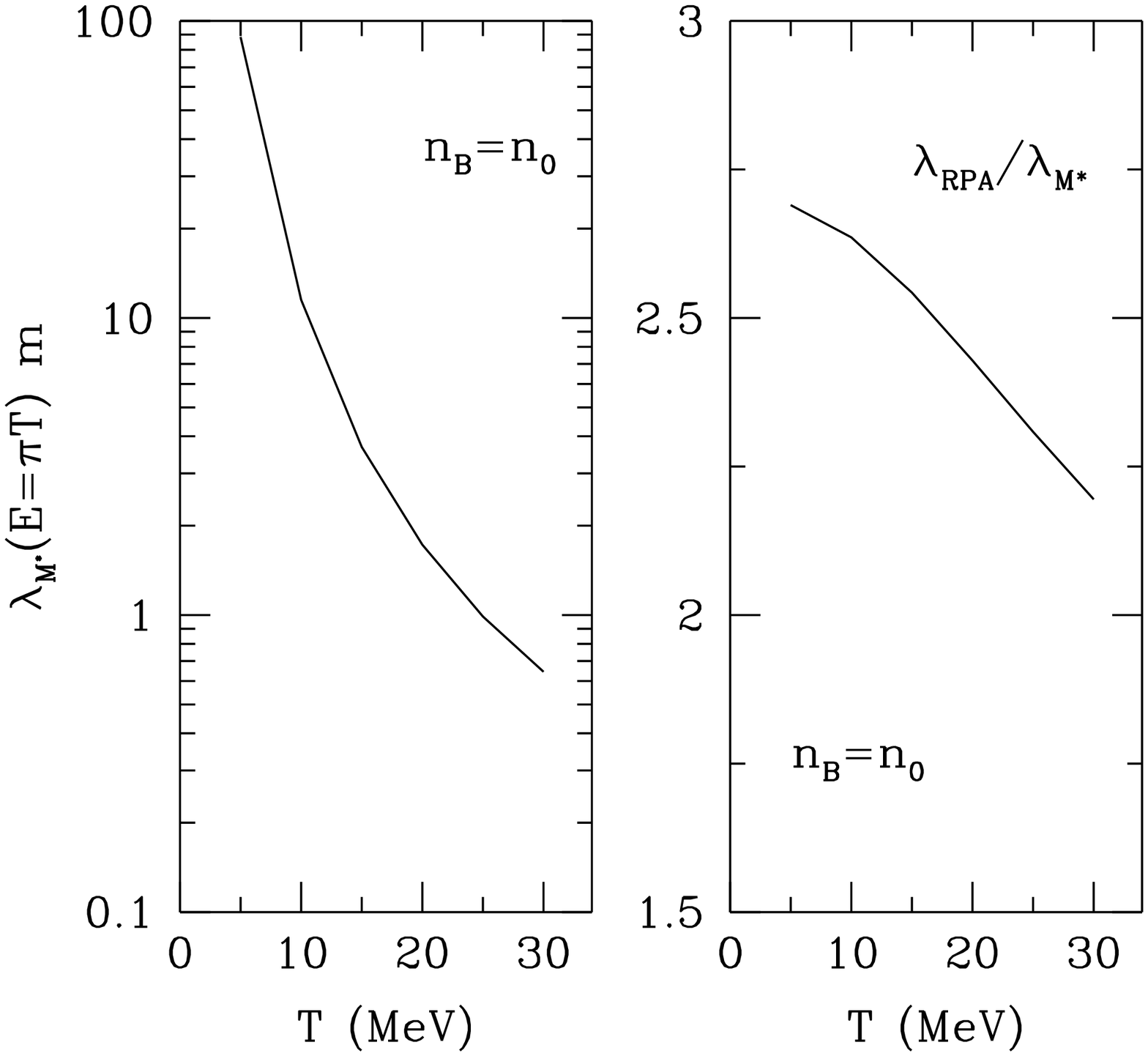}
\caption[]{}
\label{nucsig}
\end{center}
\end{figure}

\newpage
\begin{figure}
\begin{center}
\epsfxsize=6.0in
\epsfysize=7.0in
\epsffile{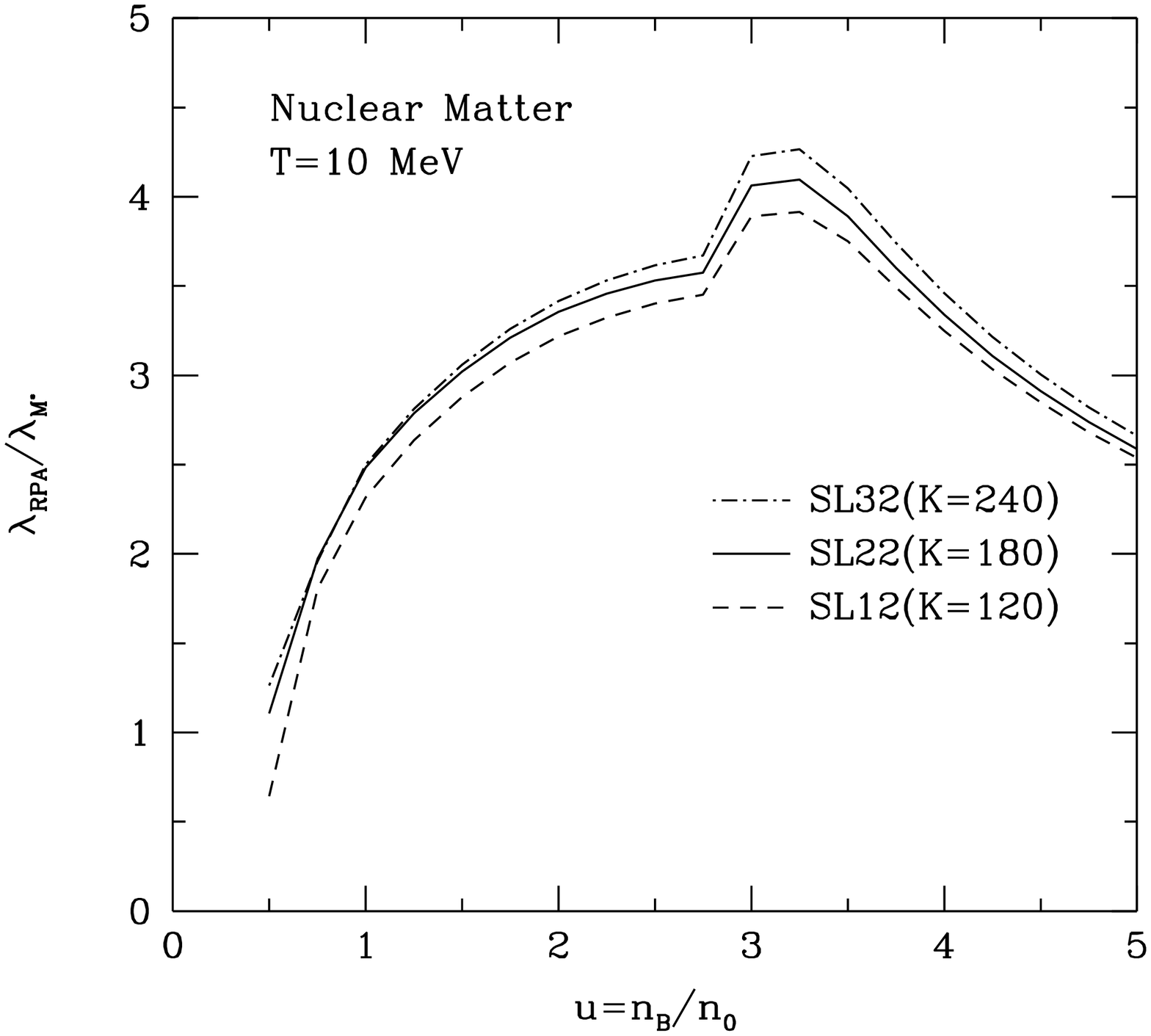}
\caption[]{}
\label{nucSLn2}
\end{center}
\end{figure}

\newpage
\begin{figure}
\begin{center}
\epsfxsize=6.0in
\epsfysize=7.0in
\epsffile{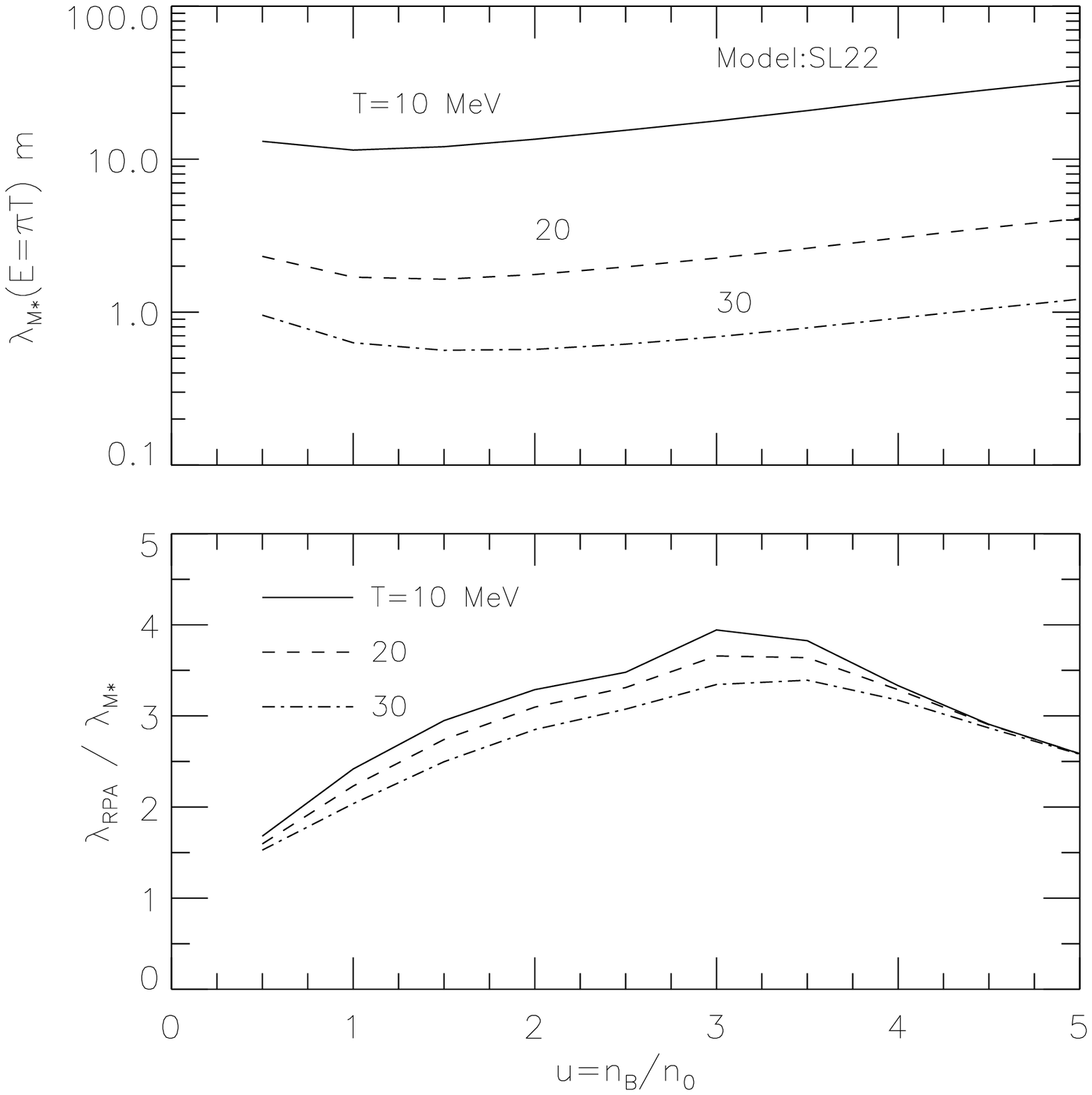}
\caption[]{}
\label{pnucsig}
\end{center}
\end{figure}

\newpage
\begin{figure}
\begin{center}
\epsfxsize=6.0in
\epsfysize=7.0in
\epsffile{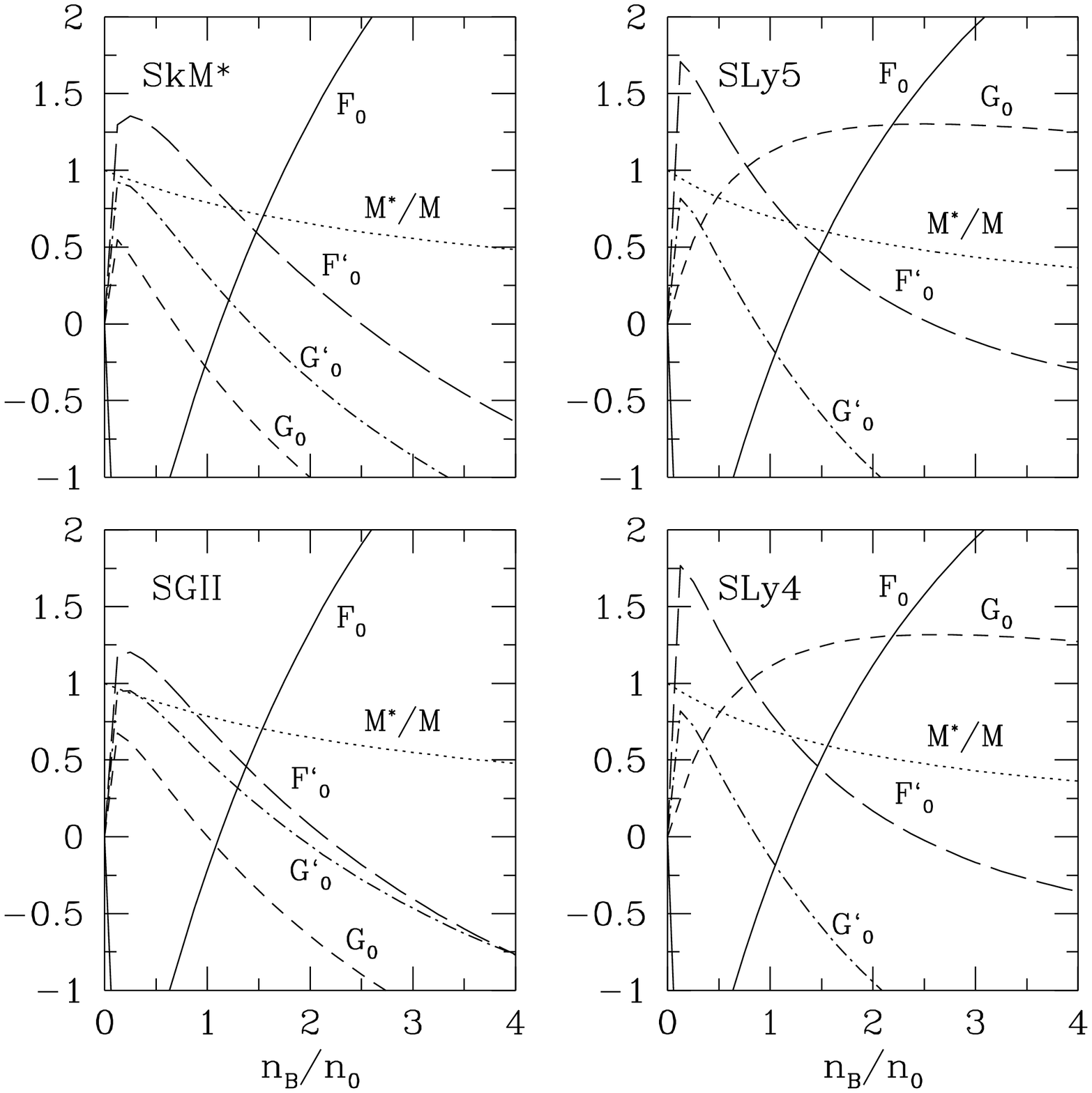}
\caption[]{}
\label{fliquid}
\end{center}
\end{figure}

\newpage
\begin{figure}
\begin{center}
\epsfxsize=6.0in
\epsfysize=7.0in
\epsffile{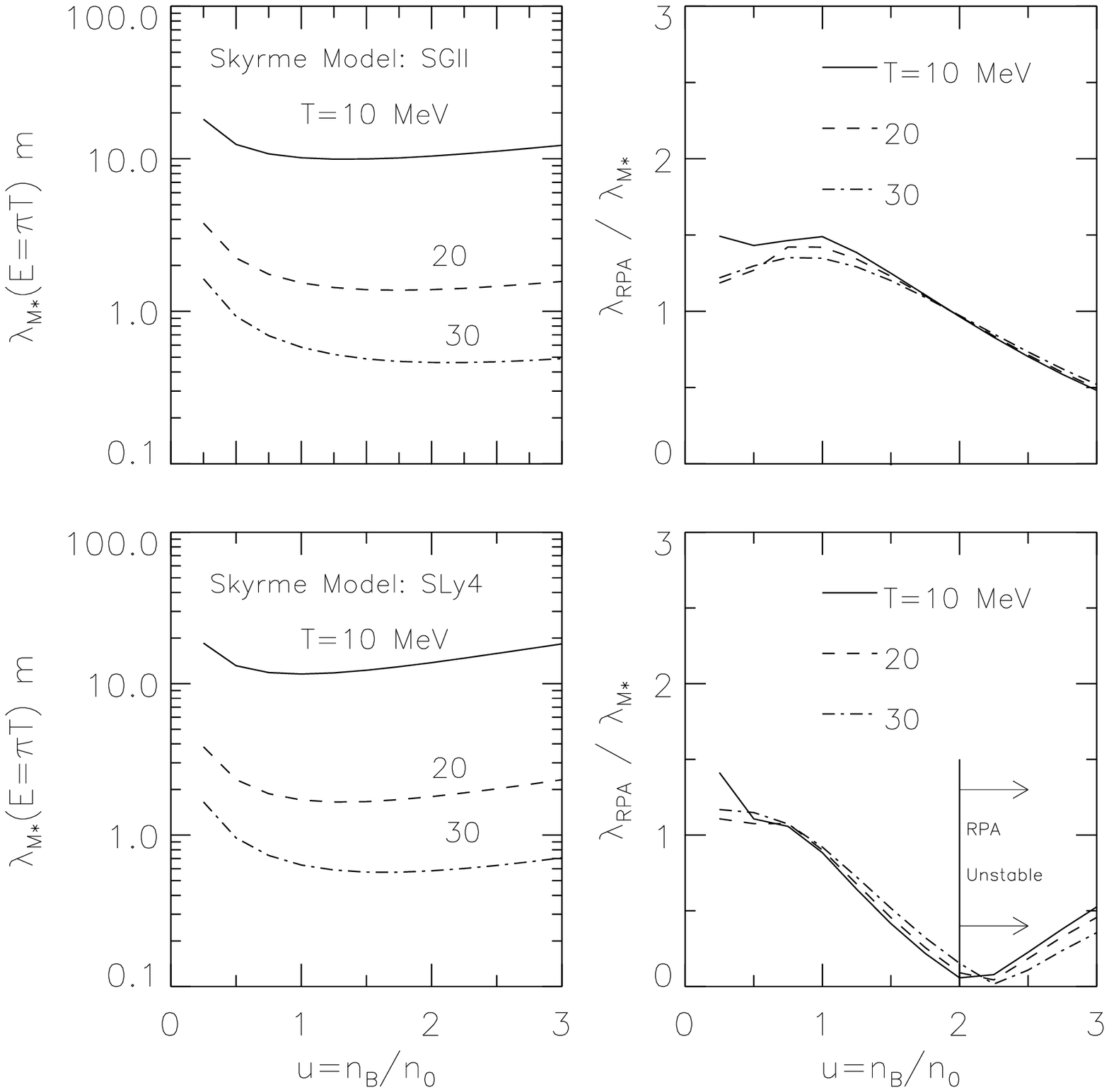}
\caption[]{}
\label{skynucsig}
\end{center}
\end{figure}

\newpage
\begin{figure}
\begin{center}
\epsfxsize=6.0in
\epsfysize=7.0in
\epsffile{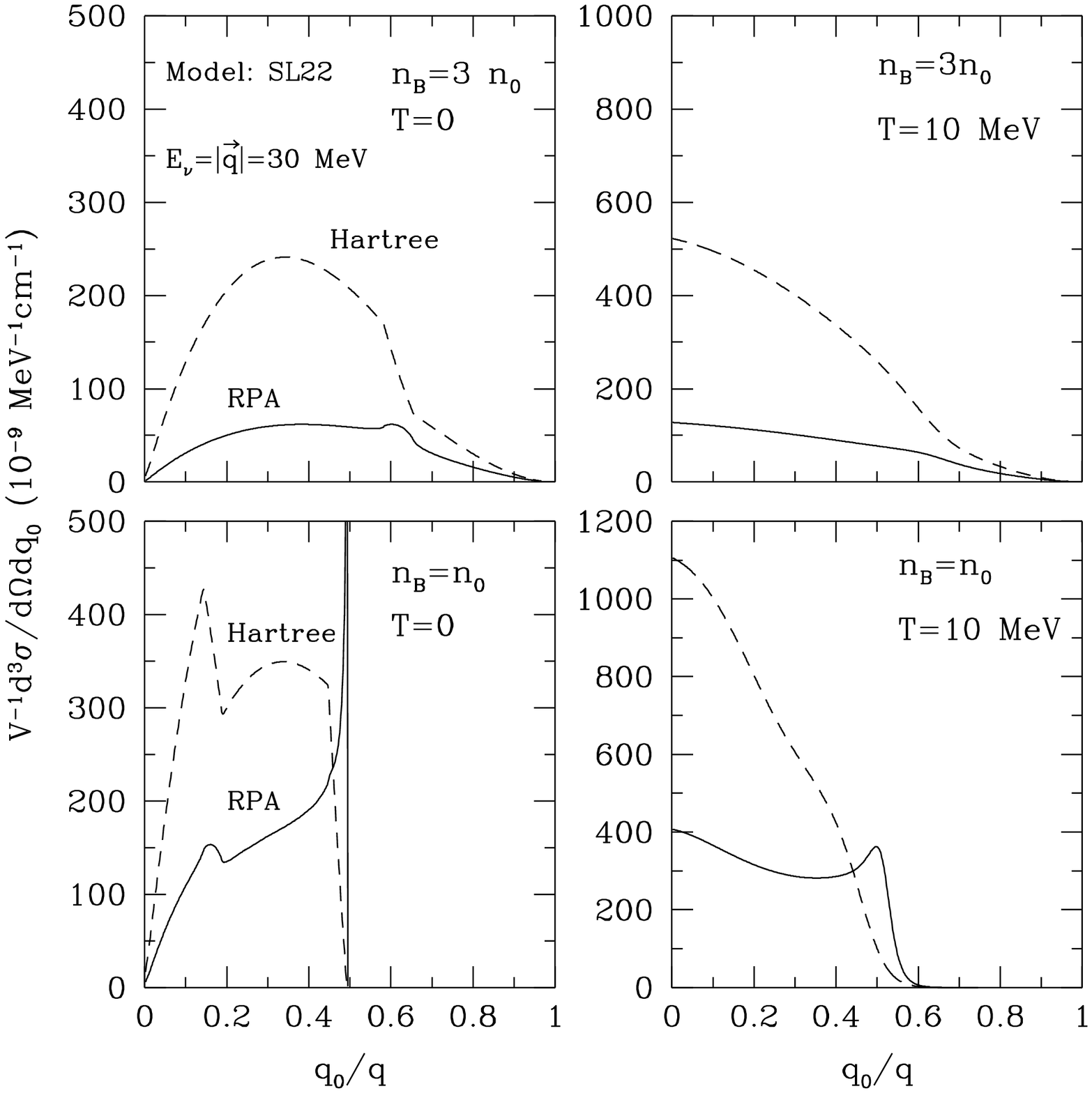}
\caption[]{}
\label{paldsig}
\end{center}
\end{figure}

\newpage
\begin{figure}
\begin{center}
\epsfxsize=6.0in
\epsfysize=7.0in
\epsffile{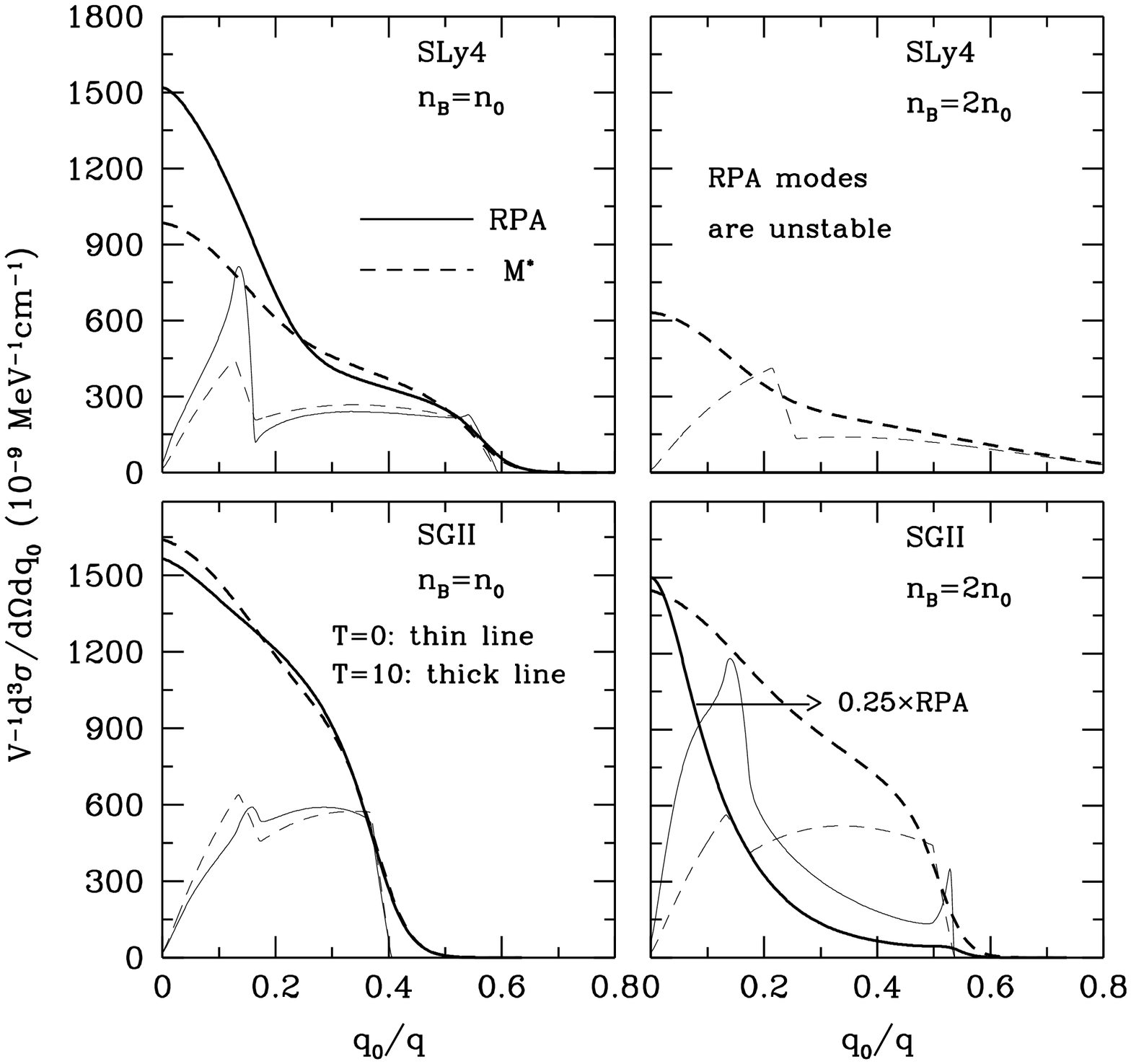}
\caption[]{}
\label{betadsig2}
\end{center}
\end{figure}

\newpage
\begin{figure}
\begin{center}
\epsfxsize=6.0in
\epsfysize=7.0in
\epsffile{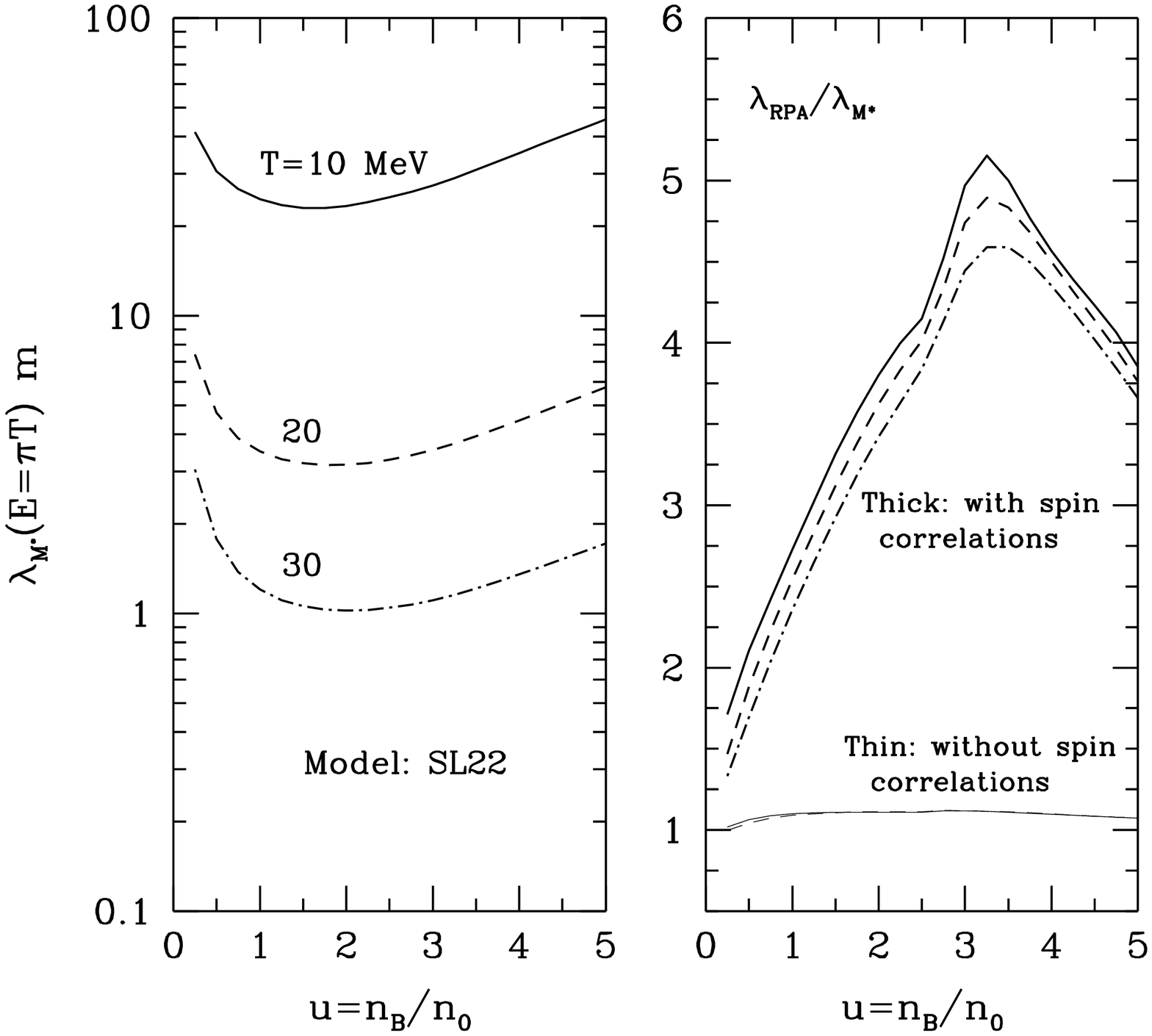}
\caption[]{}
\label{betasig}
\end{center}
\end{figure}

\newpage
\begin{figure}
\begin{center}
\epsfxsize=6.0in
\epsfysize=7.0in
\epsffile{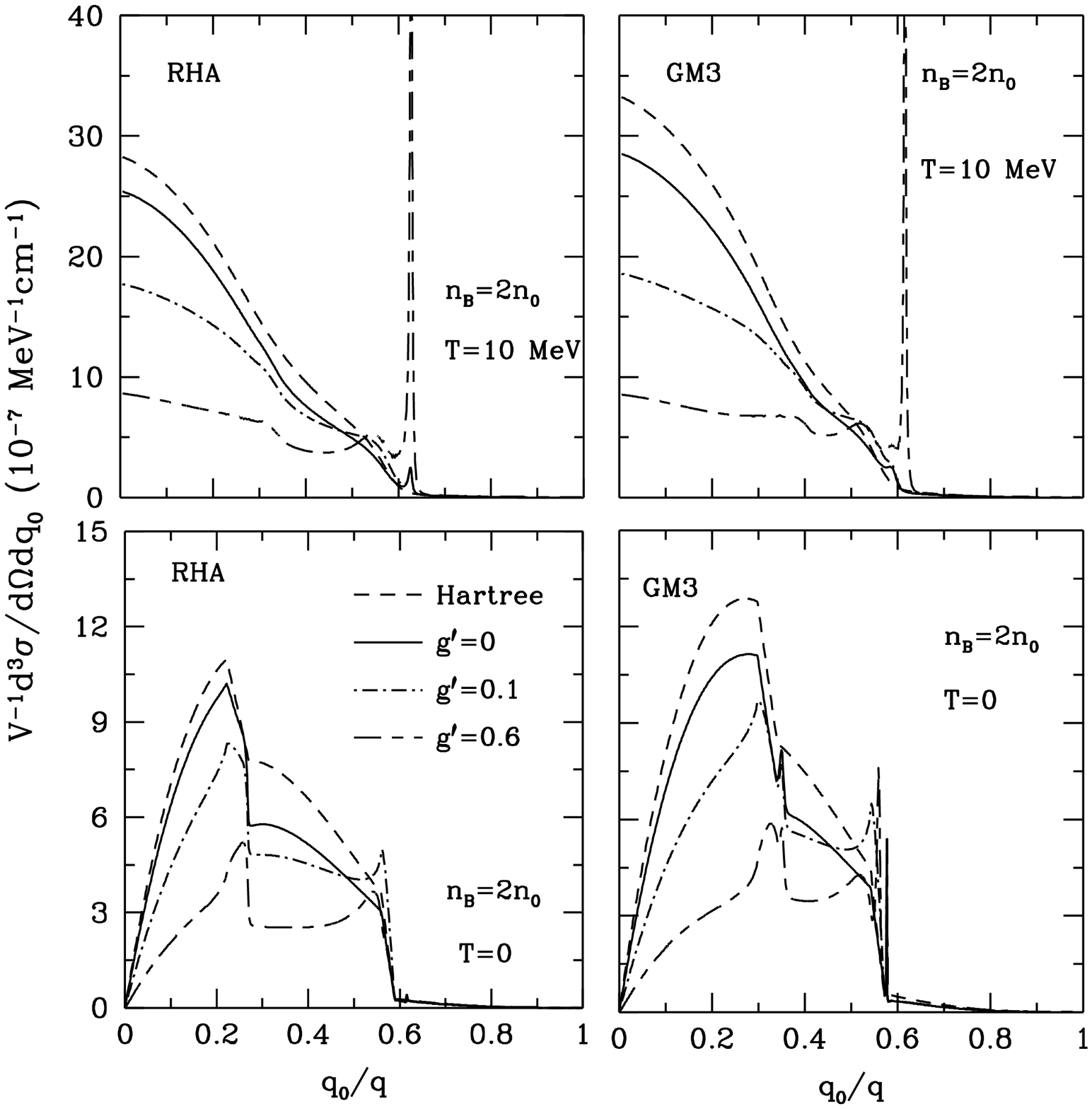}
\caption[]{}
\label{rdsig}
\end{center}
\end{figure}

\newpage
\begin{figure}
\begin{center}
\epsfxsize=6.0in
\epsfysize=7.0in
\epsffile{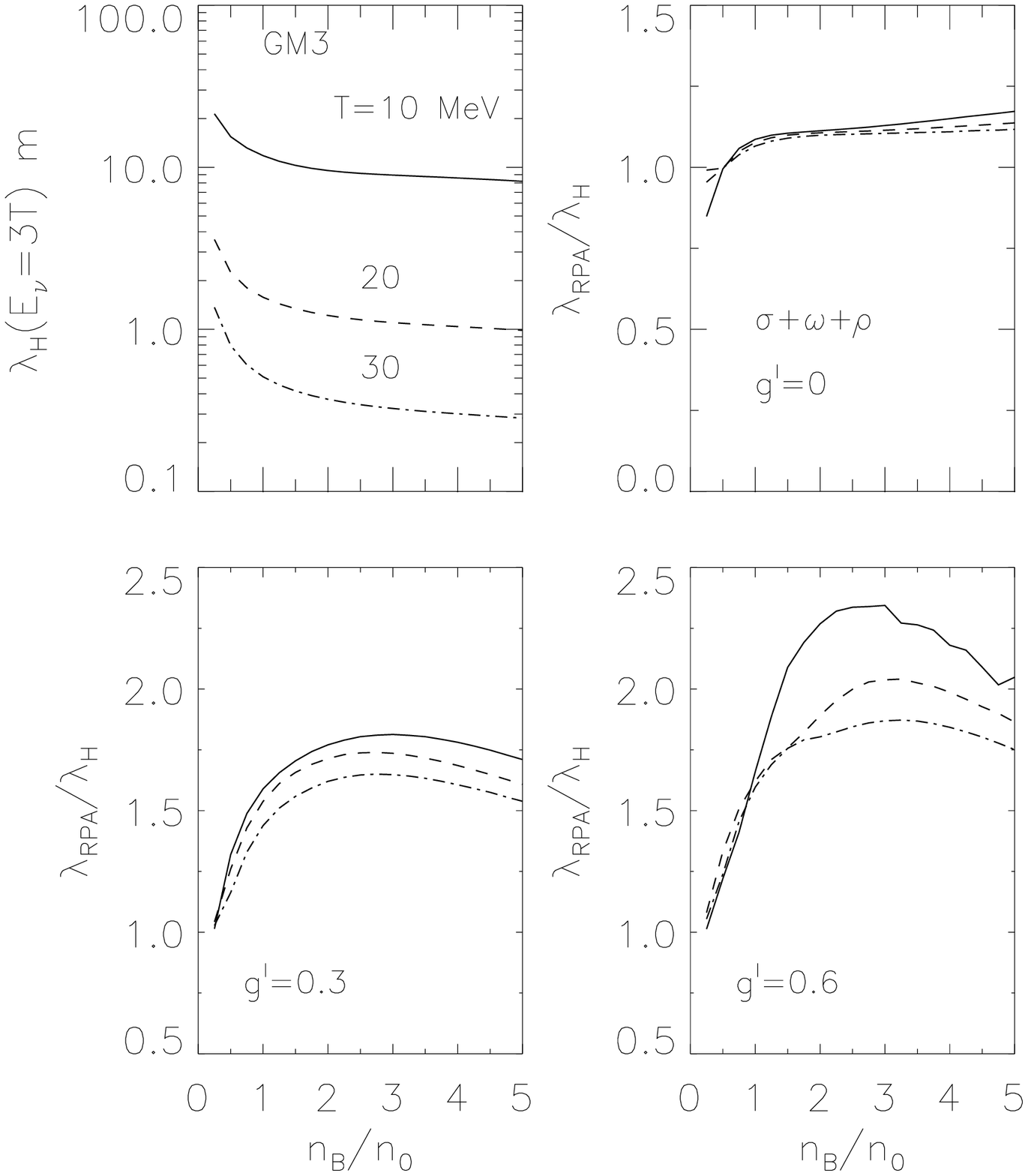}
\caption[]{}
\label{rlam}
\end{center}
\end{figure}

\newpage
\begin{figure}
\begin{center}
\epsfxsize=6.in
\epsfysize=7.in
\epsffile{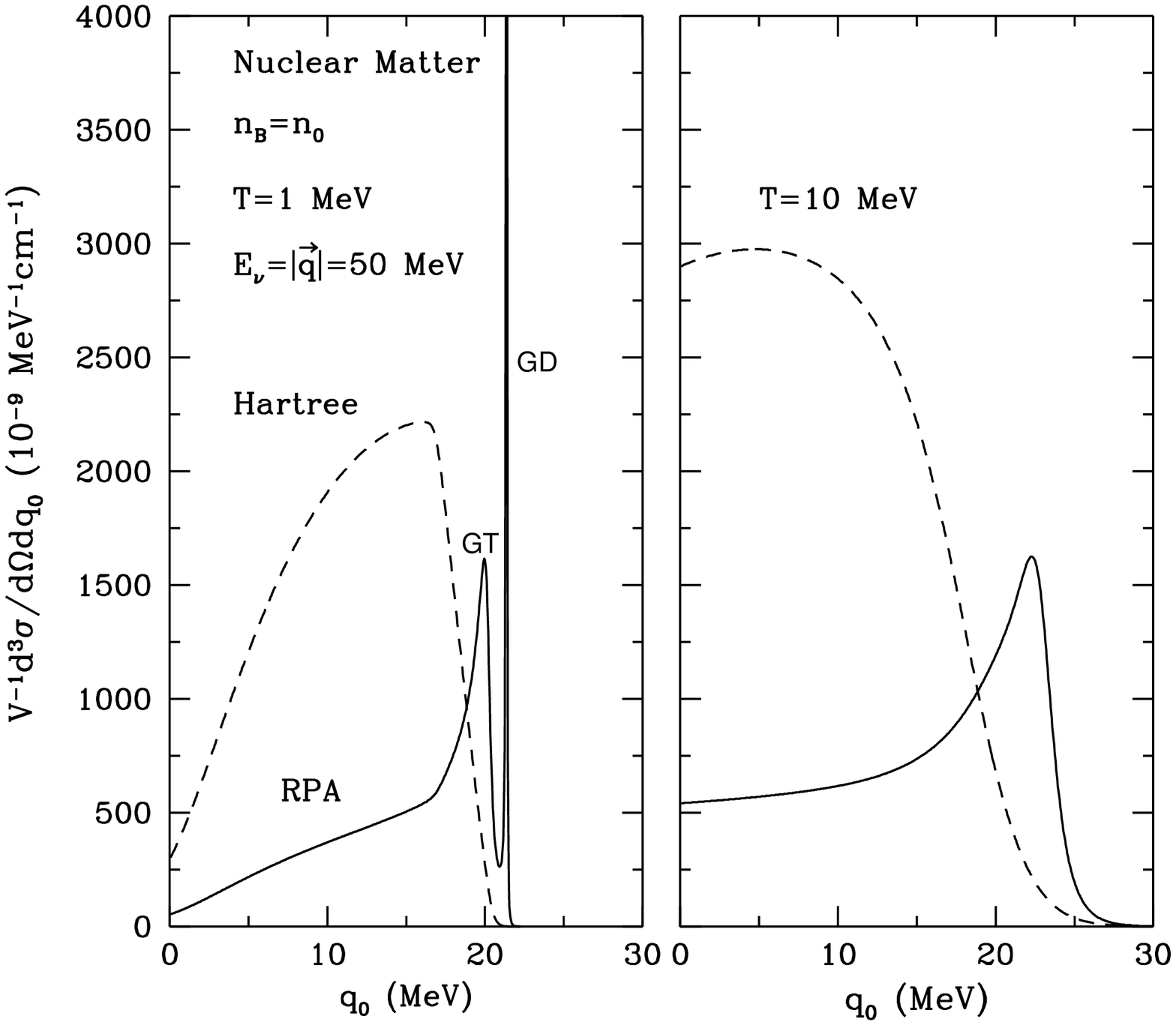}
\caption[]{}
\label{absnuc}
\end{center}
\end{figure}

\newpage
\begin{figure}
\begin{center}
\epsfxsize=6.in
\epsfysize=7.in
\epsffile{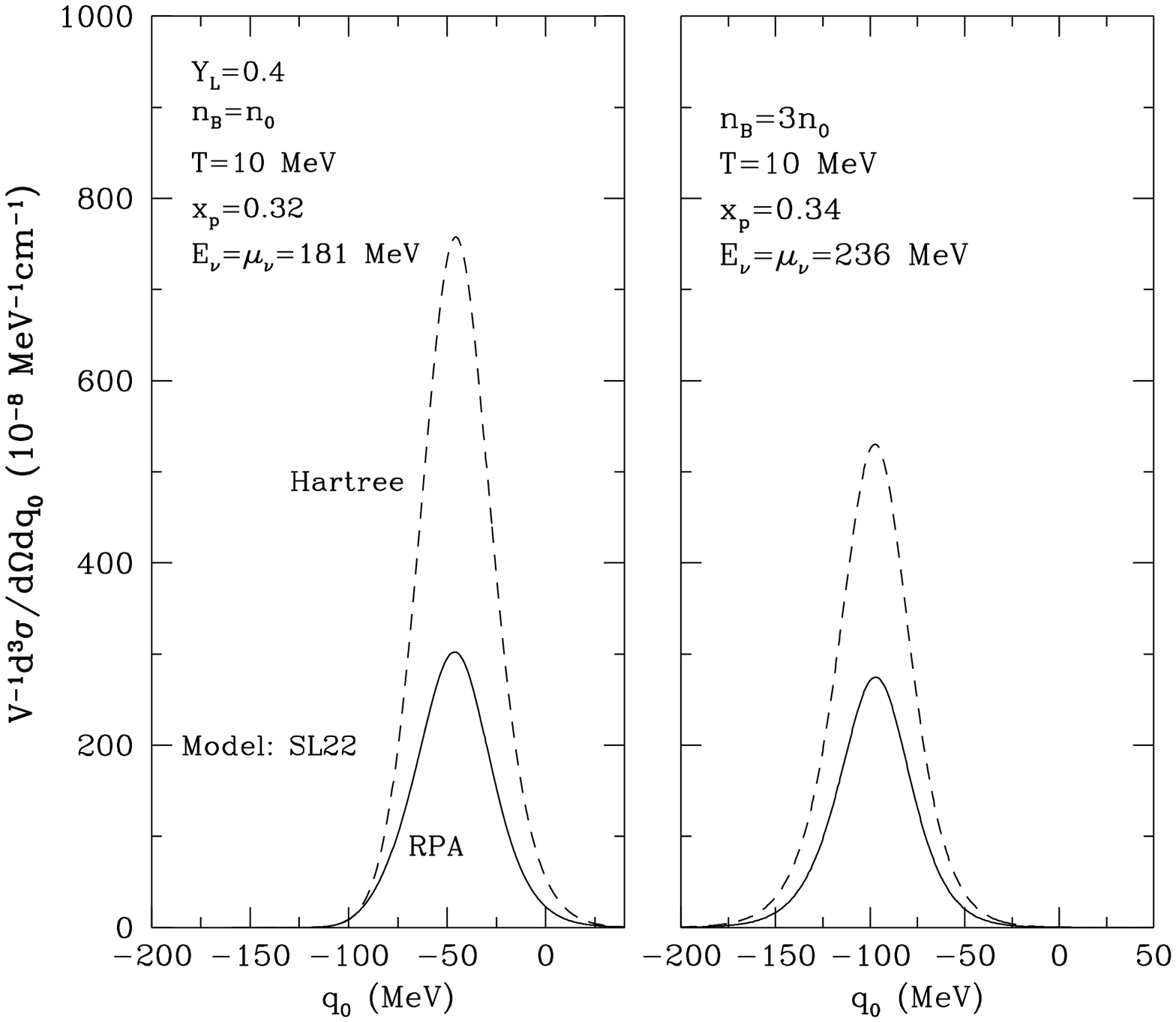}
\caption[]{}
\label{dabs_y4}
\end{center}
\end{figure}

\newpage
\begin{figure}
\begin{center}
\epsfxsize=6.in
\epsfysize=7.in
\epsffile{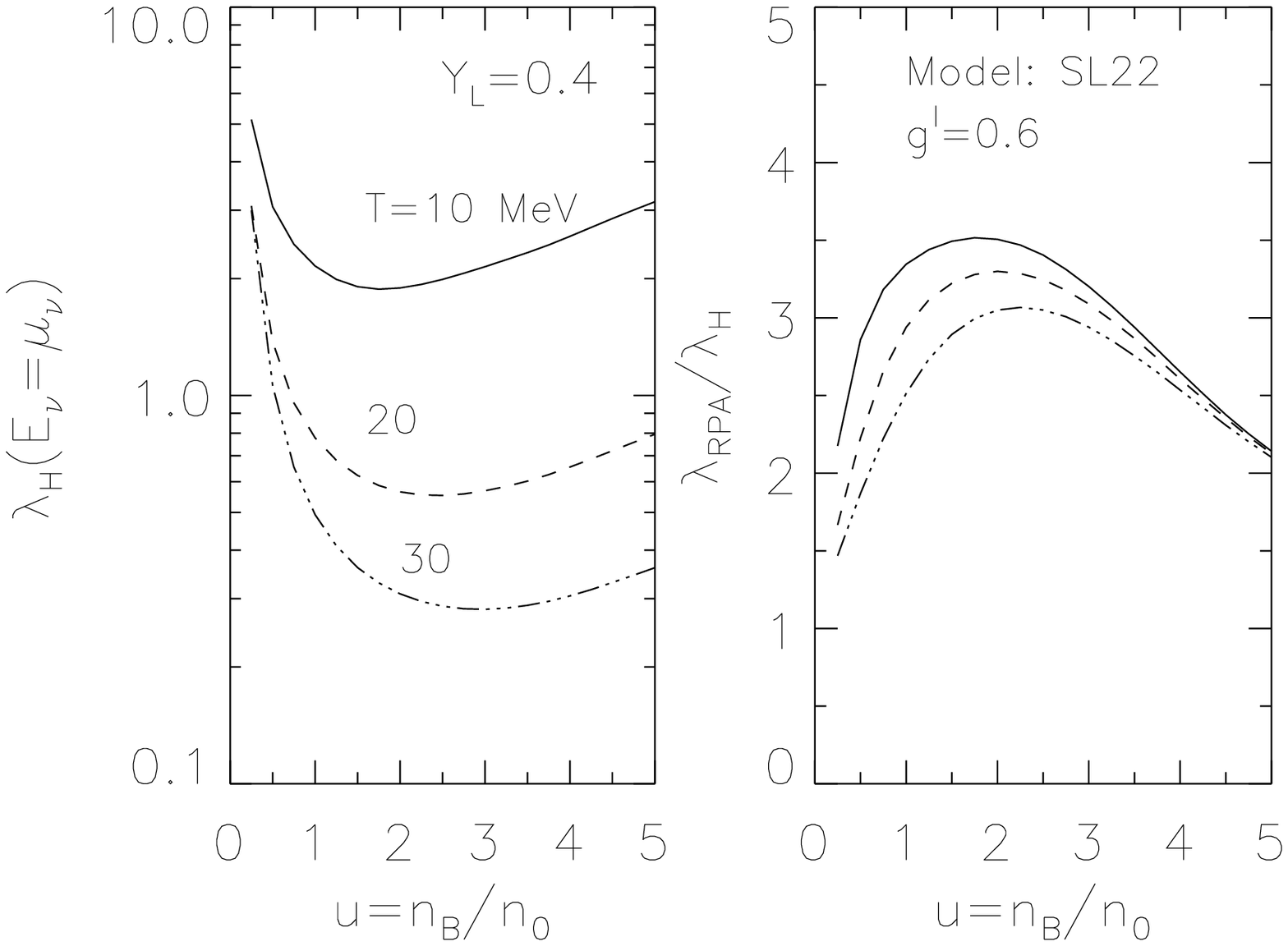}
\caption[]{}
\label{asig_yl}
\end{center}
\end{figure}

\newpage
\begin{figure}
\begin{center}
\epsfxsize=6.in
\epsfysize=7.in
\epsffile{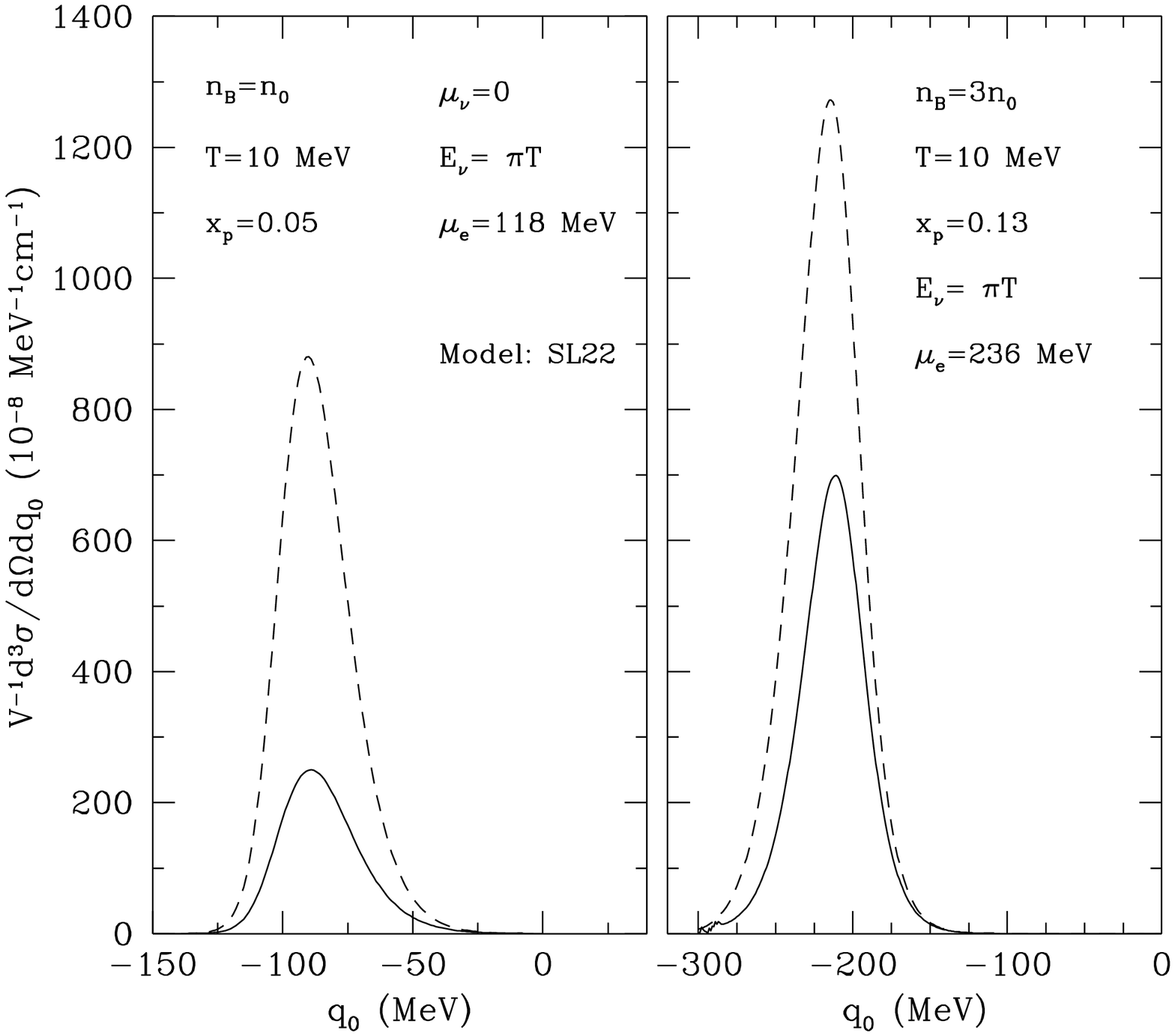}
\caption[]{}
\label{dabs_y0}
\end{center}
\end{figure}

\newpage
\begin{figure}
\begin{center}
\epsfxsize=6.in
\epsfysize=7.in
\epsffile{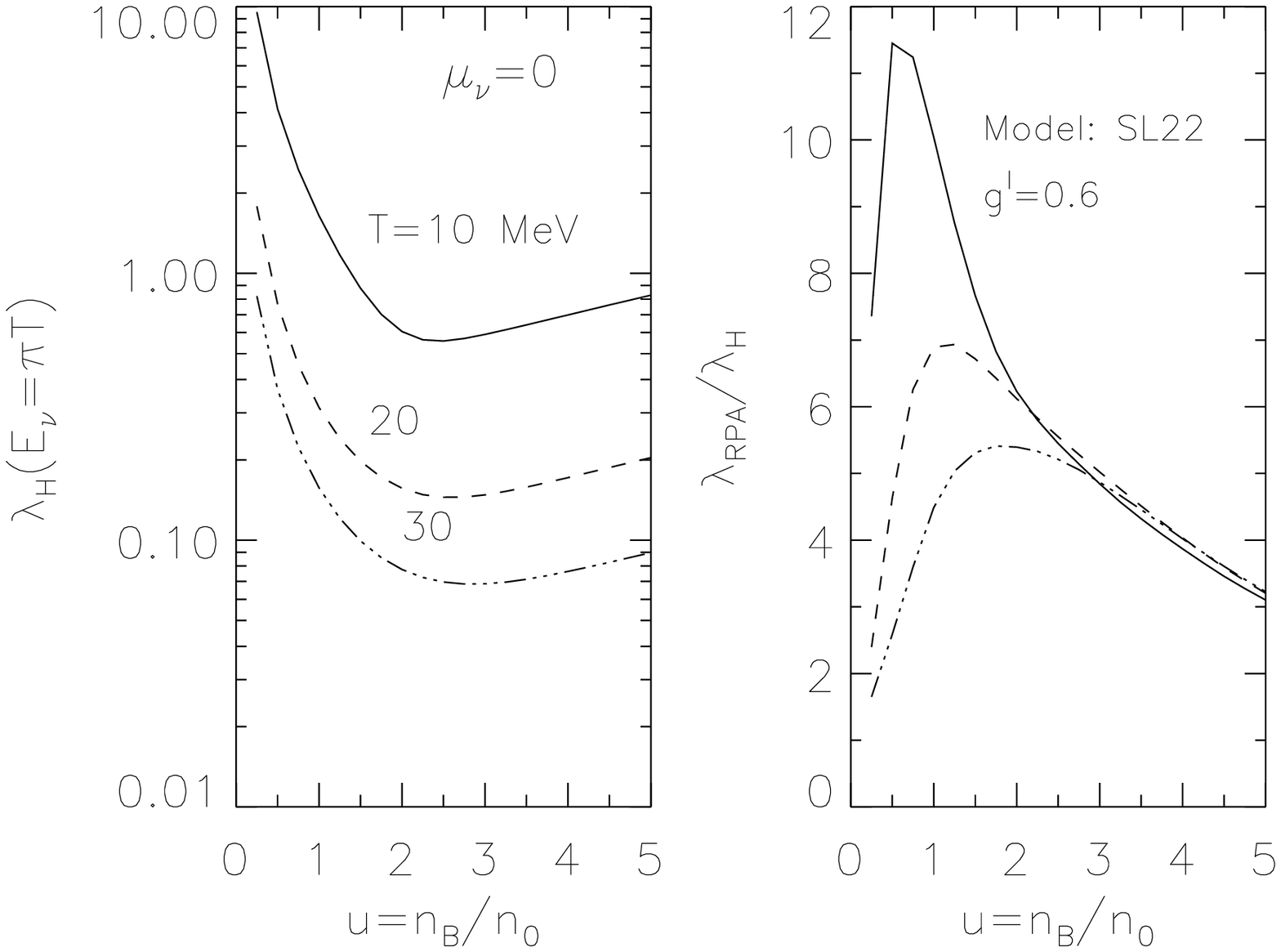}
\caption[]{}
\label{asig_y0}
\end{center}
\end{figure}

\newpage
\begin{figure}
\begin{center}
\epsfxsize=6.in
\epsfysize=7.in
\epsffile{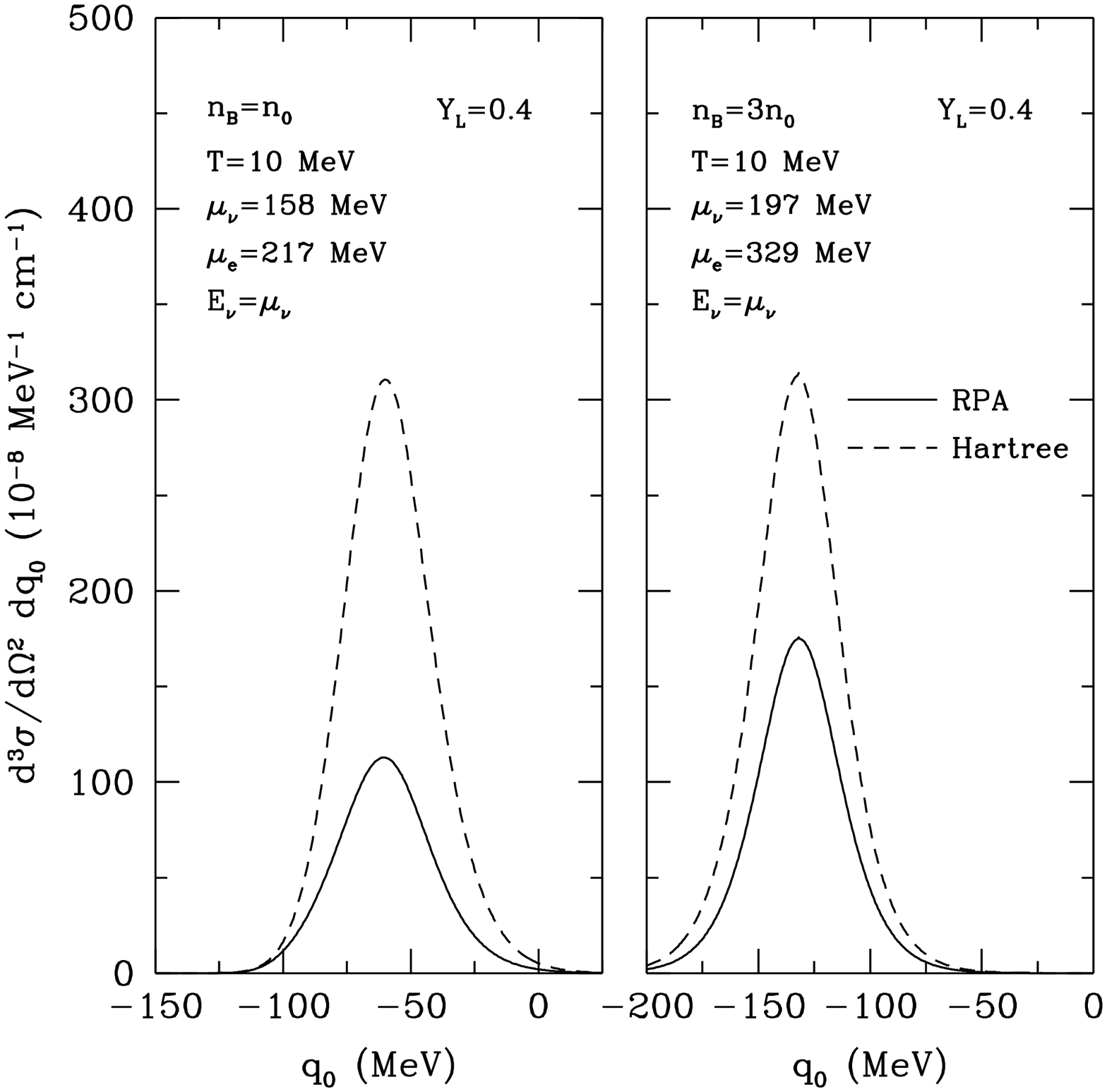}
\caption[]{}
\label{ardsig_y4}
\end{center}
\end{figure}

\newpage
\begin{figure}
\begin{center}
\epsfxsize=6.in
\epsfysize=7.in
\epsffile{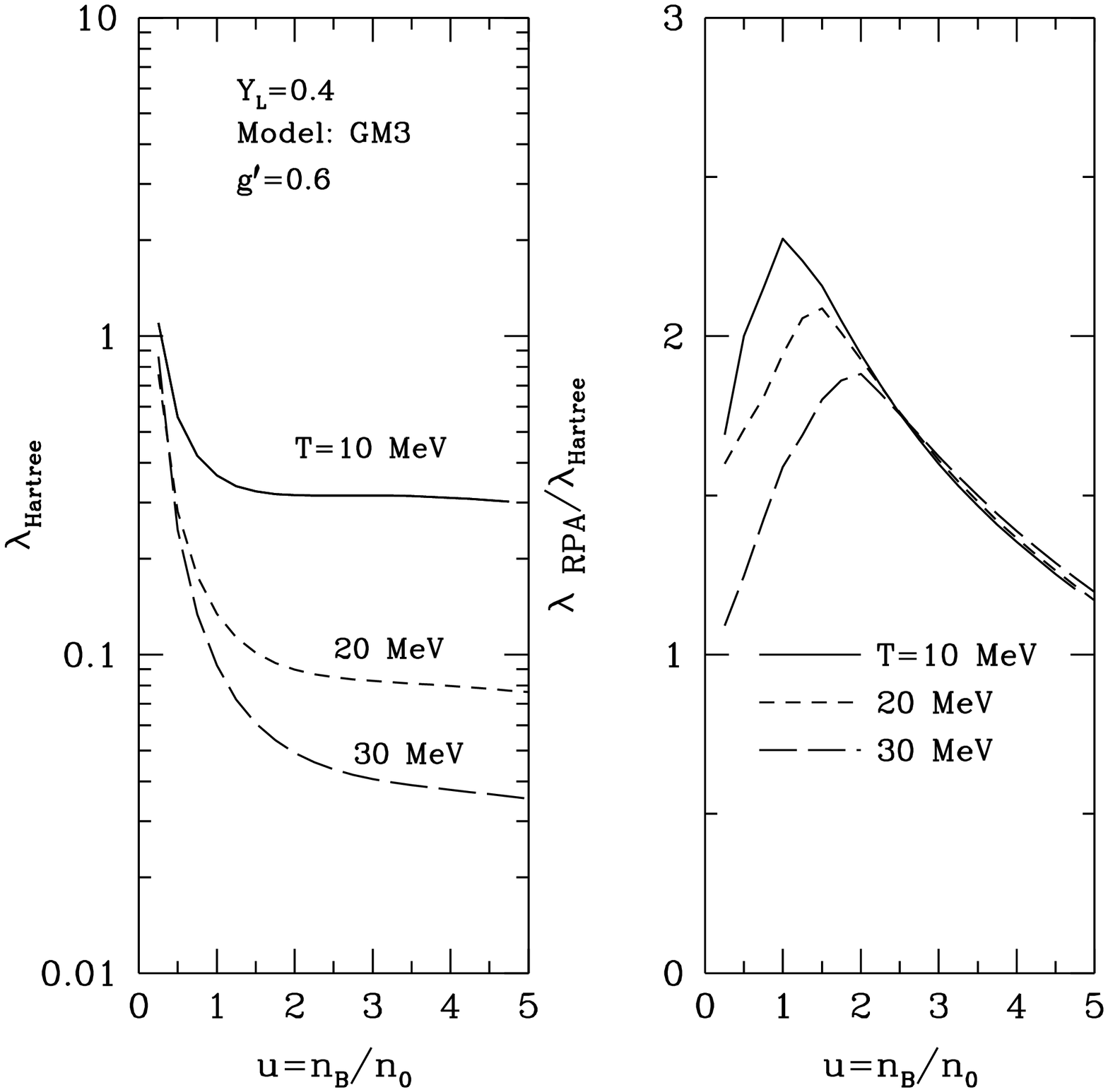}
\caption[]{}
\label{arsig_y4}
\end{center}
\end{figure}

\newpage
\begin{figure}
\begin{center}
\epsfxsize=6.in
\epsfysize=7.in
\epsffile{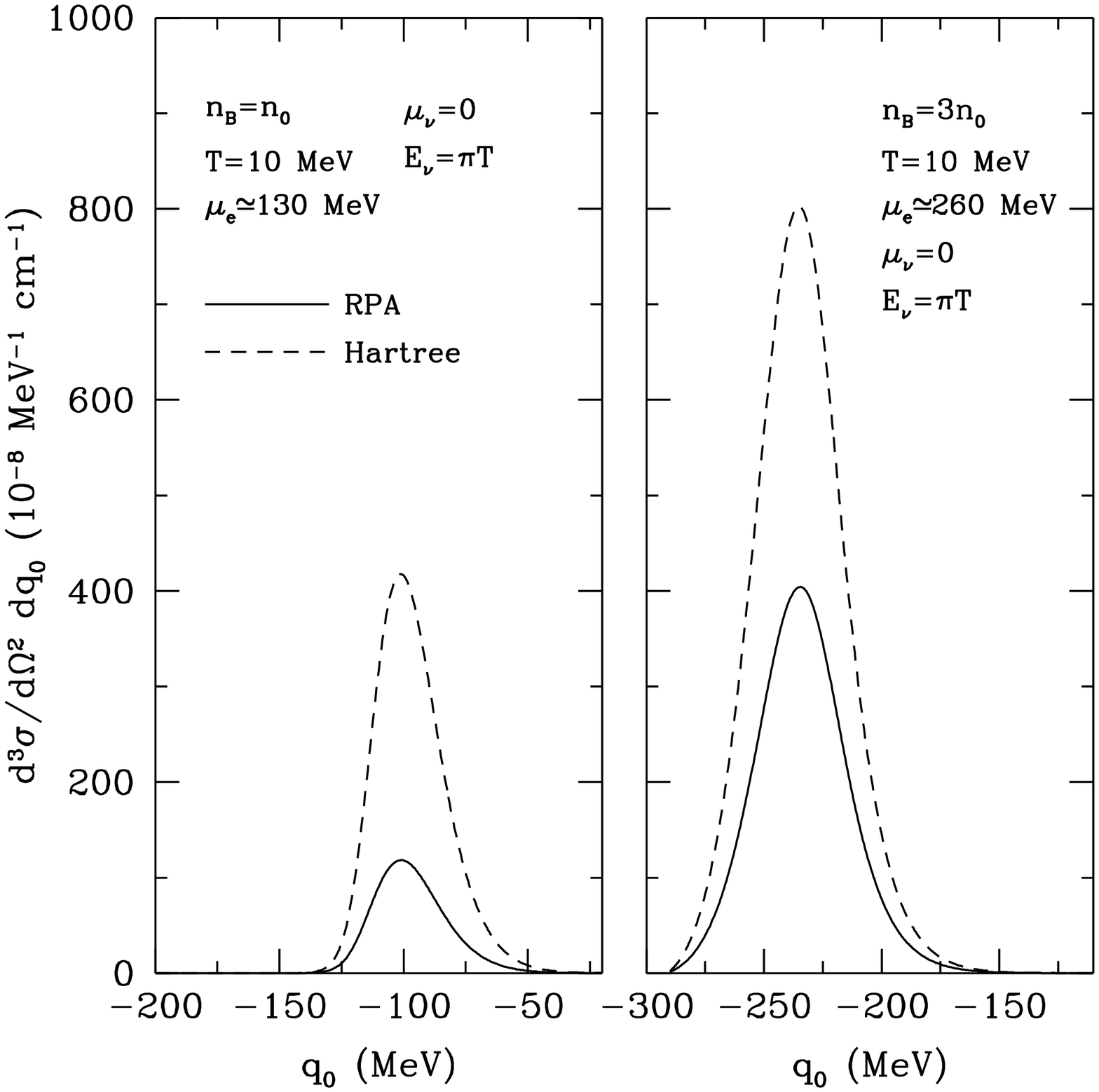}
\caption[]{}
\label{ardsig_y0}
\end{center}
\end{figure}

\newpage
\begin{figure}
\begin{center}
\epsfxsize=6.in
\epsfysize=7.in
\epsffile{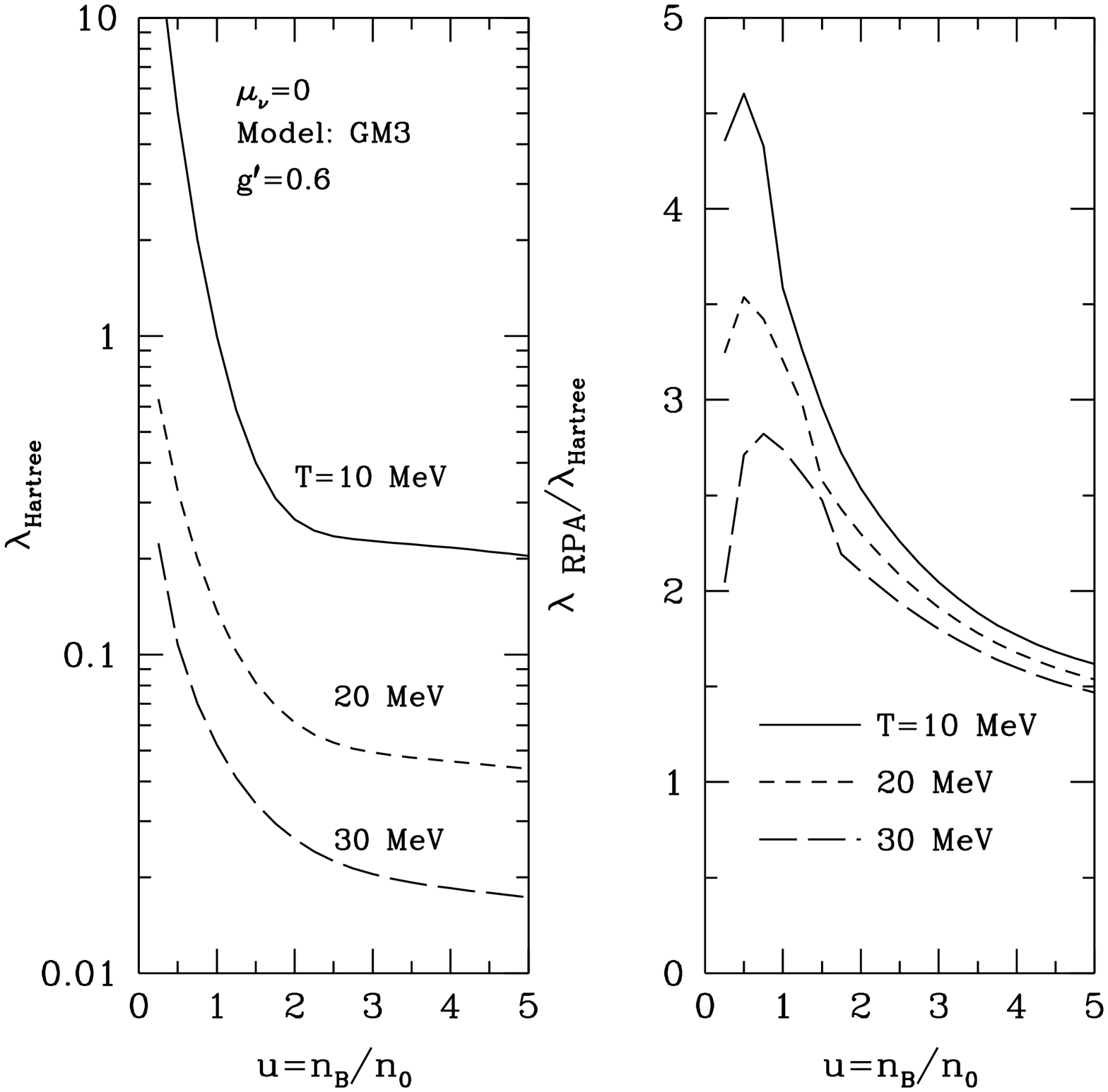}
\caption[]{}
\label{arsig_y0}
\end{center}
\end{figure}

\begin{figure}
\begin{center}
\epsfxsize=16cm
\epsfysize=20cm
\epsffile{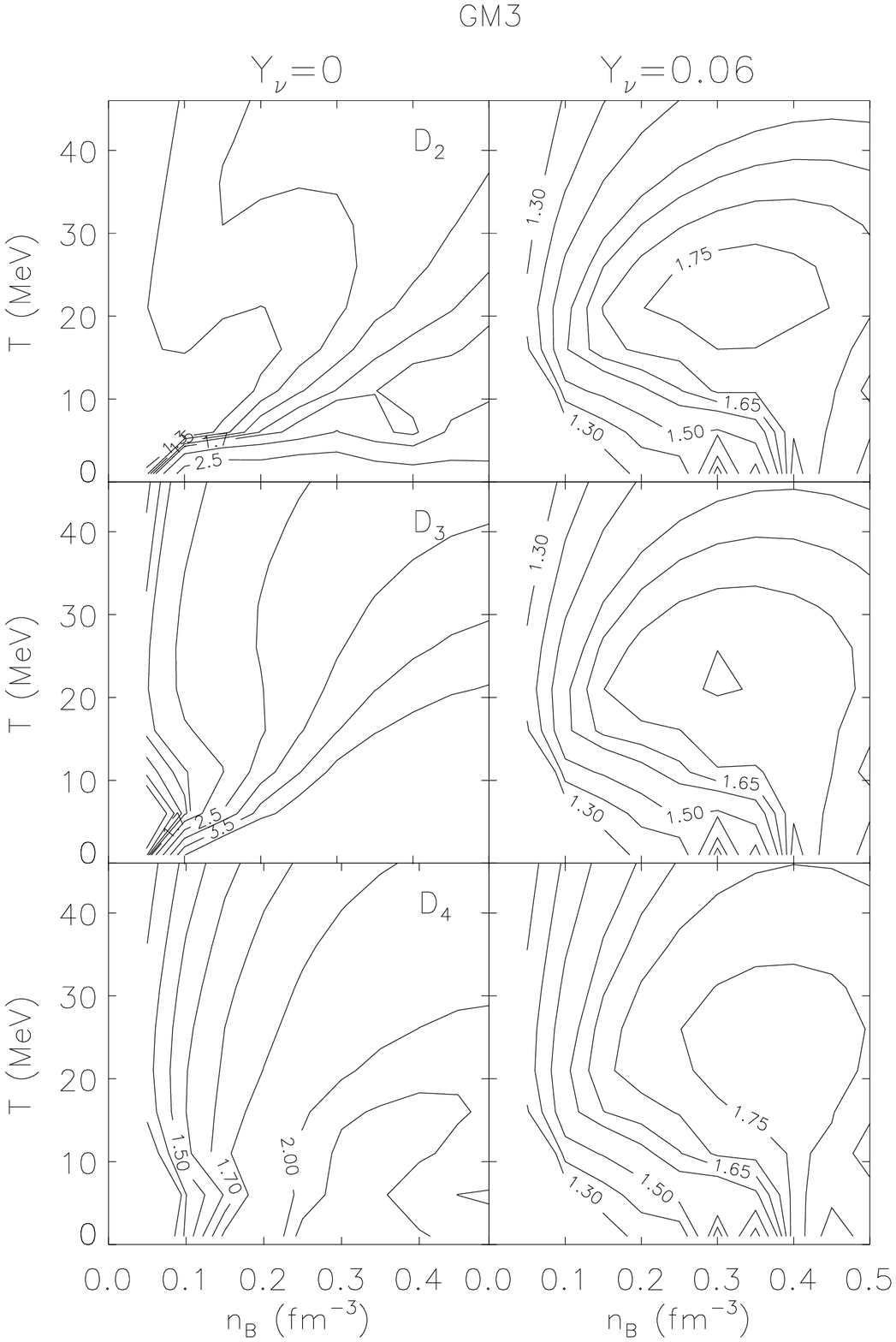}
\caption[]{}
\label{difc}
\end{center}
\end{figure}

\begin{figure}
\begin{center}
\epsfxsize=16cm
\epsfysize=20cm
\epsffile{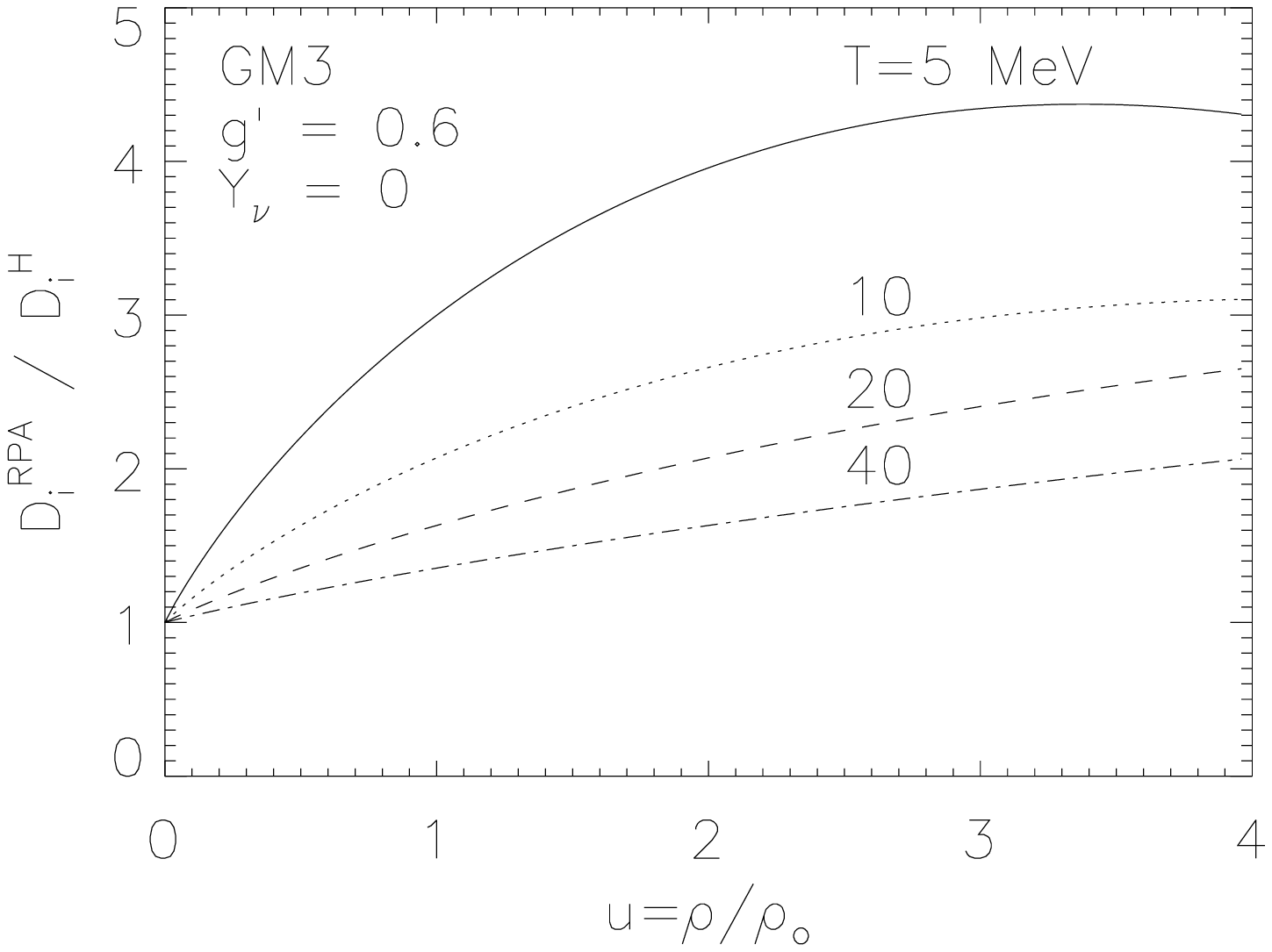}
\caption[]{}
\label{factor}
\end{center}
\end{figure}

\begin{figure}
\begin{center}
\epsfxsize=16cm
\epsfysize=20cm
\epsffile{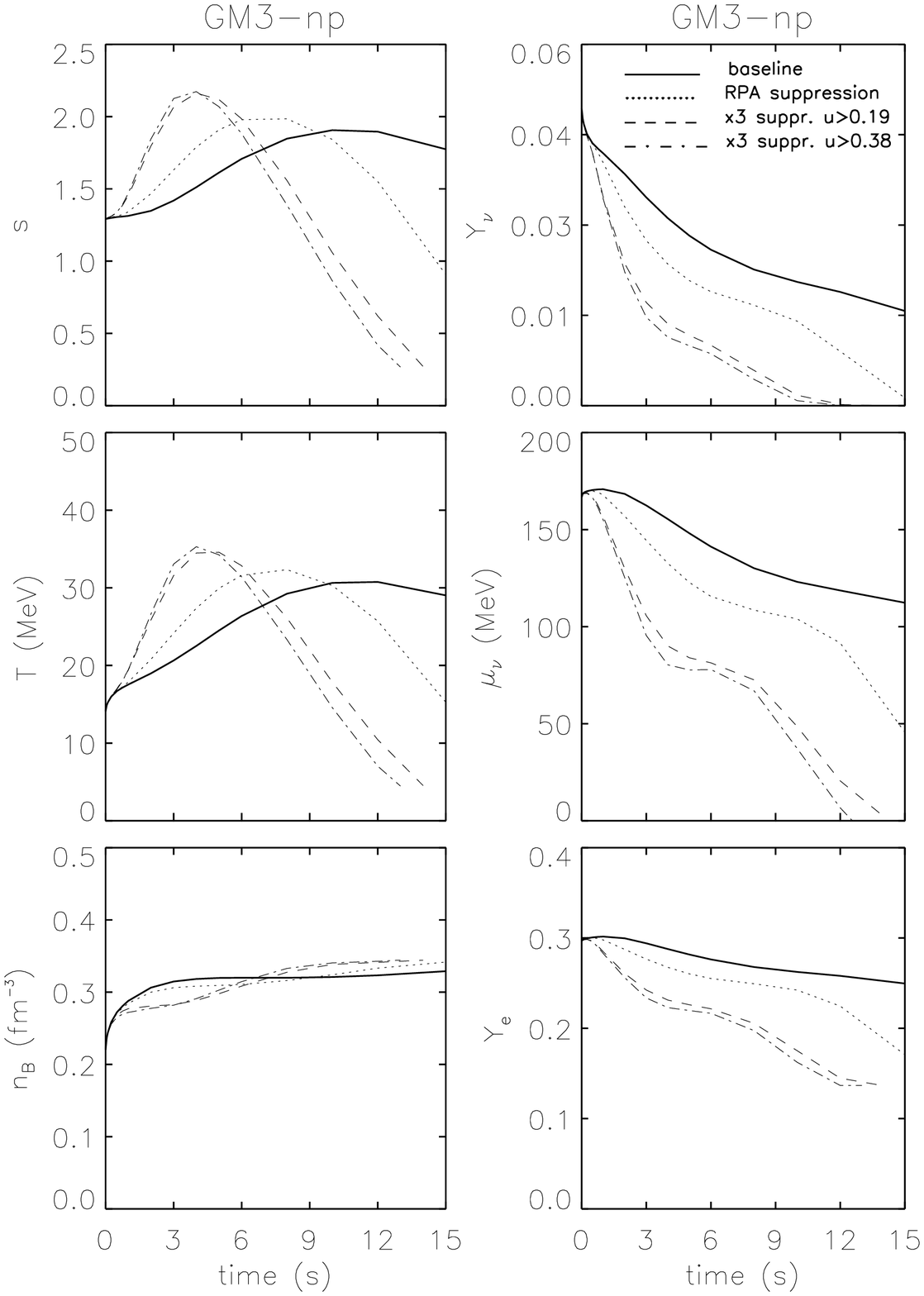}
\caption[]{}
\label{cquant}
\end{center}
\end{figure}

\begin{figure}
\begin{center}
\epsfxsize=16cm
\epsfysize=20cm
\epsffile{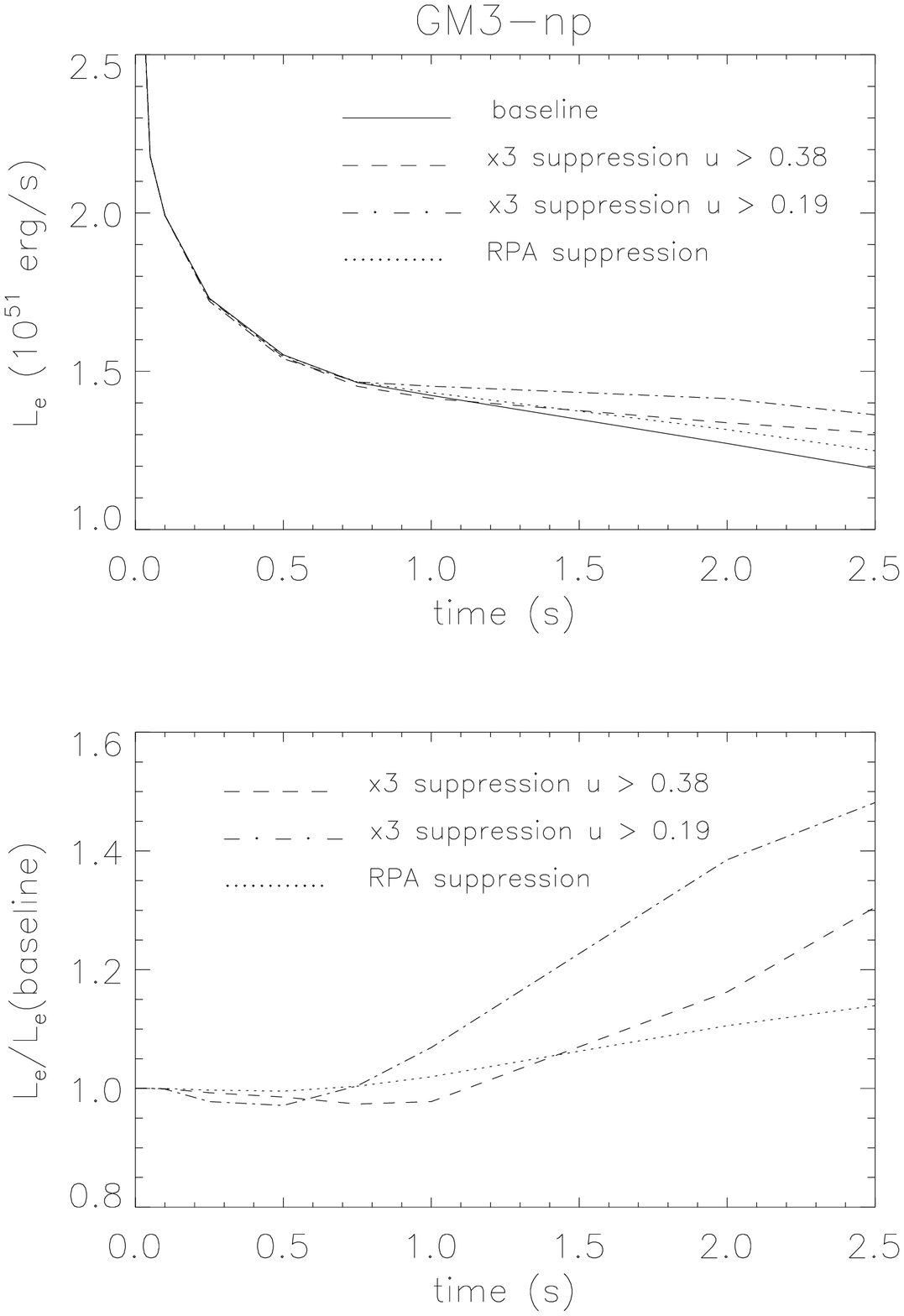}
\caption[]{}
\label{lums}
\end{center}
\end{figure}

\begin{figure}
\begin{center}
\epsfxsize=16cm
\epsfysize=20cm
\epsffile{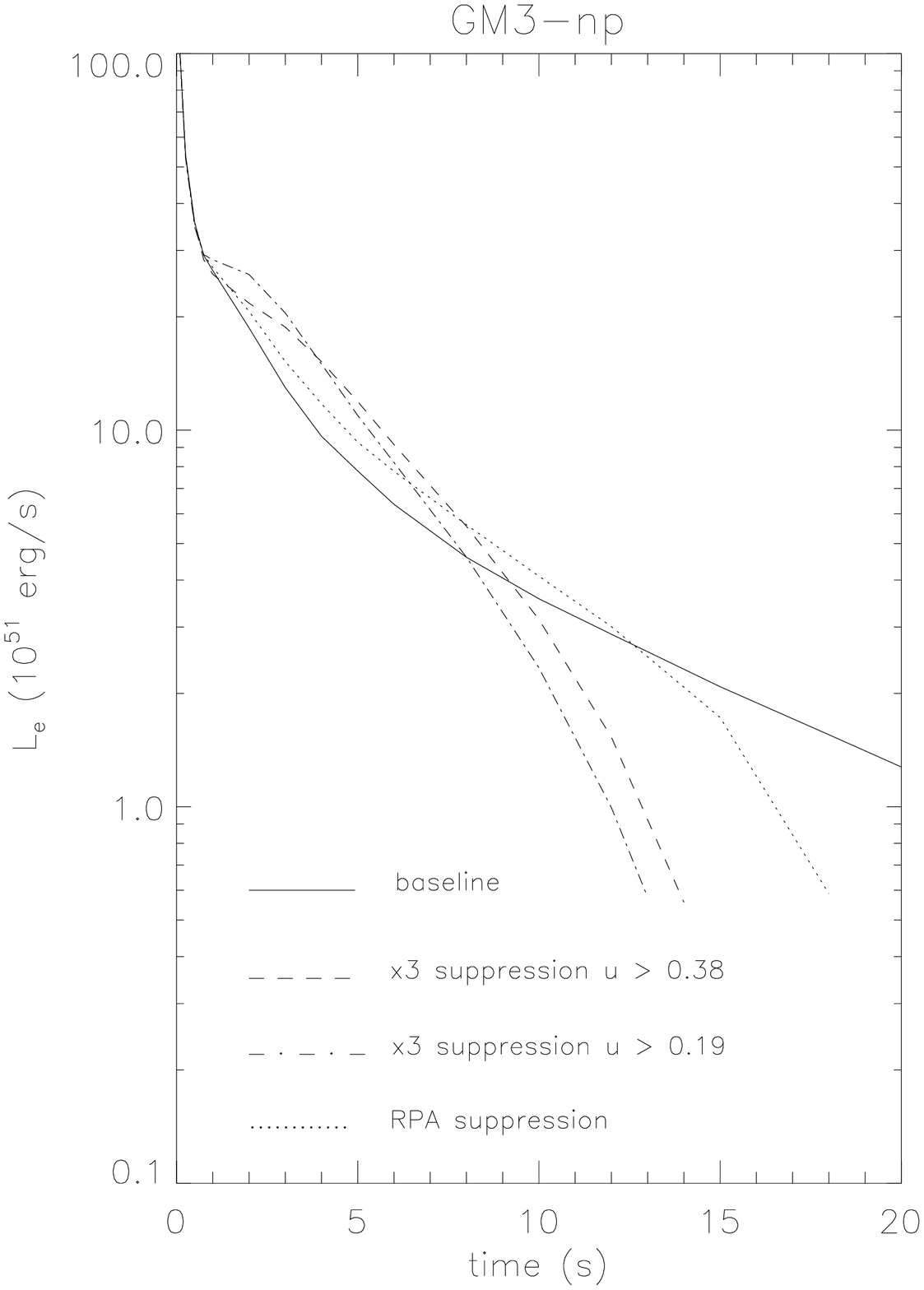}
\caption[]{}
\label{lumslong}
\end{center}
\end{figure}

\end{document}